\documentclass[epj,nopacs]{svjour}

\usepackage{cite,mcite}
\usepackage{xspace}
\usepackage{subfigure}
\usepackage{graphics}
\usepackage{graphicx}
\usepackage{url}
\usepackage{hyperref}
\def\ie{{\rm i.e.}}%
\def\eg{{\rm e.g.}}%
\def\ra{\ensuremath{\rightarrow}}%  "GOES TO" arrow.
\def\antibar#1{\ensuremath{#1\bar{#1}}}%
\def\ttbar{\antibar{t} \xspace }%
\def\Zboson{\ensuremath{Z} \xspace }
\def\Wboson{\ensuremath{W} \xspace }%
\def\epem{\ensuremath{e^+ e^-} \xspace }%
\def\Zee{\ensuremath{\Zboson\ra\ensuremath{e^+ e^-}} \xspace }
\def\Zmumu{\ensuremath{\Zboson\ra\mathrm{\mu^+ \mu^-}} \xspace }
\def\Ztautau{\ensuremath{\Zboson\ra\mathrm{\tau^+ \tau^-}} \xspace }
\def\Wenu{\ensuremath{\Wboson^{\pm}\ra e^{\pm}\nu_e} \xspace }
\def\Wmunu{\ensuremath{\Wboson^{\pm}\ra\mu^{\pm}\nu_{\mu}} \xspace }
\def\Wtaunu{\ensuremath{\Wboson^{\pm}\ra\tau^{\pm}\nu_{\tau}} \xspace }

\def\TeV{\ifmmode {\mathrm{\ Te\kern -0.1em V}}\else
                   \textrm{Te\kern -0.1em V}\fi \xspace}%
\def\GeV{\ifmmode {\mathrm{\ Ge\kern -0.1em V}}\else
                   \textrm{Ge\kern -0.1em V}\fi \xspace}%
\def\MeV{\ifmmode {\mathrm{\ Me\kern -0.1em V}}\else
                   \textrm{Me\kern -0.1em V}\fi \xspace}%

\begin{document}
\hugehead

\title{The ATLAS Simulation Infrastructure}
\author{
\begin{flushleft}
{\Large The ATLAS Collaboration}\\
\bigskip
G.~Aad$^{\rm 48}$,
B.~Abbott$^{\rm 111}$,
J.~Abdallah$^{\rm 11}$,
A.A.~Abdelalim$^{\rm 49}$,
A.~Abdesselam$^{\rm 118}$,
O.~Abdinov$^{\rm 10}$,
B.~Abi$^{\rm 112}$,
M.~Abolins$^{\rm 88}$,
H.~Abramowicz$^{\rm 152}$,
H.~Abreu$^{\rm 115}$
% ...
\\
\end{flushleft}
}
\bigskip
\institute{
\begin{flushleft}
$^{1}$ University at Albany, 1400 Washington Ave, Albany, NY 12222, United States of America\\
$^{2}$ University of Alberta, Department of Physics, Centre for Particle Physics, Edmonton, AB T6G 2G7, Canada\\
$^{3}$ Ankara University$^{(a)}$, Faculty of Sciences, Department of Physics, TR 061000 Tandogan, Ankara; Dumlupinar University$^{(b)}$, Faculty of Arts and Sciences, Department of Physics, Kutahya; Gazi University$^{(c)}$, Faculty of Arts and Sciences, Department of Physics, 06500, Teknikokullar, Ankara; TOBB University of Economics and Technology$^{(d)}$, Faculty of Arts and Sciences, Division of Physics, 06560, Sogutozu, Ankara; Turkish Atomic Energy Authority$^{(e)}$, 06530, Lodumlu, Ankara, Turkey\\
$^{4}$ LAPP, Universit\'e de Savoie, CNRS/IN2P3, Annecy-le-Vieux, France\\
$^{5}$ Argonne National Laboratory, High Energy Physics Division, 9700 S. Cass Avenue, Argonne IL 60439, United States of America\\
$^{6}$ University of Arizona, Department of Physics, Tucson, AZ 85721, United States of America\\
$^{7}$ The University of Texas at Arlington, Department of Physics, Box 19059, Arlington, TX 76019, United States of America\\
$^{8}$ University of Athens, Nuclear \& Particle Physics, Department of Physics, Panepistimiopouli, Zografou, GR 15771 Athens, Greece\\
$^{9}$ National Technical University of Athens, Physics Department, 9-Iroon Polytechniou, GR 15780 Zografou, Greece\\
$^{10}$ Institute of Physics, Azerbaijan Academy of Sciences, H. Javid Avenue 33, AZ 143 Baku, Azerbaijan\\
$^{11}$ Institut de F\'isica d'Altes Energies, IFAE, Edifici Cn, Universitat Aut\`onoma  de Barcelona,  ES - 08193 Bellaterra (Barcelona), Spain\\
$^{12}$ University of Belgrade$^{(a)}$, Institute of Physics, P.O. Box 57, 11001 Belgrade; Vinca Institute of Nuclear Sciences$^{(b)}$Mihajla Petrovica Alasa 12-14, 11001 Belgrade, Serbia\\
$^{13}$ University of Bergen, Department for Physics and Technology, Allegaten 55, NO - 5007 Bergen, Norway\\
$^{14}$ Lawrence Berkeley National Laboratory and University of California, Physics Division, MS50B-6227, 1 Cyclotron Road, Berkeley, CA 94720, United States of America\\
$^{15}$ Humboldt University, Institute of Physics, Berlin, Newtonstr. 15, D-12489 Berlin, Germany\\
$^{16}$ University of Bern,
Albert Einstein Center for Fundamental Physics,
Laboratory for High Energy Physics, Sidlerstrasse 5, CH - 3012 Bern, Switzerland\\
$^{17}$ University of Birmingham, School of Physics and Astronomy, Edgbaston, Birmingham B15 2TT, United Kingdom\\
$^{18}$ Bogazici University$^{(a)}$, Faculty of Sciences, Department of Physics, TR - 80815 Bebek-Istanbul; Dogus University$^{(b)}$, Faculty of Arts and Sciences, Department of Physics, 34722, Kadikoy, Istanbul; $^{(c)}$Gaziantep University, Faculty of Engineering, Department of Physics Engineering, 27310, Sehitkamil, Gaziantep, Turkey; Istanbul Technical University$^{(d)}$, Faculty of Arts and Sciences, Department of Physics, 34469, Maslak, Istanbul, Turkey\\
$^{19}$ INFN Sezione di Bologna$^{(a)}$; Universit\`a  di Bologna, Dipartimento di Fisica$^{(b)}$, viale C. Berti Pichat, 6/2, IT - 40127 Bologna, Italy\\
$^{20}$ University of Bonn, Physikalisches Institut, Nussallee 12, D - 53115 Bonn, Germany\\
$^{21}$ Boston University, Department of Physics,  590 Commonwealth Avenue, Boston, MA 02215, United States of America\\
$^{22}$ Brandeis University, Department of Physics, MS057, 415 South Street, Waltham, MA 02454, United States of America\\
$^{23}$ Universidade Federal do Rio De Janeiro, COPPE/EE/IF $^{(a)}$, Caixa Postal 68528, Ilha do Fundao, BR - 21945-970 Rio de Janeiro; $^{(b)}$Universidade de Sao Paulo, Instituto de Fisica, R.do Matao Trav. R.187, Sao Paulo - SP, 05508 - 900, Brazil\\
$^{24}$ Brookhaven National Laboratory, Physics Department, Bldg. 510A, Upton, NY 11973, United States of America\\
$^{25}$ National Institute of Physics and Nuclear Engineering$^{(a)}$, Bucharest-Magurele, Str. Atomistilor 407,  P.O. Box MG-6, R-077125, Romania; University Politehnica Bucharest$^{(b)}$, Rectorat - AN 001, 313 Splaiul Independentei, sector 6, 060042 Bucuresti; West University$^{(c)}$ in Timisoara, Bd. Vasile Parvan 4, Timisoara, Romania\\
$^{26}$ Universidad de Buenos Aires, FCEyN, Dto. Fisica, Pab I - C. Universitaria, 1428 Buenos Aires, Argentina\\
$^{27}$ University of Cambridge, Cavendish Laboratory, J J Thomson Avenue, Cambridge CB3 0HE, United Kingdom\\
$^{28}$ Carleton University, Department of Physics, 1125 Colonel By Drive,  Ottawa ON  K1S 5B6, Canada\\
$^{29}$ CERN, CH - 1211 Geneva 23, Switzerland\\
$^{30}$ University of Chicago, Enrico Fermi Institute, 5640 S. Ellis Avenue, Chicago, IL 60637, United States of America\\
$^{31}$ Pontificia Universidad Cat\'olica de Chile, Facultad de Fisica, Departamento de Fisica$^{(a)}$, Avda. Vicuna Mackenna 4860, San Joaquin, Santiago; Universidad T\'ecnica Federico Santa Mar\'ia, Departamento de F\'isica$^{(b)}$, Avda. Esp\~ana 1680, Casilla 110-V,  Valpara\'iso, Chile\\
$^{32}$ Institute of High Energy Physics, Chinese Academy of Sciences$^{(a)}$, P.O. Box 918, 19 Yuquan Road, Shijing Shan District, CN - Beijing 100049; University of Science \& Technology of China (USTC), Department of Modern Physics$^{(b)}$, Hefei, CN - Anhui 230026; Nanjing University, Department of Physics$^{(c)}$, 22 Hankou Road, Nanjing, 210093; Shandong University, High Energy Physics Group$^{(d)}$, Jinan, CN - Shandong 250100, China\\
$^{33}$ Laboratoire de Physique Corpusculaire, Clermont Universit\'e, Universit\'e Blaise Pascal, CNRS/IN2P3, FR - 63177 Aubiere Cedex, France\\
$^{34}$ Columbia University, Nevis Laboratory, 136 So. Broadway, Irvington, NY 10533, United States of America\\
$^{35}$ University of Copenhagen, Niels Bohr Institute, Blegdamsvej 17, DK - 2100 Kobenhavn 0, Denmark\\
$^{36}$ INFN Gruppo Collegato di Cosenza$^{(a)}$; Universit\`a della Calabria, Dipartimento di Fisica$^{(b)}$, IT-87036 Arcavacata di Rende, Italy\\
$^{37}$ Faculty of Physics and Applied Computer Science of the AGH-University of Science and Technology, (FPACS, AGH-UST), al. Mickiewicza 30, PL-30059 Cracow, Poland\\
$^{38}$ The Henryk Niewodniczanski Institute of Nuclear Physics, Polish Academy of Sciences, ul. Radzikowskiego 152, PL - 31342 Krakow, Poland\\
$^{39}$ Southern Methodist University, Physics Department, 106 Fondren Science Building, Dallas, TX 75275-0175, United States of America\\
$^{40}$ University of Texas at Dallas, 800 West Campbell Road, Richardson, TX 75080-3021, United States of America\\
$^{41}$ DESY, Notkestr. 85, D-22603 Hamburg and Platanenallee 6, D-15738 Zeuthen, Germany\\
$^{42}$ TU Dortmund, Experimentelle Physik IV, DE - 44221 Dortmund, Germany\\
$^{43}$ Technical University Dresden, Institut f\"{u}r Kern- und Teilchenphysik, Zellescher Weg 19, D-01069 Dresden, Germany\\
$^{44}$ Duke University, Department of Physics, Durham, NC 27708, United States of America\\
$^{45}$ University of Edinburgh, School of Physics \& Astronomy, James Clerk Maxwell Building, The Kings Buildings, Mayfield Road, Edinburgh EH9 3JZ, United Kingdom\\
$^{46}$ Fachhochschule Wiener Neustadt; Johannes Gutenbergstrasse 3 AT - 2700 Wiener Neustadt, Austria\\
$^{47}$ INFN Laboratori Nazionali di Frascati, via Enrico Fermi 40, IT-00044 Frascati, Italy\\
$^{48}$ Albert-Ludwigs-Universit\"{a}t, Fakult\"{a}t f\"{u}r Mathematik und Physik, Hermann-Herder Str. 3, D - 79104 Freiburg i.Br., Germany\\
$^{49}$ Universit\'e de Gen\`eve, Section de Physique, 24 rue Ernest Ansermet, CH - 1211 Geneve 4, Switzerland\\
$^{50}$ INFN Sezione di Genova$^{(a)}$; Universit\`a  di Genova, Dipartimento di Fisica$^{(b)}$, via Dodecaneso 33, IT - 16146 Genova, Italy\\
$^{51}$ Institute of Physics of the Georgian Academy of Sciences, 6 Tamarashvili St., GE - 380077 Tbilisi; Tbilisi State University, HEP Institute, University St. 9, GE - 380086 Tbilisi, Georgia\\
$^{52}$ Justus-Liebig-Universit\"{a}t Giessen, II Physikalisches Institut, Heinrich-Buff Ring 16,  D-35392 Giessen, Germany\\
$^{53}$ University of Glasgow, Department of Physics and Astronomy, Glasgow G12 8QQ, United Kingdom\\
$^{54}$ Georg-August-Universit\"{a}t, II. Physikalisches Institut, Friedrich-Hund Platz 1, D-37077 G\"{o}ttingen, Germany\\
$^{55}$ Laboratoire de Physique Subatomique et de Cosmologie, CNRS/IN2P3, Universit\'e Joseph Fourier, INPG, 53 avenue des Martyrs, FR - 38026 Grenoble Cedex, France\\
$^{56}$ Hampton University, Department of Physics, Hampton, VA 23668, United States of America\\
$^{57}$ Harvard University, Laboratory for Particle Physics and Cosmology, 18 Hammond Street, Cambridge, MA 02138, United States of America\\
$^{58}$ Ruprecht-Karls-Universit\"{a}t Heidelberg: Kirchhoff-Institut f\"{u}r Physik$^{(a)}$, Im Neuenheimer Feld 227, D-69120 Heidelberg; Physikalisches Institut$^{(b)}$, Philosophenweg 12, D-69120 Heidelberg; ZITI Ruprecht-Karls-University Heidelberg$^{(c)}$, Lehrstuhl f\"{u}r Informatik V, B6, 23-29, DE - 68131 Mannheim, Germany\\
$^{59}$ Hiroshima University, Faculty of Science, 1-3-1 Kagamiyama, Higashihiroshima-shi, JP - Hiroshima 739-8526, Japan\\
$^{60}$ Hiroshima Institute of Technology, Faculty of Applied Information Science, 2-1-1 Miyake Saeki-ku, Hiroshima-shi, JP - Hiroshima 731-5193, Japan\\
$^{61}$ Indiana University, Department of Physics,  Swain Hall West 117, Bloomington, IN 47405-7105, United States of America\\
$^{62}$ Institut f\"{u}r Astro- und Teilchenphysik, Technikerstrasse 25, A - 6020 Innsbruck, Austria\\
$^{63}$ University of Iowa, 203 Van Allen Hall, Iowa City, IA 52242-1479, United States of America\\
$^{64}$ Iowa State University, Department of Physics and Astronomy, Ames High Energy Physics Group,  Ames, IA 50011-3160, United States of America\\
$^{65}$ Joint Institute for Nuclear Research, JINR Dubna, RU - 141 980 Moscow Region, Russia\\
$^{66}$ KEK, High Energy Accelerator Research Organization, 1-1 Oho, Tsukuba-shi, Ibaraki-ken 305-0801, Japan\\
$^{67}$ Kobe University, Graduate School of Science, 1-1 Rokkodai-cho, Nada-ku, JP Kobe 657-8501, Japan\\
$^{68}$ Kyoto University, Faculty of Science, Oiwake-cho, Kitashirakawa, Sakyou-ku, Kyoto-shi, JP - Kyoto 606-8502, Japan\\
$^{69}$ Kyoto University of Education, 1 Fukakusa, Fujimori, fushimi-ku, Kyoto-shi, JP - Kyoto 612-8522, Japan\\
$^{70}$ Universidad Nacional de La Plata, FCE, Departamento de F\'{i}sica, IFLP (CONICET-UNLP),   C.C. 67,  1900 La Plata, Argentina\\
$^{71}$ Lancaster University, Physics Department, Lancaster LA1 4YB, United Kingdom\\
$^{72}$ INFN Sezione di Lecce$^{(a)}$; Universit\`a  del Salento, Dipartimento di Fisica$^{(b)}$Via Arnesano IT - 73100 Lecce, Italy\\
$^{73}$ University of Liverpool, Oliver Lodge Laboratory, P.O. Box 147, Oxford Street,  Liverpool L69 3BX, United Kingdom\\
$^{74}$ Jo\v{z}ef Stefan Institute and University of Ljubljana, Department  of Physics, SI-1000 Ljubljana, Slovenia\\
$^{75}$ Queen Mary University of London, Department of Physics, Mile End Road, London E1 4NS, United Kingdom\\
$^{76}$ Royal Holloway, University of London, Department of Physics, Egham Hill, Egham, Surrey TW20 0EX, United Kingdom\\
$^{77}$ University College London, Department of Physics and Astronomy, Gower Street, London WC1E 6BT, United Kingdom\\
$^{78}$ Laboratoire de Physique Nucl\'eaire et de Hautes Energies, Universit\'e Pierre et Marie Curie (Paris 6), Universit\'e Denis Diderot (Paris-7), CNRS/IN2P3, Tour 33, 4 place Jussieu, FR - 75252 Paris Cedex 05, France\\
$^{79}$ Lunds universitet, Naturvetenskapliga fakulteten, Fysiska institutionen, Box 118, SE - 221 00 Lund, Sweden\\
$^{80}$ Universidad Autonoma de Madrid, Facultad de Ciencias, Departamento de Fisica Teorica, ES - 28049 Madrid, Spain\\
$^{81}$ Universit\"{a}t Mainz, Institut f\"{u}r Physik, Staudinger Weg 7, DE - 55099 Mainz, Germany\\
$^{82}$ University of Manchester, School of Physics and Astronomy, Manchester M13 9PL, United Kingdom\\
$^{83}$ CPPM, Aix-Marseille Universit\'e, CNRS/IN2P3, Marseille, France\\
$^{84}$ University of Massachusetts, Department of Physics, 710 North Pleasant Street, Amherst, MA 01003, United States of America\\
$^{85}$ McGill University, High Energy Physics Group, 3600 University Street, Montreal, Quebec H3A 2T8, Canada\\
$^{86}$ University of Melbourne, School of Physics, AU - Parkville, Victoria 3010, Australia\\
$^{87}$ The University of Michigan, Department of Physics, 2477 Randall Laboratory, 500 East University, Ann Arbor, MI 48109-1120, United States of America\\
$^{88}$ Michigan State University, Department of Physics and Astronomy, High Energy Physics Group, East Lansing, MI 48824-2320, United States of America\\
$^{89}$ INFN Sezione di Milano$^{(a)}$; Universit\`a  di Milano, Dipartimento di Fisica$^{(b)}$, via Celoria 16, IT - 20133 Milano, Italy\\
$^{90}$ B.I. Stepanov Institute of Physics, National Academy of Sciences of Belarus, Independence Avenue 68, Minsk 220072, Republic of Belarus\\
$^{91}$ National Scientific \& Educational Centre for Particle \& High Energy Physics, NC PHEP BSU, M. Bogdanovich St. 153, Minsk 220040, Republic of Belarus\\
$^{92}$ Massachusetts Institute of Technology, Department of Physics, Room 24-516, Cambridge, MA 02139, United States of America\\
$^{93}$ University of Montreal, Group of Particle Physics, C.P. 6128, Succursale Centre-Ville, Montreal, Quebec, H3C 3J7  , Canada\\
$^{94}$ P.N. Lebedev Institute of Physics, Academy of Sciences, Leninsky pr. 53, RU - 117 924 Moscow, Russia\\
$^{95}$ Institute for Theoretical and Experimental Physics (ITEP), B. Cheremushkinskaya ul. 25, RU 117 218 Moscow, Russia\\
$^{96}$ Moscow Engineering \& Physics Institute (MEPhI), Kashirskoe Shosse 31, RU - 115409 Moscow, Russia\\
$^{97}$ Lomonosov Moscow State University Skobeltsyn Institute of Nuclear Physics (MSU SINP), 1(2), Leninskie gory, GSP-1, Moscow 119991 Russian Federation, Russia\\
$^{98}$ Ludwig-Maximilians-Universit\"at M\"unchen, Fakult\"at f\"ur Physik, Am Coulombwall 1,  DE - 85748 Garching, Germany\\
$^{99}$ Max-Planck-Institut f\"ur Physik, (Werner-Heisenberg-Institut), F\"ohringer Ring 6, 80805 M\"unchen, Germany\\
$^{100}$ Nagasaki Institute of Applied Science, 536 Aba-machi, JP Nagasaki 851-0193, Japan\\
$^{101}$ Nagoya University, Graduate School of Science, Furo-Cho, Chikusa-ku, Nagoya, 464-8602, Japan\\
$^{102}$ INFN Sezione di Napoli$^{(a)}$; Universit\`a  di Napoli, Dipartimento di Scienze Fisiche$^{(b)}$, Complesso Universitario di Monte Sant'Angelo, via Cinthia, IT - 80126 Napoli, Italy\\
$^{103}$  University of New Mexico, Department of Physics and Astronomy, MSC07 4220, Albuquerque, NM 87131 USA, United States of America\\
$^{104}$ Radboud University Nijmegen/NIKHEF, Department of Experimental High Energy Physics, Heyendaalseweg 135, NL-6525 AJ, Nijmegen, Netherlands\\
$^{105}$ Nikhef National Institute for Subatomic Physics, and University of Amsterdam, Science Park 105, 1098 XG Amsterdam, Netherlands\\
$^{106}$ $^{(a)}$DeKalb, Illinois  60115, United States of America\\
$^{107}$ Budker Institute of Nuclear Physics (BINP), RU - Novosibirsk 630 090, Russia\\
$^{108}$ New York University, Department of Physics, 4 Washington Place, New York NY 10003, USA, United States of America\\
$^{109}$ Ohio State University, 191 West Woodruff Ave, Columbus, OH 43210-1117, United States of America\\
$^{110}$ Okayama University, Faculty of Science, Tsushimanaka 3-1-1, Okayama 700-8530, Japan\\
$^{111}$ University of Oklahoma, Homer L. Dodge Department of Physics and Astronomy, 440 West Brooks, Room 100, Norman, OK 73019-0225, United States of America\\
$^{112}$ Oklahoma State University, Department of Physics, 145 Physical Sciences Building, Stillwater, OK 74078-3072, United States of America\\
$^{113}$ Palack\'y University, 17.listopadu 50a,  772 07  Olomouc, Czech Republic\\
$^{114}$ University of Oregon, Center for High Energy Physics, Eugene, OR 97403-1274, United States of America\\
$^{115}$ LAL, Univ. Paris-Sud, IN2P3/CNRS, Orsay, France\\
$^{116}$ Osaka University, Graduate School of Science, Machikaneyama-machi 1-1, Toyonaka, Osaka 560-0043, Japan\\
$^{117}$ University of Oslo, Department of Physics, P.O. Box 1048,  Blindern, NO - 0316 Oslo 3, Norway\\
$^{118}$ Oxford University, Department of Physics, Denys Wilkinson Building, Keble Road, Oxford OX1 3RH, United Kingdom\\
$^{119}$ INFN Sezione di Pavia$^{(a)}$; Universit\`a  di Pavia, Dipartimento di Fisica Nucleare e Teorica$^{(b)}$, Via Bassi 6, IT-27100 Pavia, Italy\\
$^{120}$ University of Pennsylvania, Department of Physics, High Energy Physics Group, 209 S. 33rd Street, Philadelphia, PA 19104, United States of America\\
$^{121}$ Petersburg Nuclear Physics Institute, RU - 188 300 Gatchina, Russia\\
$^{122}$ INFN Sezione di Pisa$^{(a)}$; Universit\`a   di Pisa, Dipartimento di Fisica E. Fermi$^{(b)}$, Largo B. Pontecorvo 3, IT - 56127 Pisa, Italy\\
$^{123}$ University of Pittsburgh, Department of Physics and Astronomy, 3941 O'Hara Street, Pittsburgh, PA 15260, United States of America\\
$^{124}$ Laboratorio de Instrumentacao e Fisica Experimental de Particulas - LIP$^{(a)}$, Avenida Elias Garcia 14-1, PT - 1000-149 Lisboa, Portugal; Universidad de Granada, Departamento de Fisica Teorica y del Cosmos and CAFPE$^{(b)}$, E-18071 Granada, Spain\\
$^{125}$ Institute of Physics, Academy of Sciences of the Czech Republic, Na Slovance 2, CZ - 18221 Praha 8, Czech Republic\\
$^{126}$ Charles University in Prague, Faculty of Mathematics and Physics, Institute of Particle and Nuclear Physics, V Holesovickach 2, CZ - 18000 Praha 8, Czech Republic\\
$^{127}$ Czech Technical University in Prague, Zikova 4, CZ - 166 35 Praha 6, Czech Republic\\
$^{128}$ State Research Center Institute for High Energy Physics, Moscow Region, 142281, Protvino, Pobeda street, 1, Russia\\
$^{129}$ Rutherford Appleton Laboratory, Science and Technology Facilities Council, Harwell Science and Innovation Campus, Didcot OX11 0QX, United Kingdom\\
$^{130}$ University of Regina, Physics Department, Canada\\
$^{131}$ Ritsumeikan University, Noji Higashi 1 chome 1-1, JP - Kusatsu, Shiga 525-8577, Japan\\
$^{132}$ INFN Sezione di Roma I$^{(a)}$; Universit\`a  La Sapienza, Dipartimento di Fisica$^{(b)}$, Piazzale A. Moro 2, IT- 00185 Roma, Italy\\
$^{133}$ INFN Sezione di Roma Tor Vergata$^{(a)}$; Universit\`a di Roma Tor Vergata, Dipartimento di Fisica$^{(b)}$ , via della Ricerca Scientifica, IT-00133 Roma, Italy\\
$^{134}$ INFN Sezione di  Roma Tre$^{(a)}$; Universit\`a Roma Tre, Dipartimento di Fisica$^{(b)}$, via della Vasca Navale 84, IT-00146  Roma, Italy\\
$^{135}$ R\'eseau Universitaire de Physique des Hautes Energies (RUPHE): Universit\'e Hassan II, Facult\'e des Sciences Ain Chock$^{(a)}$, B.P. 5366, MA - Casablanca; Centre National de l'Energie des Sciences Techniques Nucleaires (CNESTEN)$^{(b)}$, B.P. 1382 R.P. 10001 Rabat 10001; Universit\'e Mohamed Premier$^{(c)}$, LPTPM, Facult\'e des Sciences, B.P.717. Bd. Mohamed VI, 60000, Oujda ; Universit\'e Mohammed V, Facult\'e des Sciences$^{(d)}$4 Avenue Ibn Battouta, BP 1014 RP, 10000 Rabat, Morocco\\
$^{136}$ CEA, DSM/IRFU, Centre d'Etudes de Saclay, FR - 91191 Gif-sur-Yvette, France\\
$^{137}$ University of California Santa Cruz, Santa Cruz Institute for Particle Physics (SCIPP), Santa Cruz, CA 95064, United States of America\\
$^{138}$ University of Washington, Seattle, Department of Physics, Box 351560, Seattle, WA 98195-1560, United States of America\\
$^{139}$ University of Sheffield, Department of Physics \& Astronomy, Hounsfield Road, Sheffield S3 7RH, United Kingdom\\
$^{140}$ Shinshu University, Department of Physics, Faculty of Science, 3-1-1 Asahi, Matsumoto-shi, JP - Nagano 390-8621, Japan\\
$^{141}$ Universit\"{a}t Siegen, Fachbereich Physik, D 57068 Siegen, Germany\\
$^{142}$ Simon Fraser University, Department of Physics, 8888 University Drive, CA - Burnaby, BC V5A 1S6, Canada\\
$^{143}$ SLAC National Accelerator Laboratory, Stanford, California 94309, United States of America\\
$^{144}$ Comenius University, Faculty of Mathematics, Physics \& Informatics$^{(a)}$, Mlynska dolina F2, SK - 84248 Bratislava; Institute of Experimental Physics of the Slovak Academy of Sciences, Dept. of Subnuclear Physics$^{(b)}$, Watsonova 47, SK - 04353 Kosice, Slovak Republic\\
$^{145}$ Stockholm University: Department of Physics$^{(a)}$; The Oskar Klein Centre$^{(b)}$, AlbaNova, SE - 106 91 Stockholm, Sweden\\
$^{146}$ Royal Institute of Technology (KTH), Physics Department, SE - 106 91 Stockholm, Sweden\\
$^{147}$ Stony Brook University, Department of Physics and Astronomy, Nicolls Road, Stony Brook, NY 11794-3800, United States of America\\
$^{148}$ University of Sussex, Department of Physics and Astronomy
Pevensey 2 Building, Falmer, Brighton BN1 9QH, United Kingdom\\
$^{149}$ University of Sydney, School of Physics, AU - Sydney NSW 2006, Australia\\
$^{150}$ Insitute of Physics, Academia Sinica, TW - Taipei 11529, Taiwan\\
$^{151}$ Technion, Israel Inst. of Technology, Department of Physics, Technion City, IL - Haifa 32000, Israel\\
$^{152}$ Tel Aviv University, Raymond and Beverly Sackler School of Physics and Astronomy, Ramat Aviv, IL - Tel Aviv 69978, Israel\\
$^{153}$ Aristotle University of Thessaloniki, Faculty of Science, Department of Physics, Division of Nuclear \& Particle Physics, University Campus, GR - 54124, Thessaloniki, Greece\\
$^{154}$ The University of Tokyo, International Center for Elementary Particle Physics and Department of Physics, 7-3-1 Hongo, Bunkyo-ku, JP - Tokyo 113-0033, Japan\\
$^{155}$ Tokyo Metropolitan University, Graduate School of Science and Technology, 1-1 Minami-Osawa, Hachioji, Tokyo 192-0397, Japan\\
$^{156}$ Tokyo Institute of Technology, 2-12-1-H-34 O-Okayama, Meguro, Tokyo 152-8551, Japan\\
$^{157}$ University of Toronto, Department of Physics, 60 Saint George Street, Toronto M5S 1A7, Ontario, Canada\\
$^{158}$ TRIUMF$^{(a)}$, 4004 Wesbrook Mall, Vancouver, B.C. V6T 2A3; $^{(b)}$York University, Department of Physics and Astronomy, 4700 Keele St., Toronto, Ontario, M3J 1P3, Canada\\
$^{159}$ University of Tsukuba, Institute of Pure and Applied Sciences, 1-1-1 Tennoudai, Tsukuba-shi, JP - Ibaraki 305-8571, Japan\\
$^{160}$ Tufts University, Science \& Technology Center, 4 Colby Street, Medford, MA 02155, United States of America\\
$^{161}$ Universidad Antonio Narino, Centro de Investigaciones, Cra 3 Este No.47A-15, Bogota, Colombia\\
$^{162}$ University of California, Irvine, Department of Physics \& Astronomy, CA 92697-4575, United States of America\\
$^{163}$ INFN Gruppo Collegato di Udine$^{(a)}$; ICTP$^{(b)}$, Strada Costiera 11, IT-34014, Trieste; Universit\`a  di Udine, Dipartimento di Fisica$^{(c)}$, via delle Scienze 208, IT - 33100 Udine, Italy\\
$^{164}$ University of Illinois, Department of Physics, 1110 West Green Street, Urbana, Illinois 61801, United States of America\\
$^{165}$ University of Uppsala, Department of Physics and Astronomy, P.O. Box 516, SE -751 20 Uppsala, Sweden\\
$^{166}$ Instituto de F\'isica Corpuscular (IFIC) Centro Mixto UVEG-CSIC, Apdo. 22085  ES-46071 Valencia, Dept. F\'isica At. Mol. y Nuclear; Univ. of Valencia, and Instituto de Microelectr\'onica de Barcelona (IMB-CNM-CSIC) 08193 Bellaterra Barcelona, Spain\\
$^{167}$ University of British Columbia, Department of Physics, 6224 Agricultural Road, CA - Vancouver, B.C. V6T 1Z1, Canada\\
$^{168}$ University of Victoria, Department of Physics and Astronomy, P.O. Box 3055, Victoria B.C., V8W 3P6, Canada\\
$^{169}$ Waseda University, WISE, 3-4-1 Okubo, Shinjuku-ku, Tokyo, 169-8555, Japan\\
$^{170}$ The Weizmann Institute of Science, Department of Particle Physics, P.O. Box 26, IL - 76100 Rehovot, Israel\\
$^{171}$ University of Wisconsin, Department of Physics, 1150 University Avenue, WI 53706 Madison, Wisconsin, United States of America\\
$^{172}$ Julius-Maximilians-University of W\"urzburg, Physikalisches Institute, Am Hubland, 97074 W\"urzburg, Germany\\
$^{173}$ Bergische Universit\"{a}t, Fachbereich C, Physik, Postfach 100127, Gauss-Strasse 20, D- 42097 Wuppertal, Germany\\
$^{174}$ Yale University, Department of Physics, PO Box 208121, New Haven CT, 06520-8121, United States of America\\
$^{175}$ Yerevan Physics Institute, Alikhanian Brothers Street 2, AM - 375036 Yerevan, Armenia\\
$^{176}$ ATLAS-Canada Tier-1 Data Centre, TRIUMF, 4004 Wesbrook Mall, Vancouver, BC, V6T 2A3, Canada\\
$^{177}$ GridKA Tier-1 FZK, Forschungszentrum Karlsruhe GmbH, Steinbuch Centre for Computing (SCC), Hermann-von-Helmholtz-Platz 1, 76344 Eggenstein-Leopoldshafen, Germany\\
$^{178}$ Port d'Informacio Cientifica (PIC), Universitat Autonoma de Barcelona (UAB), Edifici D, E-08193 Bellaterra, Spain\\
$^{179}$ Centre de Calcul CNRS/IN2P3, Domaine scientifique de la Doua, 27 bd du 11 Novembre 1918, 69622 Villeurbanne Cedex, France\\
$^{180}$ INFN-CNAF, Viale Berti Pichat 6/2, 40127 Bologna, Italy\\
$^{181}$ Nordic Data Grid Facility, NORDUnet A/S, Kastruplundgade 22, 1, DK-2770 Kastrup, Denmark\\
$^{182}$ SARA Reken- en Netwerkdiensten, Science Park 121, 1098 XG Amsterdam, Netherlands\\
$^{183}$ Academia Sinica Grid Computing, Institute of Physics, Academia Sinica, No.128, Sec. 2, Academia Rd.,   Nankang, Taipei, Taiwan 11529, Taiwan\\
$^{184}$ UK-T1-RAL Tier-1, Rutherford Appleton Laboratory, Science and Technology Facilities Council, Harwell Science and Innovation Campus, Didcot OX11 0QX, United Kingdom\\
$^{185}$ RHIC and ATLAS Computing Facility, Physics Department, Building 510, Brookhaven National Laboratory, Upton, New York 11973, United States of America\\
$^{a}$ Also at CPPM, Marseille, France.\\
$^{b}$ Also at TRIUMF, 4004 Wesbrook Mall, Vancouver, B.C. V6T 2A3, Canada\\
$^{c}$ Also at Faculty of Physics and Applied Computer Science of the AGH-University of Science and Technology, (FPACS, AGH-UST), al. Mickiewicza 30, PL-30059 Cracow, Poland\\
$^{d}$ Also at  Universit\`a di Napoli  Parthenope, via A. Acton 38, IT - 80133 Napoli, Italy\\
$^{e}$ Also at Institute of Particle Physics (IPP), Canada\\
$^{f}$ Louisiana Tech University, 305 Wisteria Street, P.O. Box 3178, Ruston, LA 71272, United States of America   \\
$^{g}$ At Department of Physics, California State University, Fresno, 2345 E. San Ramon Avenue, Fresno, CA 93740-8031, United States of America\\
$^{h}$ Currently at Istituto Universitario di Studi Superiori IUSS, V.le Lungo Ticino Sforza 56, 27100 Pavia, Italy\\
$^{i}$ Also at California Institute of Technology, Physics Department, Pasadena, CA 91125, United States of America\\
$^{j}$ Also at University of Montreal, Canada\\
$^{k}$ Also at Institut f\"ur Experimentalphysik, Universit\"at Hamburg,  Luruper Chaussee 149, 22761 Hamburg, Germany\\
$^{l}$ Also at Petersburg Nuclear Physics Institute,  RU - 188 300 Gatchina, Russia\\
$^{m}$ Also at School of Physics and Engineering, Sun Yat-sen University, China\\
$^{n}$ Also at School of Physics, Shandong University, Jinan, China\\
$^{o}$ Also at Rutherford Appleton Laboratory, Science and Technology Facilities Council, Harwell Science and Innovation Campus, Didcot OX11, United Kingdom\\
$^{p}$ Also at school of physics, Shandong University, Jinan\\
$^{q}$ Also at Rutherford Appleton Laboratory, Science and Technology Facilities Council, Harwell Science and Innovation Campus, Didcot OX11 0QX, United Kingdom\\
$^{r}$ Now at KEK\\
$^{s}$ University of South Carolina, Dept. of Physics and Astronomy, 700 S. Main St, Columbia, SC 29208, United States of America\\
$^{t}$ Also at KFKI Research Institute for Particle and Nuclear Physics, Budapest, Hungary\\
$^{u}$ Also at Institute of Physics, Jagiellonian University, Cracow, Poland\\
$^{v}$ Also at School of Physics and Engineering, Sun Yat-sen University, Taiwan\\
$^{w}$ Transfer to LHCb 31.01.2010\\
$^{x}$ Also at Dept of Physics, Nanjing University, China\\
$^{*}$ Deceased\end{flushleft}
}

\abstract{The simulation software for the ATLAS Experiment at the Large Hadron Collider is being used for large-scale production of events on the LHC Computing Grid. This simulation requires many components, from the generators that simulate particle collisions, through packages simulating the response of the various detectors and triggers. All of these components come together under the ATLAS simulation infrastructure. In this paper, that infrastructure is discussed, including that supporting the detector description, interfacing the event generation, and combining the GEANT4 simulation of the response of the individual detectors. Also described are the tools allowing the software validation, performance testing, and the validation of the simulated output against known physics processes.
} %end of abstract

\date{Submitted to EPJ C 20 May 2010}

\maketitle

\newcommand{\Geant}{{\sc Ge\-ant4}\xspace}
\newcommand{\Python}{{\sc Py\-thon}\xspace}
\newcommand{\Pythia}{{\sc Pyth\-ia}\xspace}
\newcommand{\Herwig}{{\sc Her\-wig}\xspace}
\newcommand{\Herwigpp}{{\sc Her\-wig++}\xspace}
\newcommand{\Fatras}{{Fa\-tras}\xspace}
\newcommand{\Fluka}{{\sc Flu\-ka}\xspace}
\newcommand{\Atlfast}{{ATL\-FAST}\xspace}

\clearpage

\section{Introduction}

ATLAS~\cite{techpaper}, one of the general-purpose detectors at the Large Hadron Collider~\cite{LHC}, began operation in 2008.  The detector will collect data from proton-proton collisions with center-of-mass energies up to 14~\TeV{}, as well as 5.5~\TeV{} per nucleon pair in heavy ion (Pb-Pb) collisions.  During proton-proton collisions at the design luminosity of $10^{34}$~cm$^{-2}$s$^{-1}$, beam bunches will cross every 25~ns (40~MHz) and provide on average 23 collisions per bunch crossing.  ATLAS has been designed to record up to 200 bunch crossings per second, keeping only the most interesting interactions for physics analyses, including searches for new phys\-ics.

In order to study the detector response for a wide range of physics processes and scenarios, a detailed simulation has been implemented that carries events from the event generation through to output in a format which is identical to that of the true detector.  The simulation program is integrated into the ATLAS software framework, Athe\-na~\cite{Athena}, and uses the \Geant\ simulation tool\-kit~\cite{G4,G4Update}.  The core software and large-scale production infrastructures are discussed further in Section~\ref{Overview}.

The simulation software chain is generally divided into three steps, though they may be combined into a single job: generation of the event and immediate decays (see Section~\ref{EvGen}), simulation of the detector and physics interactions (see Section~\ref{Core}), and digitization of the energy deposited in the sensitive regions of the detector into voltages and currents for comparison to the readout of the ATLAS detector (see Section~\ref{Digi}).  The output of the simulation chain can be presented in either an object-based format or in a format identical to the output of the ATLAS data acquisition system (DAQ).  Thus, both the simulated and real data from the detector can then be run through the same ATLAS trigger and reconstruction packages.

The ATLAS detector geometry used for simulation, digitization, and reconstruction is built from da\-ta\-bases containing the information describing the physical construction and conditions data.  The latter contains all the information needed to emulate a single data-taking run of the real detector (\eg\ detector misalignments or temperatures).  The same geometry and simulation infrastructure is able to reproduce the test stands and installation configurations of the ATLAS detector.  The detector description is discussed in Section~\ref{DetDesc}.  

Large computing resources are required to accurately model the complex detector geometry and physics descriptions in the standard ATLAS detector simulation.  This has led to the development of several varieties of fast simulation.  Each is best suited to a particular use-case, and they are described in Section~\ref{FastSim}.  Validation of the software, testing of the software performance, and validation of the physics performance and output of each piece of the simulation software chain is discussed in Section~\ref{Validation}.  

This paper reviews the status of the software and geometry used for large-scale production in 2008.

\section{ATLAS Offline Software Overview}
\label{Overview}

The ATLAS software framework, Athe\-na~\cite{Athena}, uses \Python\ as an object-oriented scripting and interpreter language to configure and load C++ algorithms and objects.  Rather than develop an entirely new high-energy physics data processing infrastructure, ATLAS adopted the Gaudi framework~\cite{Gaudi,Gaudi2}, originally developed for LHCb and written in C++.  Gaudi was created as a flexible framework to support a variety of applications through base classes and basic functionality.  As much as possible, the infrastructure relies on the CLHEP common libraries~\cite{CLHEP}, which include utility classes particularly designed for use in high-energy physics software (\eg\ vectors and rotations).

Athe\-na releases are divided into major projects by functionality~\cite{SoftManagement}, and all of the ATLAS simulation software (including event generation and digitization) resides in a single project.  The dependencies of the ``simulation'' project are the ``core'' project, which includes the Athe\-na framework, the ``conditions'' and ``detector description'' projects, which include all code necessary for the description of the ATLAS detector, and the ``event'' project, which includes descriptions of persistent objects.  The number of lines of code by software language for the simulation project are summarized in Table~\ref{table:code}, as calculated using cloc~\cite{CLOC} in Athe\-na release 14.4.  Lines of code in the upstream Athe\-na projects, excluding external dependencies like Gaudi and CLHEP, are summarized in Table~\ref{table:codeDep}.

\begin{table}[htb]
\caption{Numbers of files, lines of code, and lines of comments in the ATLAS simulation project, by programming language for major contributors.  External dependencies are not included.\label{table:code}}
\begin{center}
\begin{tabular}{lllll}
\hline\noalign{\smallskip}
{\bf Language} & {\bf Files} & {\bf Comment} & {\bf Code} \\
\noalign{\smallskip}\hline\noalign{\smallskip}
C++            &     930 &   24,000 &  120,000 \\
FORTRAN        &     270 &   15,000 &   42,000 \\
C/C++ Header   &   1,100 &   13,000 &   34,000 \\
Python         &     430 &   16,000 &   27,000 \\
HTML           &      62 &      130 &   15,000 \\
Bourne Shell   &     390 &    1,000 &    7,300 \\
C Shell        &     380 &      210 &    3,800 \\
XML            &      52 &    1,200 &    3,400 \\
\noalign{\smallskip}\hline\noalign{\smallskip}
\bf{Sum}       &   3,600 &   70,000 &  250,000 \\
\noalign{\smallskip}\hline
\end{tabular}
\end{center}
\end{table}

\begin{table}[htb]
\caption{Numbers of lines of code in each of the projects upstream of the ATLAS simulation project, versus the programming language.  Most projects are dominated by C++ and \Python\ code.  The most significant exception is the detector project, which contains 70,000 lines of XML and Java code.\label{table:codeDep}}
\begin{center}
\begin{tabular}{lllll}
\hline\noalign{\smallskip}
{\bf Project} & {\bf C/C++} & {\bf C/C++} & {\bf \Python} & {\bf Total} \\
              & {\bf Code} & {\bf Headers} & {\bf Code} & {\bf Code} \\
\noalign{\smallskip}\hline\noalign{\smallskip}
Core       & 390,000 &  43,000 & 240,000 & 860,000 \\
Event      & 200,000 & 110,000 &  16,000 & 350,000 \\
Conditions & 280,000 &  90,000 &  21,000 & 620,000 \\
Detector   &  38,000 &   6,100 &   8,400 & 140,000 \\
\noalign{\smallskip}\hline\noalign{\smallskip}
{\bf Sum}  & 910,000 & 250,000 & 280,000 & 2,000,000 \\
\noalign{\smallskip}\hline
\end{tabular}
\end{center}
\end{table}

All Athe\-na jobs consist of three distinct steps.  First, in the initialization step, services and algorithms are loaded on demand using dynamic library loading.  Generally, algorithms include methods to be called once per event, whereas services may be accessed many times during a single event.  The configuration and initialization is controlled within a common \Python\ infrastructure which allows introspection, particularly useful in debugging and providing help for the users.  Also, by using a scripting language for loading and configuring objects, there is no need to recompile C++ code or a script for each job.  Small modifications can be made in the scripts (also called ``fragments'' or ``job options''), or even in the midst of the job, without having to stop and recompile the libraries.  This scripting method also lightens the load on the user, since there is, under normal circumstances, no need to compile anything prior to running a job.  Each algorithm and service can be configured differently for each step of the simulation software chain, allowing maximal sharing of infrastructure among the distinct steps of the chain.  Algorithms can be added to a top list of methods to be run during the event loop.  

Second, the event loop begins.  All algorithms in the top list are run sequentially on each event.  An external generator or algorithm controlling \Geant\ may be added to this list, for example.  From these main methods, other services and algorithms can be called.  A messaging service, called throughout the jobs, controls log file outputs with different levels of verbosity.  The user may configure the total logging verbosity or configure the verbosity individually for a single algorithm, particularly useful for debugging.  

During the finalization stage of the job, all algorithms are terminated and all objects are deleted.  At this point, algorithms may output any statistics (\eg\ memory or CPU usage) they track.

These three steps comprise each Athe\-na job, but the infrastructure allows for the insertion of hooks at various places.  Each step of the ATLAS simulation chain takes advantage of this infrastructure to provide maximal flexibility for the user.  Only requested modules are loaded as plug-ins, keeping each step as light as possible in memory and as fast as possible during the event loop.

For storing data, ATLAS has adopted a scheme for separating transient from persistent objects.  Most general C++ types, immediately prior to storage, are converted to a type that requires less space.  Although, for example, energy is accumulated in the ca\-lo\-ri\-me\-ter by summing double-precision floating point numbers, at the end of each event and prior to storage, the total energy is converted into a single-precision floating point number (float).  Summing with floats was found to alter the total energy because of truncation.  For some types, more complicated storage schemes are implemented that rely on properties of the information to be stored (\eg\ where it is possible to sacrifice some accuracy).  Metadata, general property information for data collected in a file, are included in the output files for all the stages of the event simulation.  The metadata include all configuration information for the job.  Athe\-na has also adopted the POOL (Pool Of persistent Objects for LHC) file handling and persistency framework~\cite{Dullmann:2005zzb,Chytracek:2005cm,Duellmann:2003iy}.

\subsection{ATLAS Simulation Overview}

\begin{figure*}
%%\vspace{8cm}
\begin{center}
\includegraphics[height=3.15in]{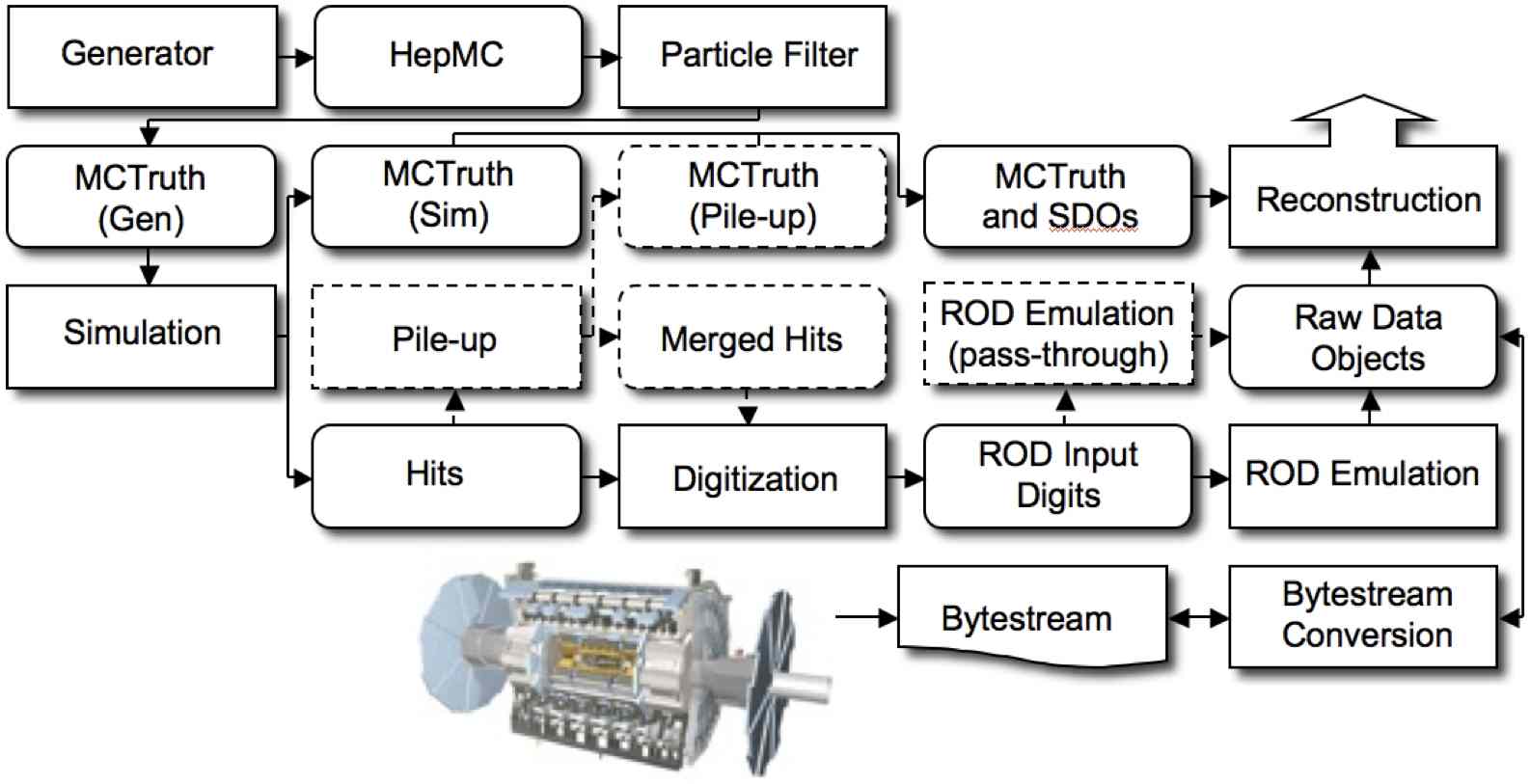}
\end{center}
\caption{The flow of the ATLAS simulation software, from event generators (top left) through reconstruction (top right).  Algorithms are placed in square-cornered boxes and persistent data objects are placed in rounded boxes.  The optional pile-up portion of the chain, used only when events are overlaid, is dashed.  Generators are used to produce data in HepMC format.  Monte Carlo truth is saved in addition to energy depositions in the detector (hits).  This truth is merged into Simulated Data Objects (SDOs) during the digitization.  Also, during the digitization stage, Read Out Driver (ROD) electronics are simulated.\label{fig:outline}}
\end{figure*}

An overview of the ATLAS simulation data flow can be seen in Figure~\ref{fig:outline}.  Algorithms and applications to be run are placed in square-cornered boxes, and persistent data objects are placed in round-cornered boxes.  The optional steps required for pile-up or event overlay (see Section~\ref{PileUpDigi}) are shown with a dashed outline.

A generator produces events in standard HepMC format~\cite{Dobbs:2001ck}.  These events can be filtered at generation time so that only events with a certain property (\eg\ leptonic decay or missing energy above a certain value) are kept.  The generator is responsible for any prompt decays (\eg\ $\Zboson$\ or $\Wboson$\ bosons) but stores any ``stable'' particle expected to propagate through a part of the detector (see Section~\ref{EvGen}).  Because it only considers immediate decays, there is no need to consider detector geometry during the generation step, except in controlling what particles are considered stable.  During this step, the run number for the simulated data set and event numbers for each event are established.  Event numbers are generally ordered in a single job, though events may be omitted because of filtering at each step.  Run numbers for simulated data sets derive from the job options used to generate the sample and mimic real run numbers used during data taking.

These generated events are then read into the simulation.  A record of all particles produced by the generator is retained in the simulation output file (see Section~\ref{GenMCTruth}), but cuts can be applied to select only certain particles to process in the simulation.  Each particle is propagated through the full ATLAS detector by \Geant.  The configuration of the detector, including misalignments and distortions, can be set at run time by the user.  The energies deposited in the sensitive portions of the detector are recorded as ``hits,'' containing the total energy deposition, position, and time, and are written to a simulation output file, called a hit file.  

In both event generation and detector simulation, information called ``truth'' is recorded for each event.  In the generation jobs, the truth is a history of the interactions from the generator, including incoming and outgoing particles.  A record is kept for every particle, whether the particle is to be passed through the detector simulation or not.  In the simulation jobs, truth tracks and decays for certain particles are stored. This truth contains, for example, the locations of the conversions of photons within the inner detector and the subsequent electron and positron tracks.  In the digitization jobs, Simulated Data Objects (SDOs) are created from the truth.  These SDOs are maps from the hits in the sensitive regions of the detector to the particles in the simulation truth record that deposited the hits' energy.  The truth information is further processed in the reconstruction jobs and can be used during the analysis of simulated data to quantify the success of the reconstruction software.

The digitization takes hit output from simulated e\-vents: hard scattering signal, minimum bias, beam halo, beam gas, and cavern back\-ground e\-vents.  Each type of event can be overlaid at a user-spec\-i\-fed rate before the detector signal (\eg\ voltage or time) is generated. The overlay (called ``pile-up'') is done during digitization to save the CPU time required by the simulation.  At this stage, detector noise is added to the event.  The first level trigger, implemented with hardware on the real detector, is also simulated in a ``pass'' mode.  Here no events are discarded but each trigger hypothesis is evaluated.  The digitization first constructs ``digits,'' inputs to the read out drivers (RODs) in the detector electronics.  The ROD functionality is then emulated, and the output is a Raw Data Object (RDO) file.  The output from the ATLAS detector itself is in ``byte\-stream'' format, which can be fairly easily converted to and from RDO file format.  The two are similar, and in some subdetectors they are almost interchangeable.  Truth information is the major exception.  It is stripped in the conversion to byte\-stream.

The simulation software chain, divided in this way, uses resources more effectively than a single-step event simulation and simplifies software validation.  Event generation jobs, typically quick and with small output files, can be run for several thousands of events at a time.  By storing the output rather than regenerating it each time, it becomes possible to run identical events through different versions of the simulation software or with different detector configurations.  The simulation step is particularly slow, and can take several minutes per event (see Section~\ref{Perf}).  Simulation jobs are therefore divided into groups of 50 or fewer events; only a few events may be completed in a single heavy ion simulation job.  Digitization jobs are generally configured to run $\sim1000$ events. This configuration eases file handling by producing a smaller number of RDO files.  Each step is partially configured based on the input files.  For example, the detector geometry used for a digitization job is selected based on the input hit file.

The ATLAS high level trigger\footnote{The ATLAS high level trigger comprises two stages: level 2, and the event filter.  Both are software triggers run with the reconstruction, and may be treated as a single unit for the purposes of this discussion.} (HLT)~\cite{HLT} and reconstruction~\cite{Reconstruction} run on these RDO files.  The reconstruction is identical for the simulation and the data, with the exception that truth information can be treated and is available only in simulated data.  During data taking, the HLT is run on byte\-stream files, however all hypotheses and ad\-di\-tion\-al test hypotheses may be evaluated by translating the RDOs into byte\-stream format.

\subsection{Large-Scale Production System}
\label{Prod}

Because of the significant time consumption of the ATLAS simulation, only minimal jobs can be completed interactively on most computers.  It is, therefore, desirable to distribute as much as possible production of the necessary simulated data for ATLAS.  Complete software releases are built and distributed to production sites and users every few weeks, providing all Athe\-na software and all external dependencies, including generators and \Geant.   These releases are patched several times with ``production caches'' before a new clean release is built and distributed.  With each release or production cache, a small set of data files are packaged that include database replicas, any necessary external data files, and some sample output files.  These sample files can be used to ensure that the locally installed release can be validated by processing events through the entire software chain, from generation through reconstruction.

Large-scale production is then done on the World-wide LHC Computing Grid (``WLCG'' or ``Grid'')~\cite{LCG}.  A single task on the Grid (\eg\ simulation of 500,000~$\ttbar$\ events) is separated into many jobs depending on the content and complexity of the task.  A job can be completed by a single CPU within the maximum allowed time for a job on the Grid (typically 2-3 days).  The output, including log files, of every Grid job is registered with the ATLAS Distributed Data Management system (DDM)~\cite{DDM}. The DDM uses DQ2~\cite{DQ2} for dataset bookkeeping, and allows users to search for datasets on the Grid, analyze them in place, and, if necessary, retrieve them. Separate Grid software controls the distribution of jobs to the various Grid sites.  In a typical task, 10 jobs are queued and run as a test sample, and only once they finish successfully are the remainder of the jobs released to the Grid.  In the case of a full chain of jobs being run (generation, simulation, and digitization), each subsequent step is automatically held in the queue until the required data are available from the previous step.  Frequently, Grid jobs are configured to run two steps (\eg\ simulation and digitization together, or digitization and reconstruction together).  About one million events per day can be produced using \Geant\ on the Grid.

On the Grid, ``job transforms'' are run, which may only include well-defined, minor modifications to some standard job configuration after the input events have been specified.  A task is given a random number seed, and each job increments the seed in sequence.  The modifications to a generation job also may include a configuration file for the selected generator to be run.  These configuration files are included with each release and may not be arbitrarily modified by the user during production.  The modification to a simulation job may include detector geometry and conditions and specially designed job options fragments that are included with each release.  These fragments are typically constructed for a very specific purpose, for example a non-standard vertex smearing, simulation of cavern background, or propagation and late decay of long-lived exotic particles.  Many of these modifications can be chained to provide maximal flexibility to the user, but if two fragments are sufficiently complex such chaining becomes impossible.  The modifications to a digitization job may include geometry and conditions versions, ca\-lo\-ri\-me\-ter sampling fraction, trigger configuration, and noise control.  These modifications are discussed further in the subsequent sections.

%\clearpage

\section{Event Generation Overview}
\label{EvGen}

Event generation consists of the production of a set of particles which is passed to either full or fast detector simulation. Event generation runs within the Athe\-na framework, but most of the generators themselves are written and maintained by authors external to ATLAS. The AT\-LAS-spe\-ci\-fic implementation, therefore, consists mostly of a set of interface packages. These are designed to be as simple as practicable and wherever possible to be factorized from the external packages. This is essential to allow rapid feedback and bug reporting to the authors of the external packages. Most of the well-understood and thoroughly debugged generators are written in FORTRAN.  Their interfaces transfer the event information, mostly contained in FORTRAN common blocks, into an object format that can be used by the ATLAS software. This ensures that any downstream algorithms are shielded from details specific to an individual generator. Events can either be stored as POOL files for later use or passed to simulation in the same Athe\-na job.

Details of the framework and comments specific to each generator are listed below. Large-scale production has been run with \Pythia~\cite{Sjostrand:2006za} (including an ATLAS variant, PythiaB~\cite{PythiaB,PythiaB2}, used for production of events with B-had\-rons), \Herwig~\cite{Herwig6,Corcella:2000bw,Herwig65}, Sherpa~\cite{Gleisberg:2003xi}, Hijing~\cite{Gyulassy:1994ew}, Alpgen~\cite{Mangano:2002ea}, MC\-@\-NLO~\cite{Frixione:2003ei}, and AcerMC~\cite{Kersevan:2004yg}. Tauola~\cite{Jadach:1990mz} and Photos~\cite{Barberio:1990ms} are routinely used to handle tau decays and photon emission. EvtGen~\cite{Lange:2001uf} is used for B-decays in cases where the physics is sensitive to details of the B hadron decays\footnote{\Pythia\ remains the default for current inclusive production, but EvtGen is likely to be used by default for the long-term production.}. ISAJET~\cite{Paige:2003mg} is used for generating supersymmetric particles in conjunction with \Herwig. The newer C++ generators \Pythia\ 8~\cite{Sjostrand:2007gs} and \Herwigpp~\cite{Bahr:2008pv} are being tested. Both produce events in the HepMC format, so no translation is needed. They can be passed directly to simulation. As these new generators evolve and undergo extensive testing and validation, they are expected to enter the production shortly and eventually supersede their FORTRAN predecessors.  Some production was also done with MadGraph~\cite{MadGraph} (vector boson scattering), CHARYBDIS~\cite{CHARYBDIS} (black hole event generation), and CompHep~\cite{CompHep,CompHep2} (specific exotic physics models).  Discussion of the generation of cavern background, beam halo, and beam gas events follow in Sections~\ref{CavernBackground} and~\ref{BeamHalo}.  Single particle generators are also used to generate cosmic ray events and single particle events for performance studies and calibration of the detector.

Each generated event contains the particles from a single interaction with a vertex located at the geometric origin. Modifications to account for the beam properties are applied to the event before it is passed to \Geant\ (see Section~\ref{SimIn}). Particles with a proper lifetime $c\tau>10$~mm are considered stable by the event generator. They can propagate far enough to interact with detector material before decaying. Their decays are handled by the simulation. Any particles with $c\tau<10$~mm are decayed by the event generator, and their interactions with material or curving in the magnetic field of ATLAS are ignored.

\subsection{Generator Framework}
\label{sec-frame}

Many external generator packages assume that the parameters for a particular job are set via a main program. This would require recompilation to change parameters. The Athe\-na generator interfaces allow for the passing of all relevant parameters at run time, permitting a fixed software release to be used to produce different physics configurations. During initialization, the relevant parameters are passed via \Python\ fragments. The combination of the fragments, random number seeds, and the software release uniquely identifies the resulting data\footnote{Since pseu\-do-random number generators are chip architecture dependent, jobs are exactly reproducible only when run on the same type of processor (\eg\ Intel or AMD).}. The Athe\-na event manager is run for each event, and a run number and an event number are created; then the event generator is asked to produce an event. This event is created in memory in the format specific to the generator itself. The event must then be mapped into a common format so that subsequent algorithms are independent of the generator used.

ATLAS uses the HepMC event record~\cite{Dobbs:2001ck}, initially developed by the ATLAS collaboration but now supported by WLCG~\cite{LCG}. This is a set of C++ classes which holds the full event as produced by the generator. Stable particles are used as input to simulation; unstable ones can be of use in physics studies and diagnostics. Each event generator produces a very large number of stable particles (\eg\ muon, kaons, pions, electrons, photons), a much larger number of unstable particles (\eg\ gluons, quarks, B mesons, heavy hyperons), and, possibly, other objects (\eg\ ``strings'' or ``color singlet clusters'') specific to an individual generator. The HepMC record consists of a connected tree, navigation inside of which retains information on the event history including the parents of unstable particles. There is an important caveat here: the event generators are modeling quantum processes, and the event record has the structure of a classical decay chain. It is inevitable that compromises must be made and difficulties can arise from an over-literal interpretation of the tree structure. A very simple example is provided by events containing an $\epem$\ pair. The parent of the $\epem$\ pair cannot be uniquely specified, as the pair may arise from an intermediate $\Zboson$\ boson, photon, or quantum interference. The HepMC event record is also used to contain the particle information from secondaries produced by interactions in the detector. This is discussed below in the section on Monte Carlo truth (see Section~\ref{GenMCTruth}).  Information about all interacting partons (\eg\ momentum fractions $x_1$ and $x_2$) is saved, so that parton distribution function reweighting can be done without rerunning the event generation.

The FORTRAN generators usually use the HEPEVT common block~\cite{HEPEVT} to store the information. Unfortunate\-ly, the different generators use slightly different structures.  A separate translation into HepMC is needed for each one. The C++ generators such as \Herwigpp\ produce output in the HepMC format. No translation is required and the integrity of the HepMC event record is the responsibility of the generator authors.

\subsection{General Purpose Generators}

General purpose generators produce complete events starting from a pro\-ton-pro\-ton, pro\-ton-nu\-cle\-us or nu\-cle\-us-nu\-cle\-us initial state. They are used standalone or with specialized  generators that improve the description of certain final states. They have many parameters, some of which are related to fundamental parameters such as the QCD coupling constant and electroweak parameters, and some of which describe the models used to parametrize long distance QCD, soft QCD, and electroweak processes.

\subsubsection{\Pythia and PythiaB}

\Pythia~\cite{Sjostrand:2006za} and \Herwig\ (see below) in their FORTRAN versions have been tested, used, and validated over many years in $e^+e^-$ and hadron colliders. They start with a hard scattering process calculated to lowest order in QCD.  They then add ad\-di\-tion\-al QCD and QED radiation in a shower approximation which is most accurate when the radiation is emitted at small angle. The approximation is poorest in those cases with a large number of widely separated emissions of comparable energy.  In addition, \Pythia\ use a model for hard and soft scattering processes in a single event in order to simulate underlying activity.  This model is used in the simulation of minimum bias events.  While other generators may be used for specific final states, \Pythia\ and \Herwig\ are the benchmarks. 

ATLAS uses \Pythia\ 6.4. There are two models of QCD radiation in \Pythia. By default, ATLAS uses the showering model introduced in \Pythia\ 6.3. This showering model is believed to better match the theoretical description of QCD showers.  It produces somewhat more jet activity~\cite{Moraes:2007rq,BadPythia2}, resulting in ``busier'' events than the older model which was used, for example, for detailed simulations at the Tevatron (see, for example,~\cite{Tevatron1,Tevatron2}). In this model, the multiple scatters which make up the underlying event are interleaved with the parton shower according to the hard scale of the scatter or the emission.  At the end of the shower, a phenomenological model is used to combine the quarks and gluons into had\-rons. This hadronization model, which has many parameters, has been tuned by comparison with data in $\epem$, $ep$, and hadron colliders~\cite{Field:2005sa,Norrbin:2000zc}.  The underlying event model was retuned within ATLAS~\cite{PythiaAtlasTune} to recover an acceptable description of the Tevatron data~\cite{PythiaAtlasTevatron1,PythiaAtlasTevatron2}.  \Pythia\ contains a very large number of built-in processes, and new ones can be added by modifying the code.  Hard scattering events can also be generated in a separate program in a standard format and fed into \Pythia\ for the addition of a parton shower and hadronization.  \Pythia\ is the default generator in ATLAS: many hundreds of millions of events have been generated using it. Its ease of use, speed, and robustness make it an ideal choice for the default. It is supplemented by other generators, either to obtain some estimate of the uncertainties, or when specialized generators are expected to give a better physical description in certain final states.

PythiaB~\cite{PythiaB,PythiaB2} is an AT\-LAS-spe\-ci\-fic modification of \Pythia\ aimed at the efficient generation of events related to B-phys\-ics. In \Pythia, most high $p_T$ bottom quarks are produced in the QCD shower of a high $p_T$ light quark or gluon from a hard scattering process.  Most showers do not produce such a $b\overline{b}$ pair, so using \Pythia\ to generate B-phys\-ics events is inefficient. PythiaB reuses those QCD showers that contain a $b$- or $c$-quark, hadronizing them several times to increase the probability of producing a b-hadron. Since the probability producing a $b$- or $c$-quark in a parton shower is low, this procedure results in more efficient procedure of making $b$-hadron events without introducing any bias in the distribution of $b$-had\-rons within the event. If a specific decay mode is then required the $b$-hadron decay can be forced using a modified $b$-hadron decay table, either in \Pythia\ itself or via EvtGen.

\subsubsection{\Herwig}
ATLAS uses \Herwig\ 6.5~\cite{Herwig6,Corcella:2000bw,Herwig65}, the last release of the FORTRAN \Herwig\ package which is now superseded by \Herwigpp\ (see below). It is a flexible generator with a large number of built in processes and has been tuned to agree with the Tevatron data~\cite{PythiaAtlasTevatron1,PythiaAtlasTevatron2}. In particular, most of the generation of supersymmetric processes is done with \Herwig\ using the ISAWIG package~\cite{Herwig6,Corcella:2000bw,Herwig65} with the particle spectra and decay modes generated by ISAJET. ATLAS uses \Herwig\ with the Jimmy~\cite{Jimmy} implementation of the underlying event.

\subsubsection{Sherpa}
Sherpa~\cite{Gleisberg:2003xi} is a generator written in C++ which implements the CKKW duplicate removal prescription~\cite{Catani:2001cc} to match fixed-order QCD matrix elements to QCD showers. It uses an interface to \Pythia's hadronization model and produces complete events. It is expected to give better approximations for final states with large numbers of isolated jets than generators such as \Pythia\ and \Herwig\ based on pure QCD showering. Sherpa generates underlying events using a simple multi-parton interaction model based on that of \Pythia. For each new process to be generated, Sherpa must be recompiled to incorporate the specific libraries for the process of interest. On the Grid, this implies either recompiling Sherpa at the production site or deploying updated libraries for new production jobs.  Instead, Sherpa is run locally to produce event files in Sherpa's native format. These files are then translated into the HepMC format with an ad\-di\-tion\-al Athe\-na Grid job.  It is also possible to run Sherpa entirely within an Athe\-na job. 

\subsubsection{Hijing}
\label{Hijing}

Hijing~\cite{Gyulassy:1994ew} is a dedicated generator for the production of heavy ion events at all impact parameters. In a dense nuclear environment, such as appears in central collisions, a particle produced in a primary collision can re-interact several times as it propagates.  Hijing models the propagation. It is also the only generator that can be used for proton-nucleus collisions occurring in beam-gas interactions. Hijing uses the \Pythia\ hadronization model. 

\subsubsection{Single Particle Generators}
\label{SingleParticle}

A single particle event generator is frequently used for calibrating the detector, testing, and evaluating the reconstruction efficiencies.  Although unphysical, these generators produce events with a single primary particle, for example a muon, electron, or charged pion, at a specified energy, position, and momentum direction.  A range may also be specified for either the energy or direction.  No underlying event, proton remnants, or other primary interactions are included when these events are generated.

A specialized single particle generator is used to produce cosmic ray events.  Single muons are generated at the earth's surface in a square region (typically 600~m by 600~m) above the ATLAS detector and with the standard cosmic ray $p_T$ spectrum~\cite{CosmicGen1,CosmicGen2}.  The upper and lower energy cutoffs of the spectrum are configurable.  Those muons pointing to a sphere of configurable size (typically 20~m) centered at the geometric origin are propagated through the bedrock and the ATLAS cavern during simulation.

\subsection{Specialized Generators}

Specialized generators do not produce complete events which can be passed directly to simulation. Rather, they are run in conjunction with one of the general purpose generators to improve the accuracy for specific decays or specific final states.  Several of these specialized generators are ``Les Houches'' type generators.  That is, they are run standalone using unmodified code from the generator author and produce an ASCII file containing partonic four-vectors in the ``Les Houches'' format~\cite{LhaExt,Alwall:2006yp}.  Athe\-na uses a common interface that reads in these files and prepares them for processing in \Pythia\ or \Herwig~\cite{LhaExt}.

\subsubsection{ISAJET}
The FORTRAN generator ISAJET~\cite{Paige:2003mg} is not used in large-scale production. However, it is used in conjunction with \Herwig\ for generation of supersymmetric events. Here, the ISASUGRA component of ISAJET is used to generate consistent sets of masses and decay modes for supersymmetric models. These are then loaded into \Herwig\ using the ISAWIG translation package, and \Herwig\ then generates complete final states.

\subsubsection{Photos and Tauola}

ATLAS uses the dedicated tau decay package Tauola to handle tau decays~\cite{Jadach:1990mz}. General purpose generators are set to treat tau leptons as stable.  The events are passed to Tauola for decay. Because Tauola is a FORTRAN package, the events are extracted from the HEPEVT record. The Tauola interface is dependent on the generator that produced the tau, because helicities and helicity correlations are passed in generator-dependent formats.  The original generator's results must be replaced, so both the input and output formats of Tauola are in fact generator-dependent. Special attention is paid to the polarization of the tau. In certain cases, for example the decay $\Wtaunu$, the polarization is known for the tau. In others, such as $\Ztautau$, there is a correlation between the polarization of the taus.

Photos handles electromagnetic radiation~\cite{Barberio:1990ms}. It is used by Tauola, and, therefore, Tauola cannot be used without Photos. Photos is also used to improve the description of electromagnetic radiation in, for example, the decay $\Wenu$, where radiation distorts the electron energy distribution.  In these cases the final state electromagnetic radiation is switched off in the general purpose generator, usually \Herwig\ or \Pythia, to avoid double counting.

\subsubsection{EvtGen}

EvtGen~\cite{Lange:2001uf}, originally developed by the CLEO collaboration, provides a more complete description of B meson and hadron decays than that provided by
defaults in \Pythia\ or \Herwig. Recent modifications have been made to handle $B_S$ and b-baryon decays, incorporating measurements from the Tevatron, BaBar, and Belle.  In particular, EvtGen incorporates the best measurements of branching ratios and has theoretical models for unmeasured decay modes. It includes angular correlations, which impact the acceptance for certain decay modes of B mesons and baryons. It has been used for ATLAS studies involving the prospects for measurements of exclusive B decays.

\subsubsection{Alpgen}

Alpgen~\cite{Mangano:2002ea} is a ``Les Houches'' type generator enabling more sophisticated generation of certain final states. \Herwig\ or \Pythia\ is then used to perform the hadronization and produce final (and initial) state QCD radiation. Alpgen is targeted at final states with several well-separated hadronic jets where the fixed order QCD matrix element is expected to give a better approximation than the shower approximation of \Pythia\ or \Herwig.  Alpgen is used, for example, to generate final states containing a $\Wboson$\ or $\Zboson$\ and many jets.  Alpgen also provides an algorithm to prevent double counting by event rejection. The Athe\-na interface package includes the methods needed to pass events through \Herwig\ or \Pythia\ and veto those events that would contribute to double counting. This process can be very inefficient for final states with large numbers of jets, and generation time can be significant. 

\subsubsection{MC@NLO}

MC\-@\-NLO~\cite{Frixione:2003ei}, which is also a ``Les Houches'' type generator, runs standalone to produce ASCII files which are then processed by \Herwig\ running inside of Athe\-na. MC\-@\-NLO uses fundamental (hard scattering) processes evaluated at next to leading order in QCD perturbation theory. It is used, for example, to generate top events as it gives a better representation of the transverse momentum~($p_T$) distribution of top quarks than \Pythia\ or \Herwig. MC\-@\-NLO includes one loop corrections, with the consequence that events appear with negative and positive weight which must be taken into account when they are used. Any resulting distribution will contain entries from both types of event, and, given sufficient statistics, the result will by physical (\ie\ positive)\footnote{An alternative tool, POWHEG~\cite{POWHEG}, implements essentially the same physics and produces events with only positive weight.  Once it includes all the processes that MC\-@\-NLO does and has been validated, it is expected to take the place of MC\-@\-NLO.}. MC\-@\-NLO has been used for large-scale production of top, $\Wboson$\ and $\Zboson$\ events. Only the parts of MC\-@\-NLO needed to read these events and process them via \Herwig\ are included in Athe\-na releases.

\subsubsection{AcerMC}

AcerMC~\cite{Kersevan:2004yg} is a ``Les Houches'' type generator aimed primarily at the production of $\Wboson$\ or $\Zboson$\ bosons with several jets, including jets originating from b-quarks. A partonic final state is obtained by running it standalone and making an external ASCII file. Only the parts needed to read these events and process them via \Pythia\ are included in Athe\-na releases.

\subsection{New C++ Generators}

\subsubsection{\Pythia\ 8}

\Pythia\ 8~\cite{Sjostrand:2007gs} is a rewrite of the FORTRAN \Pythia\ in C++ with new and expanded physics models.  It provides a new user interface, transverse-momentum-ordered showers, and interleaving with multiple interactions. The program is under intensive tests and it will require some further tunings before it can replace the Pythia6  code as a leading generator. It is, however, interfaced to Athe\-na and used for generator studies in ATLAS.  It includes support for both ``Les Houches'' and HepMC event formats.

\subsubsection{\Herwigpp}

\Herwigpp~\cite{Bahr:2008pv} is the C++ based replacement for \Herwig. It contains only important processes from the Standard Model, the universal extra dimensions model, and supersymmetric models (whose details are specified via Supersymmetric Les Houches Accord model files~\cite{SLHA1,SLHA2}).  Ad\-di\-tion\-al hard scattering processes can be used via ``Les Houches'' input from specialized generators, and ad\-di\-tion\-al decay models can be added by users.

\Herwigpp\ will soon be used for generation of some Standard Model processes, notably $\Wboson$\ and $\Zboson$\ production.  It will also be used for supersymmetric processes, because it includes full spin correlations and QCD radiation in the supersymmetric decay chains.  The current version of \Herwigpp\ also incorporates an underlying event model based on the extension of Jimmy~\cite{Jimmy} to include soft scatters~\cite{HerwigppMB} and can thus potentially generate minimum bias physics.

\subsection{Parton Distribution Functions}

Parton distribution functions (PDFs) are used to describe the substructure of the proton and are used by all the event generators as external inputs. ATLAS uses the Les Houches Accord PDF Interface (LHAPDF~\cite{Bourilkov:2006cj}) library which is a replacement for PDFLIB~\cite{PlothowBesch:1992qj} which provides a large repository of PDFs. CTEQ~\cite{Tung:2007bm} PDFs are used by default (MC\-@\-NLO uses NLO PDFs, and all other generators use LO PDFs).  There is a correlation between the PDFs and the tuning of parameters connected to initial state radiation~\cite{PDFISR,TevForLHC}: inconsistent results can be obtained by varying the PDFs in isolation.  Therefore, when a new set of PDFs is used, the parameters of the event generator are retuned to produce consistent results~\cite{Moraes:2007rq}.

\subsection{Monte Carlo Truth}
\label{GenMCTruth}

The entire connected tree of the HepMC event record is stored as the Monte Carlo truth.  Only the stable particles are propagated by the simulation. The various status codes and event history provided by the individual generators are retained within the HepMC event record. Unfortunate\-ly, much of this information is specific to a particular generator. Only status codes 1 (stable) and 2 (unstable) have a general meaning: the remaining values are used differently by the individual generators. As remarked in Section~\ref{sec-frame}, there can be ambiguities resulting from the attempt to represent a quantum process by a classical tree. Some filters have been provided to select HepMC particles that, for example, are stable at the generator level or are non-interacting (\eg\ neutrinos). 

When the simulation is run, the HepMC tree from the event generator is copied, and some particles resulting from decays within, or interactions simulated by, \Geant\ are added to the copy (see Section~\ref{MCTruth}). In this way, a complete event including both the generator and simulation information is provided. In order to ensure consistency, a particle decayed by \Geant\ but considered stable by the generator (such as a $K_S$) has its status code changed when the copy is made. A particle that has status code 2 after simulation will be identified as stable at the generator level, if the decay took place in \Geant. \Geant\ secondaries are distinguished from those from the generators by an offset applied to their numerical identifier. The resulting Monte Carlo truth record can be large and account for a significant fraction ($\sim$30\%) of the disk space used by a simulated event after reconstruction.  

\subsection{Default Parameters, Tuning and Bug Fixing}

The generator authors define default parameters. In some cases, however, these parameters are not tuned for use at the Large Hadron Collider and are superseded by parameters obtained by comparisons to data.  The criteria for a particle to be considered stable are modified for use in ATLAS, for example. Once high-energy data appear, it is expected that retuning of the parameters will occur. These tunings can be made by varying parameters at run time. Once a new tuning is available, it can be loaded as a \Python\ fragment at run time or hard coding the values into the generator interfaces. In either case, the tuning becomes available as part of the next Athe\-na software release and will be enabled by default.  The settings can be overridden if needed or the previous defaults re-established. It is important to note that the parameters are often not independent and a complete set must be used.  Arbitrary adjustments of a few parameters may result in inconsistent results. One of the most important sets of tunings is concerned with structure of minimum bias events and spectator processes in a hard scattering event: the underlying event.  At present, these tunings are obtained for both \Pythia\ and \Herwig~\cite{Moraes:2007rq} by first tuning to the Tevatron data and then extrapolating.  The extrapolation from the Tevatron relies on the models used by \Pythia\ and \Herwig.  This extrapolation has had testing from comparisons of the Tevatron data at~630 and 1800~\GeV~\cite{Field:2005sa,Skands}. A high priority task for the ATLAS simulation as data accumulates is the testing of these tunings and changing of the parameters as needed.

%%\clearpage

\section{ATLAS Detector Description}
\label{DetDesc}

The ATLAS detector is described in detail in Ref.~\cite{techpaper}, but its main features will be summarized here.  We discuss the geometry used in the simulation, which as much as possible matches the as-built detector.  A cut-away view of the entire detector is shown in Figure~\ref{fig:atlas}.  ATLAS comprises several concentric components.  The subdetectors are:

\begin{figure*}
%\vspace{8cm}
\begin{center}
\includegraphics[width=5.8in]{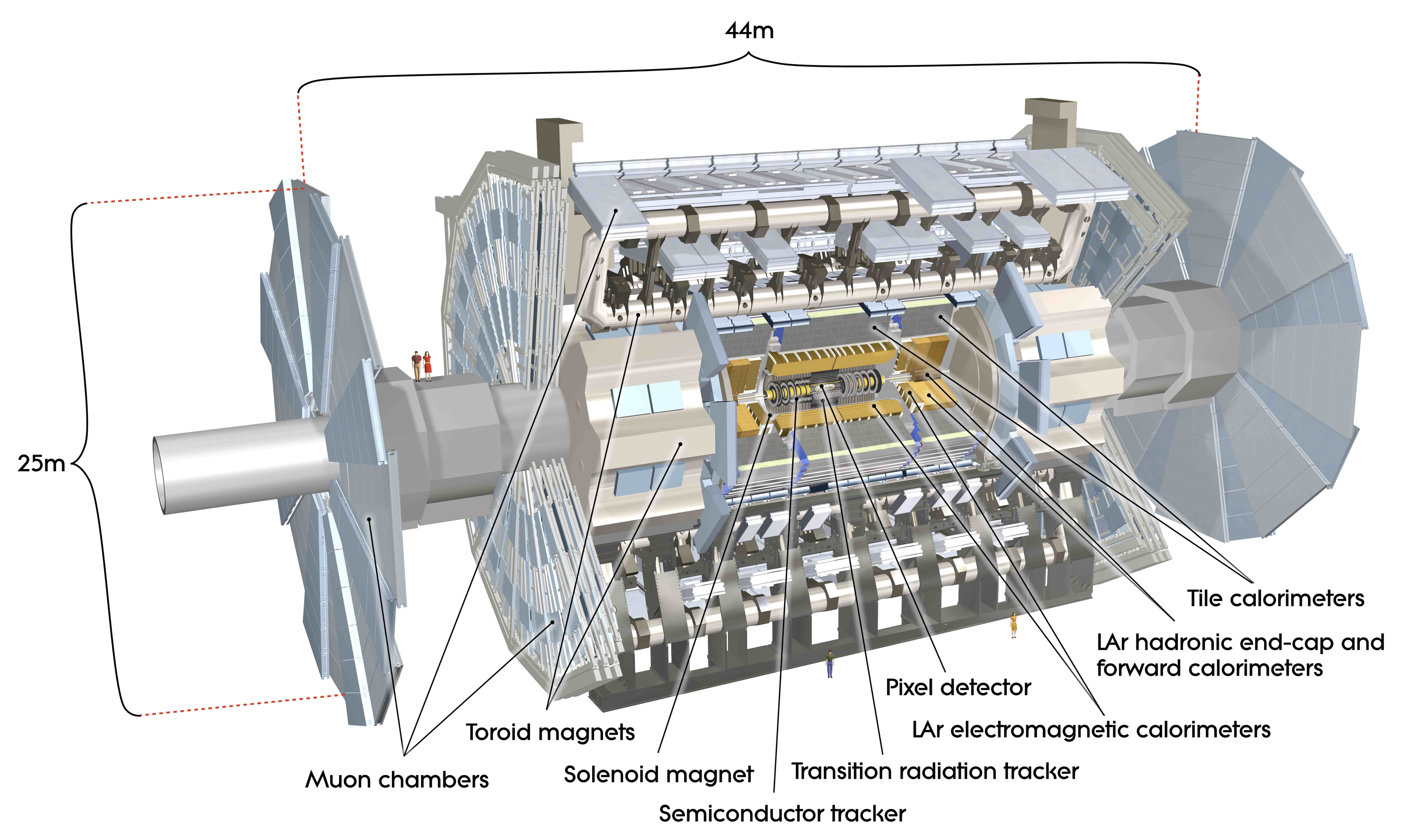}
\end{center}
\caption{ATLAS detector view.\label{fig:atlas}}
\end{figure*}

\begin{itemize}
\item A Beam Conditions Monitor~(BCM) and Beam Loss Monitor used for detecting dangerous conditions and triggering an abort in the detector system.  The BCM is located 1.84~m from the interaction point (IP) at $|\eta|\sim4.2$ \footnote{Pseu\-do\-rapidity, $\eta \equiv -\ln{ \tan( \theta/2 ) }$, where $\theta$ is the polar angle measured from the beam pipe.  The other coordinate variables used are typically $r$, $z$ and $\phi$, where the $x$-axis points towards the center of the LHC ring, the $y$-axis points up, the $z$-axis defines a right-handed coordinate system, $r \equiv \sqrt{x^2+y^2}$, and $\phi$ is the azimuthal angle defined such that $\phi=0$ along the $x$-axis.}.
\item A tracking detector composed of a fine granularity pixel detector with three layers covering $|\eta|<2.5$, a silicon strip tracker (SCT) with eight layers determining four space points covering $|\eta|<2.5$, and a transition radiation tracker (TRT) which has 32 space points on a typical track, covering $|\eta|<2.0$.
\item Hermetic calorimetry composed of liquid argon (LAr) electromagnetic calorimetry covering $|\eta|<3.2$, scintillating tile hadronic calorimetry in the barrel ($|\eta|<1.7$), sampling LAr hadronic calorimetry in the endcap ($1.5<|\eta|<3.2$), and LAr electromagnetic and hadronic forward calorimetry covering $3.2<|\eta|<4.9$.
\item Four different types of muon chambers, two of which are high precision (monitored drift tubes, and cathode strip chambers) and two of which have a rapid response for the muon trigger (thin gap chambers, and resistive plate chambers), covering $|\eta|<2.7$.
\item Luminosity detectors, including a zero-degree ca\-lo\-ri\-me\-ter that sits 140~m from the interaction point, a detector that performs a luminosity measurement using Cherenkov integration (LUCID), and an absolute luminosity detector for ATLAS.
\end{itemize}

The ATLAS magnetic field is formed by a solenoid, providing a 2.0~T uniform magnetic field in the tracking subdetectors, and a toroidal magnet system, composed of a barrel and two endcap toroid magnets.  In the inner detector, the field has small $\phi$- and $z$-asymmetries due to the toroid field and perturbations from the iron nearby.  The field in the toroidal system has approximate $z$- and eight-fold $\phi$-symmetry and provides on average 2.5~Tm of bending power in the barrel and 5~Tm in the endcap.  During a simulation run, the field map required about 30~MB of memory.

In the standard production simulation (see Section~\ref{Prod}) the luminosity detectors are not included.  They can be simulated in dedicated jobs, but keeping particles in such a high pseu\-do\-rapidity region increases simulation time by approximately 50\% per unit of pseu\-do\-rapidity ($|\eta|$) per event (see Section~\ref{SimIn}).  

Several layouts of the complete detector are available, including those that were used for recording cosmic ray events while the detector was being completed.  Test stands are also supported with the same infrastructure.  All these layouts are described in Section~\ref{AltLay}. As much as possible, the details of the detector geometry are preserved in the simulation layout.  Some approximations are necessary for describing dead materials, for example bundles of cables and cooling pipes in the service areas of the detector.  In these cases, the description only aims to match the general distribution of the material, including inhomogeneities in $\phi$.

\subsection{Simulated Detector Geometry}
\label{SimGeo}

The geometry structure can be viewed in terms of solids, basic shapes without a position in the detector; logical volumes, solids with ad\-di\-tion\-al properties (\eg\ name or material); and physical volumes, individual placements of logical volumes. Table~\ref{table:geometry} shows the number of materials, solids, logical volumes, physical volumes, and total volumes created when constructing various pieces of the ATLAS detector. Not all volumes are equivalent, however: in the case of repeating structures, as in the sampling portion of the LAr calorimetry in particular, it is possible to define a single logical volume that is repeated in hundreds of physical volumes (known as volume parameterization). Because of nesting, one can also define dependencies that create many total volumes from the physical volumes used. In other cases, a single volume can correspond to a piece of shielding or support with a complex shape. One can see in this table the complexity of the ATLAS detector, with hundreds of materials and hundreds of thousands of physical volumes.  Such a detailed detector description is crucial for accurately modeling, for example, missing transverse energy, track reconstruction efficiencies, and ca\-lo\-ri\-me\-ter response.

Table~\ref{table:geomem} shows the number of physical volumes contained in each detector subsystem and the memory required to build each using the GeoModel library (see Section~\ref{GeoModel})~\cite{GeoModelRef}.  As expected, the two are correlated, although differences in volume complexity invalidate a direct correspondence.  The entire geometry must be translated into a \Geant\ equivalent, so the total memory required for the geometry of the entire ATLAS detector is almost 300~MB (see Section~\ref{Validation}).

\begin{table*}[htb]
\caption{Numbers of materials, solids, logical volumes, physical volumes, and total volumes required to construct various pieces of the ATLAS detector.  ``Inner Detector'' here includes the beampipe, BCM, pixel tracker, SCT, and TRT.\label{table:geometry}}
\begin{center}
\begin{tabular}{llllll}
\hline\noalign{\smallskip}
{\bf Subsystem} & {\bf Materials} & {\bf Solids} & {\bf Logical Vol.} & {\bf Physical Vol.} & {\bf Total Vol.} \\
\noalign{\smallskip}\hline\noalign{\smallskip}
Beampipe         &  43 &    195 &    152 &     514 &       514 \\
BCM              &  40 &    131 &     91 &     453 &       453 \\
Pixel            & 121 &  7,290 &  8,133 &   8,825 &    16,158 \\
SCT              & 130 &  1,297 &  9,403 &  44,156 &    52,414 \\
TRT              &  68 &    300 &    357 &   4,034 & 1,756,219 \\
LAr Calorimetry  &  68 &    674 &    639 & 106,519 &   506,484 \\
Tile Calorimetry &   8 & 51,694 & 35,227 &  75,745 & 1,050,977 \\
\noalign{\smallskip}\hline\noalign{\smallskip}
\noalign{\smallskip}\hline\noalign{\smallskip}
Inner Detector   & 243 & 12,501 & 18,440 &  56,838 & 1,824,614 \\
Calorimetry      &  73 & 52,366 & 35,864 & 182,262 & 1,557,459 \\
Muon System      &  22 & 33,594 &  9,467 &  76,945 & 1,424,768 \\
\noalign{\smallskip}\hline\noalign{\smallskip}
{\bf ATLAS TOTAL} & 327 & 98,459 & 63,769 & 316,043 & 4,806,839 \\
\noalign{\smallskip}\hline
\end{tabular}
\end{center}
\end{table*}

\begin{table}[htb]
\caption{Numbers of physical volumes and memory required to build various pieces of the ATLAS detector in GeoModel.  Here ``calorimetry'' is simply the sum of the liquid argon and tile calorimetry.\label{table:geomem}}
\begin{center}
\begin{tabular}{llllll}
\hline\noalign{\smallskip}
{\bf Subsystem}   & {\bf Phys. Volumes} & {\bf Memory [kB]}\\
\noalign{\smallskip}\hline\noalign{\smallskip}
Inner Detector    &  56,838 & 22,268 \\
Calorimetry       & 182,262 & 44,116 \\
Muon System       &  76,945 & 31,524 \\
\noalign{\smallskip}\hline\noalign{\smallskip}
{\bf ATLAS TOTAL} & 316,043 & 97,908 \\
\noalign{\smallskip}\hline
\end{tabular}
\end{center}
\end{table}

In creating such a complex, dense geometry, removing volume overlaps and touching surfaces provides a particular challenge.  Any overlap of more than 1~picometer and any place in which two volume faces touch can lead to stuck tracks during the simulation, a situation in which a track in \Geant\ may not know in which volume it belongs.  These stuck tracks result in a loss of the event, but they can be overcome by introducing small gaps between volumes, at the cost of an extra step for each particle moving through the transition region.

Many layouts are available corresponding to the various revisions of material.  The material budget is constantly updated, so that the geometry description is as realistic as possible.  During any major updates of detector geometry, the subdetectors are generally required to make all changes backwards-compatible so that all older geometries can be configured and run as normal.  This requirement allows for a fair comparisons between software releases with consistent geometries. During any job, the user may choose to enable or disable portions of the detector.  Each subdetector is responsible for including any necessary materials and elements for its own construction.  In this way, only required elements and materials are used during simulation, and memory loads are reduced when not using the entire ATLAS detector.  The switches for disabling portions of the detector generally correspond to the highest level of the tree-structure in the detector geometry (\ie\ entire subdetectors, not pieces).

It is possible to apply detector ``conditions'' modifications to each chosen geometry layout. The detector misalignment can be configured by selecting misaligned layouts either for each subdetector or for the full detector at once. Each ATLAS subdetector sits within a well defined envelope, allowing each to shift and distort without colliding with any other. In digitization and reconstruction jobs, conditions may include detector information beyond misalignments (\eg\ dead channels). The infrastructure is in place to record detector conditions in a database and, at run time, allow the user to select conditions from a specific data taking run. Conditions and geometry versions selected by the user can be transferred from the simulation jobs to the digitization and reconstruction jobs so that no ad\-di\-tion\-al user interaction is required.  These default versions may at any time be overridden by job options.

In order to study the penalties of a poor material description on jet resolution and missing transverse energy bias, a special geometry layout with material distortions was created~\cite{CSC}. Material distortions correspond to ad\-di\-tion\-al material added to half of the detector ($y>0$) to approximate a poor material description.

\subsection{Databases and Configuration}

Two da\-ta\-bases are used to construct the detector geometry chosen by the user: one to store basic constants (the ATLAS Geometry database), and one to store various conditions data (\eg\ calibrations, dead channel, misalignments) for the specific run chosen (ATLAS Conditions database)~\cite{Databases}. At CERN, large (terabytes) Oracle da\-ta\-bases are used, primarily because they are well supported and straightforward to update. With any stable software release, a small subset of data needed for Athe\-na jobs is replicated from Oracle into SQLite~\cite{SQLite1,SQLite2,SQLite3} - file based da\-ta\-bases - and is distributed to the production centers. The large I/O requirements of production jobs can overwhelm a central Oracle server and are better handled by relatively small SQLite files. These files can also be replicated to individual production nodes for local and rapid access. The database replica version to be used can be chosen at run time for each Grid job.

Both the geometry and conditions da\-ta\-bases support versioning of the data. The data are organized in a tree consisting of branch and leaf nodes. The nodes in this tree can be ``tagged,'' and one can create a hierarchy of the tags. Such tag hierarchies are uniquely identified by the tag of the root node, which is usually referred to as top level geometry or conditions tag.

A geometry database stores all fundamental constants for detector construction. Volume dimensions, rotations, and positions, as well as element and material properties including density, are all stored as database entries. New detector-specific tags may be created for inclusion in a global ATLAS geometry tag, where different tags generally correspond to different detector geometry revisions. At run time, the user can select a global geometry tag as well as detector-specific geometry tags to create the desired geometry. In addition to constants for detector construction, the geometry database contains links to external data files that may store, for example, magnetic field maps. These files are shipped with software distribution kits to production sites. By using links through the database, it is possible to select a magnetic field map based on the chosen geometry layout. The selection of field map based upon the name provided in the database, for example, can be overridden with job options. 

A separate conditions database stores detector conditions data which are indexed by intervals of validity and tags. The entire detector may be optionally misaligned with a global misalignment tag, and the user may configure the job to use specific misalignment versions for each subdetector. The global misalignment is used frequently to study the performance of the entire ATLAS detector with misalignments of the expected as-built magnitude.  The detector-specific misalignments allow studies of the effects of misalignment of a single subdetector assuming ideal alignment of the remainder of ATLAS. The inner detector, for example, has completed an alignment challenge, wherein simulated data was produced with misalignments, and the analysis group was challenged to align the detector as with data. The tile ca\-lo\-ri\-me\-ter, on the other hand, does not use any misalignments in its geometry. A variety of misalignments have been used in the lead-up to data taking in order to speed the process of global detector alignment and improve early physics studies.

During data collection, the alignment constants of the detector are recorded periodically in the central conditions database. The user is able to recreate the misalignment conditions for a specific run by selecting an alignment version, again by subdetector if desired, at run time.

\subsection{GeoModel and Translations}
\label{GeoModel}

The ATLAS simulation, digitization, and reconstruction each run in distinct jobs, but they must be able to use the same detector geometry.  Therefore, a complete geometry description is maintained that can be used by each step and is not specific to any.  By using the geometry da\-ta\-bases, it is already possible to read identical detector constants and run conditions.

For these reasons, ATLAS uses GeoModel~\cite{GeoModelRef}, a library of basic geometrical shapes, to describe and construct the detector.  GeoModel contains geometry features similar to those of \Geant: basic volumes can be constructed, rotated, and shifted in space; subvolumes can be placed inside a volume; boolean volumes can be made by adding or subtracting primitives; volumes can be parameterized and repeated.  For the digitization and reconstruction, this detector description is entirely sufficient to place hits, reconstruct tracks and objects, and complete all necessary calculations.  

The GeoModel descriptions of most ATLAS subsystems are built using constants in the geometry database. However, a translator has been constructed that parses an XML description of a detector's geometry and builds a transient representation from GeoModel primitives at run time. This generic package can translate any valid XML description of detector geometry into GeoModel format. It has been used for describing the geometry of the muon system's rather complicated dead material.

For the simulation, the geometry is translated entirely from the GeoModel to the \Geant\ format.  All volumes and subvolumes are translated, constructed, and properly placed within the ``world volume'' (the volume allocated for the detector, at the edge of which particles cease to be simulated).  All information tied to GeoModel, including position, rotation, and dimensions, are also translated into a \Geant\ equivalent.  Once the geometry has been translated, all subsystems rely solely on their \Geant\ description.  The GeoModel geometry is currently maintained in memory for the entirety of the job, though it may be released to ease memory pressure in the future.  As shown in Table~\ref{table:geomem}, this release can be expected to save 100~MB of memory.  Sensitive detectors and particle range cuts (see Section~\ref{SimOpt}), for example, are tied to the \Geant\ geometry by volume name and can be added at any time after the geometry has been constructed. Each change in detector description is particularly weighty in simulation, because any ad\-di\-tion\-al volumes must be built both in GeoModel and in the \Geant\ geometry.  

\subsection{Alternate Layouts}
\label{AltLay}

\begin{figure*}
%\vspace{8cm}
\begin{center}
\includegraphics[height=4.25in]{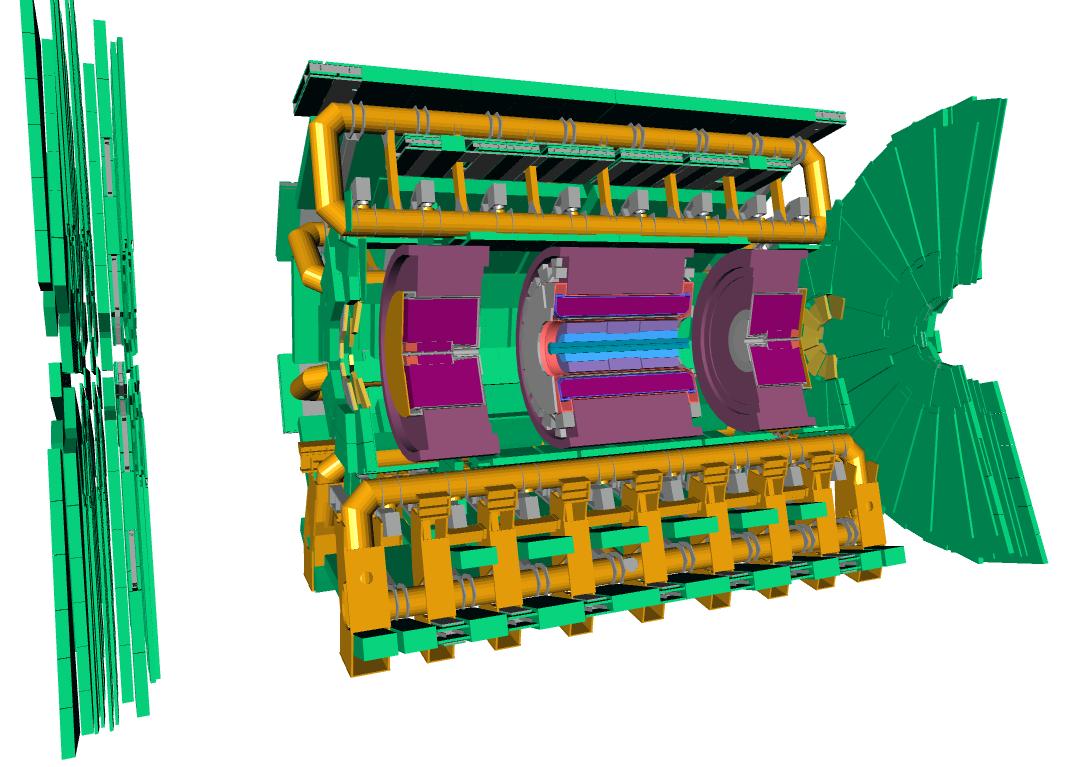}
\end{center}
\caption{Commissioning layout of the detector used for cosmic ray data taking during 2008.  The endcap toroidal magnets and beampipe are not yet installed.  The ca\-lo\-ri\-me\-ter endcaps (purple) are shifted by 3.1~m and the muon endcaps (green) are shifted to provide access to the inner detector during installation.  The barrel toroid magnets are shown in yellow, and the inner detector is shown in blue.\label{fig:comm}}
\end{figure*}

In addition to the standard detector layouts, several commissioning layouts are available to the user for simulation of cosmic ray data taking.  During detector assembly, cosmic ray data were taken for several runs using as many subdetectors as were available.  Some of these subdetector configurations included ca\-lo\-ri\-me\-ter endcaps shifted out of position while the inner detector was being accessed and were missing large portions of the beam pipe that had not yet been installed.  One such commissioning layout is shown in Figure~\ref{fig:comm}.  For studies of cosmic rays and cavern background, it is possible to simulate the ATLAS cavern surrounding the detector as well as the bedrock surrounding the two shafts leading down from the surface.  

Several different magnetic field configurations are also available for some of the full detector layouts.  Fields with the toroidal magnets on and solenoid off or solenoid on and toroidal magnets off are provided.  These magnet configurations have already been used for some cosmic ray data taking runs and may be used for brief periods during high-energy collisions as well.  Field maps have also been constructed that reflect the as-built misalignments of the magnet system, for example a vertical shift of a 1.6~mm in the solenoid.

\begin{figure*}
%\vspace{8cm}
\begin{center}
\includegraphics[width=5.95in]{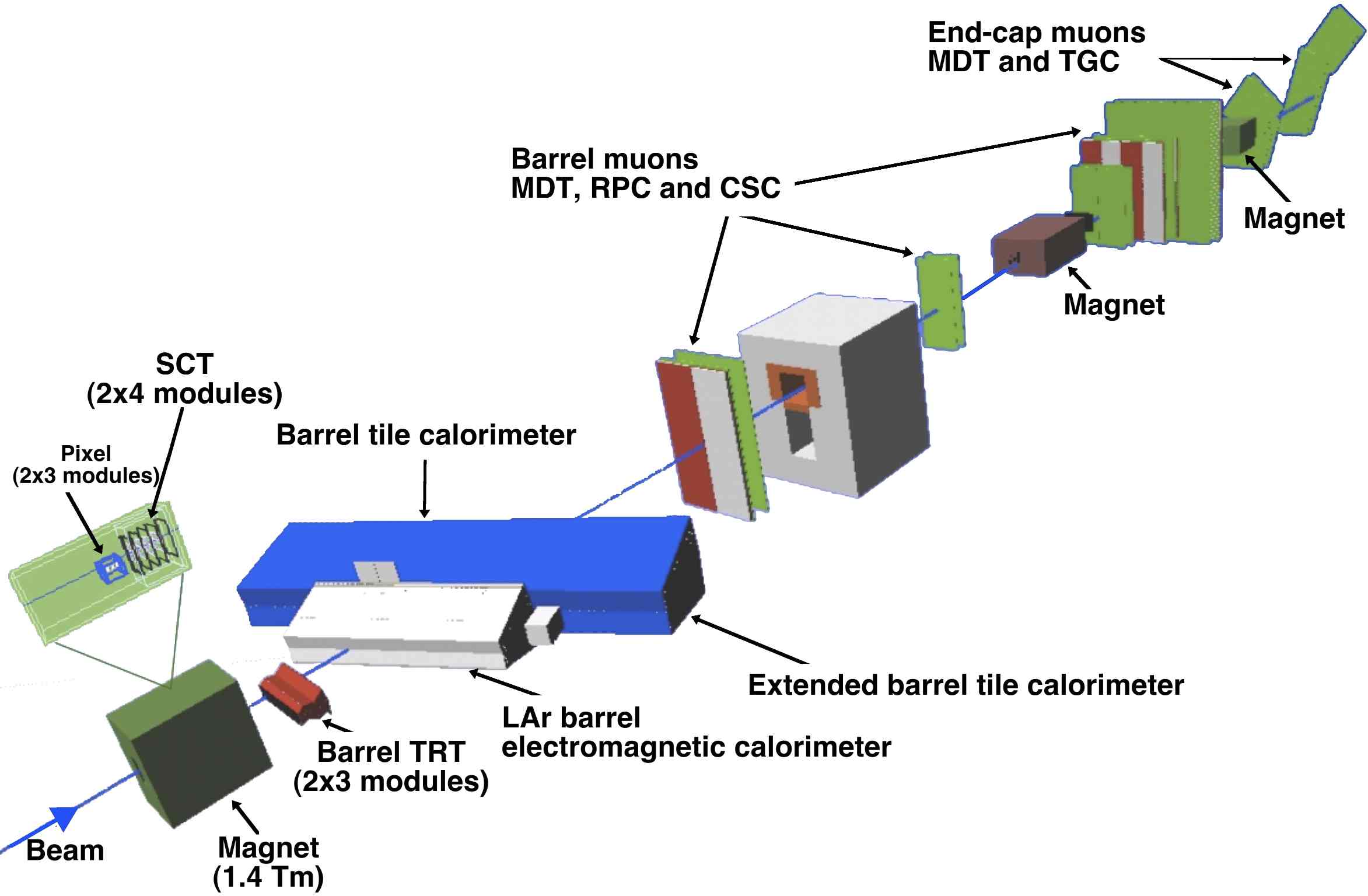}
\end{center}
\caption{Combined test beam setup from 2004.\label{fig:ctb}}
\end{figure*}

There are also several test stand layouts that were constructed to model test beam and standalone cosmic ray runs.  A sample of these various test stands, including subsystems, incident particles, and energies, are listed in Table~\ref{table:tb}.  A combined test beam run was taken with a wedge of the full detector during 2004~\cite{CTBgeometry}, and standalone test beams were constructed for the muon detectors~\cite{MuonTB1,MuonTB2}, tile ca\-lo\-ri\-me\-ter~\cite{TileTB}, and liquid argon ca\-lo\-ri\-me\-ter subsystems~\cite{LArTB1,LArTB2}.  The combined test beam setup is shown in Figure~\ref{fig:ctb}.  Cosmic ray data were also collected with various pieces of the inner detector~\cite{PixelTest,SCTTest} and with the muon chambers both prior to and after installation.

\begin{table*}[htb]
\caption{Examples of test stands for ATLAS simulated using \Geant.\label{table:tb}}
\begin{center}
\begin{tabular}{llllll}
\hline\noalign{\smallskip}
{\bf Subsystem} & {\bf Incident Particle} & {\bf Energy} \\%% & {\bf Events Recorded} \\ Location
\noalign{\smallskip}\hline\noalign{\smallskip}
Hadronic Endcap ca\-lo\-ri\-me\-ter & $e^{+/-}$, $\pi^{+/-}$, $\mu^{+/-}$ & 6-245~\GeV{} \\%%& \\ %% H6
Electromagnetic Barrel ca\-lo\-ri\-me\-ter & $e^{+/-}$ & 10-245~\GeV{} \\%%& \\ %% H8
Electromagnetic Endcap ca\-lo\-ri\-me\-ter & $e^{+/-}$ & 10-200~\GeV{} \\%%& \\ %% H6
Combined Endcap ca\-lo\-ri\-me\-ter & $e^{+/-}$, $\pi^{+/-}$, $\mu^{+/-}$ & 6-200~\GeV{} \\%%& \\ %% H6
Hadronic Barrel ca\-lo\-ri\-me\-ter & $\pi^{+/-}$, $p$ & 5-350~\GeV{} \\%%& \\ %% ?
Entire detector endcap wedge & $e^{+/-}$, $\pi^{+/-}$, $\mu^{+/-}$ & 1-350~\GeV{} \\%%& \\ %% H8
Muon Detectors & $\mu^{+/-}$ & 20-350~\GeV{} \\
Silicon Pixel Tracker Endcap & Cosmic Rays & 0.5-200~\GeV{} \\
Silicon Strip Tracker Barrel & Cosmic Rays & 0.5-200~\GeV{} \\
\noalign{\smallskip}\hline
\end{tabular}
\end{center}
\end{table*}

All test stand and commissioning layouts are available as a part of the same geometry infrastructure and can be selected at run time for simulation.  By maintaining all detector configurations as a part of a common infrastructure, it is possible to ensure consistency between, for example, the test beam and full detector simulation.  Conclusions drawn from analysis of the test beam simulated data are generally still valid for the full detector simulation.  The extensive tuning of the detector simulation and digitization on test beam data can be applied directly to the full detector.  As many common elements as possible are kept between the two, including \Geant\ version and physics list (see Section~\ref{Core}).

%%\clearpage

\section{Core Simulation}
\label{Core}

The standard simulation of ATLAS relies on the \Geant\ particle simulation tool\-kit.  \Geant\ provides models for physics and infrastructure for particle transportation through a geometry, but several AT\-LAS-spe\-ci\-fic pieces are provided as user-code.  The detector geometry itself is constructed in the \Geant\ format, and all particle scoring (done in ``sensitive detector'' classes) are done on the Athe\-na side.  Each subsystem's scoring is optimized and tailored to store only what is necessary for accurately reproducing the performance of that particular subsystem~\cite{SCTSim,TRTSim,LArSim,TileSim,MuonSim,EMCalSim,HECSim,FCALSim}.  Athe\-na code is necessary to add to the Monte Carlo truth record.  Physics models are chosen and parameters optimized for the ATLAS detector.  The results shown in this paper used \Geant\ version 8.3 with official patch \#2 and two modifications: updates for boundary represented volumes and a patch to the G4Tubs code.  The software is continuously evolving, and ATLAS has moved to newer Geant4 versions since the writing of this paper.

The \Geant\ Collaboration and ATLAS Simulation Group have benefitted from 15 years of close collaboration.  Frequently, new \Geant\ features have allowed faster or more realistic simulation of the ATLAS detector.  Feature requests from the ATLAS collaboration have helped drive the development of \Geant.  The ATLAS simulation has also provided one of the more complicated test-beds for the \Geant\ tool\-kit, and \Geant\ has been extensively evaluated and validated during large-scale simulation production.

In order to provide \Python\ flexibility to the \Geant\ simulation, an ad\-di\-tion\-al layer of infrastructure is necessary.  ``Standard'' \Geant\ simulation typically runs from compiled C++, and in order to modify any of the parameters or the geometry used in the simulation it is necessary to recompile.  The Framework for ATLAS Detector Simulation (FADS)~\cite{FADS} wraps several \Geant\ classes in order to allow selection and configuration without recompilation of any libraries.  Since a \Python\ interface is used for configuration, all the usual introspection capabilities of \Python\ may be employed.  FADS wraps \Geant\ base-classes for volumes, materials, and sensitive detectors for hit processing as well as \Geant\ physics process definitions.  These wrappers serve a dual purpose: first, they ease translations between the \Geant\ and Athe\-na standards of geometry, hits, and particle storage.   Second, FADS can catalogue the options available to the user, loading only those that will be needed for the desired simulation configuration while still providing all possibilities without any recompilation.  Through FADS, a user is free to select a physics list (see Section~\ref{PhysicsList}) for use during the simulation.  The user may also modify the physics list by adding particles or processes not included in the \Geant\ tool\-kit but included in the FADS catalogues, for instance in the simulation of long-lived exotic particles.  Similarly, the detector description is configured with \Python\ dictionaries and FADS catalogues before it is built in \Geant and may be modified by the user.  For example, sensitive detectors may be assigned to any volume in the detector.  Range cuts (see Section~\ref{SimOpt}) may also be added in the \Python\ and FADS layer prior to their being applied to any constructed \Geant\ geometry. Once the \Python\ configuration is complete, FADS objects are translated into their \Geant\ equivalents and loaded.  Even after this translation, they can be modified through the standard \Geant\ user interface.  

In order to fit into the Athe\-na framework, a service for \Geant\ and an algorithm that calls the service during the event loop have been implemented~\cite{G4Atlas}.  The service wraps the event loop of \Geant\ and provides a few ad\-di\-tion\-al handles for user-configuration in the \Python\ layer.  The service also takes care of initialization and finalization of each \Geant\ event.  The generated events are translated from HepMC format into the standard \Geant\ event format prior to each event, and at the end of each event an analysis is done to ensure that the simulation finished without errors. Most of the functionality of the standard \Geant\ run manager is included in this service, so that any Athe\-na-specific modifications (\eg\ event translation from HepMC to \Geant\ format) to the usual \Geant\ event sequence can be made.  The service also provides for interaction with \Geant\ through its standard user interface.

This section describes the possible user inputs, initialization, output, and various parameters of the simulation.  Several useful features, including visualization, are also described.

\subsection{Simulation Input}
\label{SimIn}

The ATLAS simulation offers a choice for event generation.  Events can be read from a file produced by any of the generators described in Section~\ref{EvGen}; one of the external generators can be configured and run concurrently; or commands can be provided for a single particle generator.  The single particle generator can produce particles by the particle PDG identifier~\cite{PDG} with a configurable position and momentum.  Neutral and charged geantinos (pseu\-do-part\-i\-cles without any interactions) are available for making material depth maps of the detectors and for debugging.  The user may also choose to skip a certain number of events at the beginning of an input file, allowing 20 simulation jobs of 50 events each to a 1000 event input file without overlap or repetition.

\begin{figure*}
\begin{center}
\includegraphics[width=6.2in]{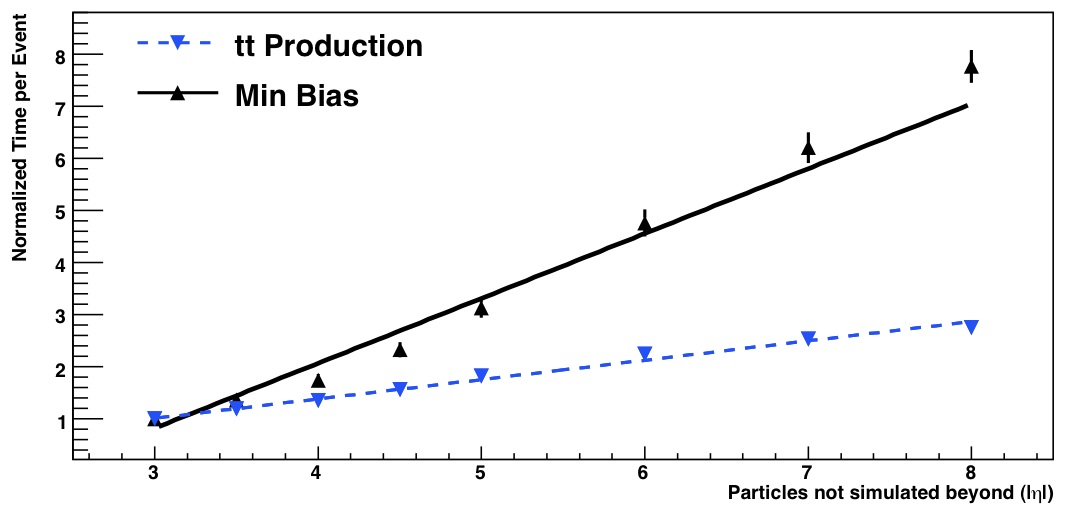}
\end{center}
\caption{CPU time per event increases with varying cuts on the $|\eta|$ of primary particles.  The time is normalized to CPU time for simulation of all primaries inside of $|\eta|<3.0$ and increases between three-fold and eight-fold for simulation of all primaries in $|\eta|<8.0$.  The average of 200 simulated $\ttbar$\ and minimum bias events was taken.  Linear fits are overlaid.\label{fig:cpueta}}
\end{figure*}

Several cuts and transformations can then be made to the event.  The vertex position is smeared to represent the luminous region within ATLAS\footnote{During early data taking, the beams will collide head-on.  Therefore, no crossing angle is added to the simulated events for the time being.}.  It can be shifted if the user desires, but both the shift and smear are given initial default values that represent ideal collisions within the ATLAS detector.  The generated event can be rotated in any direction, though only rotation in $\phi$ is physical.  Primary particles are only passed through the detector simulation if they are within a specified range in $\eta-\phi$.  By default, primary particles with $|\eta| \geq 6.0$ are not simulated to save time.  This cut was chosen to ensure consistent response in the forward ca\-lo\-ri\-me\-ter: a cut at $|\eta|=6.0$ allows a sufficient number of particles to scatter back into the forward detector from high-$\eta$ without requiring an unacceptable amount of CPU time.  Generally, an increase in one unit of pseu\-do\-rapidity corresponds to an increase of 40-120\% in CPU time, so that it is not possible to simulate the very forward detectors like LUCID during a standard simulation job.  Figure~\ref{fig:cpueta} shows the $\eta$-dependence of the CPU time per event.  The increase is approximately three-fold in $\ttbar$\ events and eight-fold in minimum bias events from simulating particles in $|\eta|<3.0$ to simulating particles in $|\eta|<8.0$.  The difference between the two types of events is primarily because the majority of activity in the minimum bias events is forward, and there is considerable central ($\eta<3.0$) activity in the $\ttbar$\ events.

At run time, either through job options or in the production system described in Section~\ref{Prod}, seeds for pseu\-do-ran\-doom number generators to be used by \Geant, Athe\-na, and any particle generators can be set.  Different pseu\-do-ran\-doom number generators may be configured for each.  Since all random number seeds can be controlled, a single job is entirely reproducible.  The seeds can also be written to a file or read from a file, providing an ad\-di\-tion\-al level of reproducibility\footnote{Because of caching in \Geant, it is not possible to reproduce an individual event or track without starting from the beginning of the job.}.

The user must also select a layout for the detector.  As described in Section~\ref{DetDesc}, several layouts of the full detector and various test stands are available.  Combined test beam simulation requires such a different configuration that an independent but similar core drives the \Python\ configuration and loading of user job options.  The distinction is made at run time based on the detector or test stand configuration selected.  The layout of the detector determines what other options are available to the user at run time.  During simulation of the entire ATLAS detector, several ad\-di\-tion\-al options are available.  For example, a neutron time cut (see Section~\ref{SimOpt}) may be enabled.  The magnetic field may be enabled and the field map may be selected.  

The user may optionally select a set of run conditions for the simulation job, through which all options and flags are set and a pre-defined job is run.  This option is particularly useful for testing and debugging.

\subsection{Simulation Initialization}

Although the initialization in a standard Athe\-na job occurs in a single step, for an ATLAS simulation job the initialization is broken into three steps to allow ad\-di\-tion\-al user intervention.  Table~\ref{table:coreInit} summarizes the processes that occur at each one of the three steps. The division of the initialization is such that most modifications to the simulation conditions can be accomplished in job options alone (\ie\ without code modification). Normally, the user provides job options and allows the initialization to progress unhindered. Some parts of the job, for example the detector layout, are only loaded after the initialization has begun.  In order to modify volumes after the layout is loaded, the user must intervene during the initialization.  Only certain commands are effective at each stage of the initialization, since some parts of the \Geant\ simulation have been loaded and created while others have yet to be translated from dictionaries.

\begin{table*}[htb]
\caption{Initialization sequence for the ATLAS simulation.  Dividing the necessary configuration into several distinct steps allows user intervention at critical points.\label{table:coreInit}}
\begin{center}
\begin{tabular}{lp{4.9in}}
\hline\noalign{\smallskip}
{\bf Init. Stage} & {\bf Processes} \\
\noalign{\smallskip}\hline\noalign{\smallskip}
1 & External modules loaded, job properties locked, metadata written, event generation configured, hit file initialized, \Geant\ service created \\
2 & Detector, physics regions, range cuts created, GeoModel geometry translated, truth strategies initialized, magnetic field loaded, physics list selected, user actions initialized \\
3 & Fast simulation models assigned, physics regions constructed, sensitive detectors assigned, \Geant\ run manager and physics models initialized, recording envelopes and visualization initialized \\
\noalign{\smallskip}\hline
\end{tabular}
\end{center}
\end{table*}

Stage one of the initialization occurs as soon as Athe\-na is started.  Several external \Python\ modules are loaded that provide basic functionality for any Athe\-na job.  The job properties provided by the user are read during this phase and are locked.  Once the job properties are locked, any significant modification to the running of a simulation job must be done by directly accessing the affected services and algorithms.  This saves propagation of changes in the case of a late modification to a job property.  Metadata that will be stored with the hit output file are gathered.  External dependencies that require early initialization are loaded, providing a service for GeoModel, a service for database interaction, and a service for frozen showers (see Section~\ref{FastG4Sim}).  The event generation mode (reading external events, generating events from an external generator, or generating single particles on-the-fly) is determined, and any necessary configuration is included for the generator.  A stream is opened for the output hit file, if necessary, and hit containers for each enabled subdetector are added to the new file.  Finally, a service is created to interface with and control \Geant, although at this point \Geant\ is not fully initialized.

Stage two of the initialization begins with the construction of the detector in \Python\ dictionaries.  Dictionaries of physics regions, range cuts, and volumes in which to apply step limitation (see Section~\ref{SimOpt}) are constructed, and all key properties are assigned to the detector facilities or built in dictionaries for later addition to the geometry.  Each enabled piece of the ATLAS detector or test setup is then recursively constructed in GeoModel according to the parameters specified in the geometry and conditions da\-ta\-bases.  After each subdetector has been constructed in GeoModel, it is translated recursively into an equivalent \Geant\ geometry.  After this point in the initialization all volumes and regions are available to the user for modification.  Next, the Monte Carlo truth strategies (see Section~\ref{MCTruth}) are added to the simulation.  The magnetic field is then loaded.  Under normal circumstances, the field is a map loaded from an external data file, the name of which is specified in the geometry database according to the geometry layout selected.  The user may optionally choose to load data from one of the magnetic field test configurations, rather than the standard ATLAS magnetic field, or to create a new, basic magnetic field.  The physics list (see Section~\ref{PhysicsList}) to be used for simulation is also set at this point.

User actions are then initialized.  \Geant\ allows a user to insert pieces of code in various places throughout the simulation event loop, including after each step, when each track is queued (``stacked''), before and after each event, before and after each track is simulated, and before and after each run\footnote{Here, run is used in the \Geant\ sense to refer to a finite set of events within a simulation job.  Several runs may comprise a job, and each run may include an arbitrary number of events.}.  By default, ATLAS includes user actions that monitor simulation time, memory, and the number of tracks generated during each event.  A neutrino cut (see Section~\ref{SimOpt}) is also implemented as a track-stacking action.  Whenever a new track is queued, its type is checked.  The LAr ca\-lo\-ri\-me\-ters also use end-of-event actions for merging hits to save space prior to storage.  All truth storage strategies (see Section~\ref{MCTruth}) are implemented as stepping actions that store interesting interactions based on the type of process, detector region, and energies of the particles involved.  Users may also configure their own actions and add them to the simulation in the same way.  Examples have been constructed for integrating interaction lengths or radiation lengths through the detector when making geantino maps, for stopping or killing particles if certain conditions are met, and for turning on ad\-di\-tion\-al output only under specific conditions in order to study a bug or issue without having to sift through enormous log files.

Stage three of the initialization completes the job prep\-a\-ra\-tion.  During this stage, the fast simulation models are built and added to the volumes to which they have been assigned.  Any physics regions that will be used are constructed.  Sensitive detectors are built and assigned to the regions of the detector that are to be made sensitive (\ie\ in which hits will be stored).  \Geant's run manager and physics models are initialized.  Recording envelopes are added (see Section~\ref{MCTruth}), and any visualization that has been enabled by the user is initialized (see Section~\ref{Vis}).

Once the initialization is complete and all the necessary elements have been loaded into memory, the event loop begins.

\subsection{Monte Carlo Truth Information}
\label{MCTruth}

The \Geant\ simulation adds to the Monte Carlo truth record already defined during generation (see Section~\ref{GenMCTruth}).  Far too many secondary tracks are produced during detector simulation to store  information for every interaction.  Only those interactions which are of greatest relevance to physics analyses are saved, according to several saving rules (``strategies''). Most are applicable only to the inner detector.  For each interaction that satisfies any of the storage criteria, the incoming particle, step information, vertex, and outgoing particles are included in the truth record.  Later in the software chain, individual track segments are recombined so that, for example, a single electron that undergoes several brems\-strah\-lung events along its path is counted as only one ``true'' particle.

The strategies include (with all cuts on kinetic energy)\footnote{For the most recent production, cuts are applied on transverse momentum, $p_T > 100$~\MeV, rather than on kinetic energy.  The lower cut allows for a study of tracking performance in minimum bias events where it may be possible to reconstruct tracks down to only a few hundred~\MeV.}:

\begin{itemize}
\item In the inner detector, brems\-strah\-lung vertices are stored if the primary electron or muon has an energy above 500~\MeV{} and the photon produced has an energy above 100~\MeV{}.
\item In the inner detector, ionization vertices are stored if the primary particle has an energy above 500~\MeV{} and the electron generated has an energy above 100~\MeV{}.
\item In the inner detector, hadronic interaction vertices are stored if the primary particle has an energy above 500~\MeV{}.
\item In the inner detector, decay vertices are stored if the decaying particle has an energy above 500~\MeV{}.
\item In the inner detector, the conversions of photons above 500~\MeV{} are stored.
\item In the ca\-lo\-ri\-me\-ter, muon brems\-strah\-lung vertices are stored if the primary muon's energy was above 1~\GeV{} and the photon generated is above 500~\MeV{}.
\end{itemize}

All cuts and regions of applicability are made configurable, so that any energy cut-offs can be modified and a strategy can be assigned to any volume in the simulation.  ad\-di\-tion\-al rules could be constructed, for example, for tracking shower development within the ca\-lo\-ri\-me\-ter, but many would consume too much CPU time and disk space for use in standard simulation jobs.

Standard simulation jobs also define several volumes that are used to record all particles escaping part of the detector.  All tracks above 1~\GeV{} are typically recorded at the end of the inner detector, the end of the ca\-lo\-ri\-me\-ter, and the end of the muon system (and the end of the ATLAS world volume).  It is possible for the user to configure the simulation at run time to add ad\-di\-tion\-al volumes to the list of these recording volume.  In each case, tracks are saved as they exit each volume.

\subsection{Physics List}
\label{PhysicsList}

Physics lists include all numerical models that describe the particles' interactions in the \Geant\ simulation.  Models are generally good for a single type of interaction and over a limited energy range.  The \Geant\ collaboration provides several combinations of these models that have been tailored to various scenarios as standard physics lists that ship with each distribution. In order to enhance reproducibility and ensure that validated combinations of models are used, only those physics lists provided by the \Geant\ collaboration are used by the ATLAS simulation.  One exception is allowed, namely transition radiation.  Transition radiation is crucial for the tracking portion of the inner detector and is added to each physics list.  

There are several physics lists that are used by ATLAS:

\begin{description}
\item[QGSP\-\_BERT] - the physics list used for all simulation production after 2008.  The list includes the Quark-Gluon String Precompound model (QGSP) and the Bertini intranuclear cascade model (BERT)~\cite{G4} as part of the hadronic physics package.  The electromagnetic physics package includes step-limiting Multiple Coulomb Scattering (MSC).
\item[QGSP\-\_EMV] - the physics list used for simulation production before 2008.  This list included the QGSP model, but without the Bertini cascade.  The MSC of this list was not allowed to limit the step, so it is labeled an electromagnetic variant (EMV).
\item[QGSP\-\_BERT\_HP] - the physics list used for neutron fluence studies and comparisons with the \Fluka\ simulation package~\cite{FLUKA}.  This list includes the QGSP and Bertini models, step-limiting MSC, and ad\-di\-tion\-al ``high-precision'' low-energy neutron physics models.
\end{description}

A step limitation process that controls the maximum allowed step length of a charged particle was added in the inner detector when using the QGSP\-\_EMV physics list. It helped the simulation to better reproduced test beam and cosmic ray data.  The step-limiting MSC that is a part of QGSP\-\_BERT was found to agree equally well with data, and therefore the step limitation was removed from simulation with QGSP\-\_BERT.

These physics lists were studied in detail for each subdetector~\cite{SimOptGroup}.  Table~\ref{table:physhits} shows the number of steps, number of hits in sensitive detector regions, and number of secondary particles with kinetic energy above 50~\MeV{} and 1~\GeV{} within several regions of the detector and for the whole of ATLAS using both the QGSP\-\_EMV and QGSP\-\_BERT physics lists.  Sensitive detectors to record calibration hits (described in Section~\ref{HitFiles}) are included in the cryostat of the LAr ca\-lo\-ri\-me\-ter.  The average was taken of 50~$\ttbar$\ events, where there were on average 482 primary (generator-level) particles per event.  The ca\-lo\-ri\-me\-ter clearly dominates the total number of steps and hits in sensitive detector for both physics lists.  The muon system, though it has a comparable number of hits, consists mostly of shielding and therefore has far fewer hits in sensitive detector regions. The numbers of steps divided into different process types for QGSP\-\_EMV and QGSP\-\_BERT are listed in Tables~\ref{table:physstepsEMV} and~\ref{table:physstepsBERT}.  In both cases, transportation processes dominate the inner detector simulation, while electromagnetic physics and transportation dominate the ca\-lo\-ri\-me\-ter and the muon system.

Simulation time was also examined for each physics list.  Simulation using the QGSP\-\_BERT physics list consumes $\sim2.5$~times more CPU time than does simulation with the QGSP\-\_EMV physics list.  However, applying a neutron time cut (see Section~\ref{SimOpt}) with the QGSP\-\_BERT list reduces simulation time by more than 30\%.  Simulation with QGSP\-\_BERT\_HP requires approximately five times more CPU time than QGSP\-\_EMV.  Therefore, the QGSP\-\_BERT\_HP physics list cannot be used for standard simulation.

\begin{table*}[htb]
\caption{Number of steps, number of hits in sensitive detector (SD) regions, and number of secondary particles with kinetic energy above 50~\MeV{} and 1~\GeV{} within several regions of the detector and for the whole of ATLAS, using both the QGSP\-\_EMV and QGSP\-\_BERT physics lists.  The average was taken of 50~$\ttbar$\ events, where the average number of primary tracks per event was 482.  Sensitive detectors to record calibration hits are included in the cryostat of the LAr ca\-lo\-ri\-me\-ter.\label{table:physhits}}
\begin{center}
\begin{tabular}{lllll}
\hline\noalign{\smallskip}
{\bf QGSP\-\_EMV} & {\bf Steps} & {\bf Hits in SD} & {\bf Sec. above 50~\MeV{}} & {\bf Sec. above 1~\GeV{}}\\
\noalign{\smallskip}\hline\noalign{\smallskip}
Inner Detector & $1.80\times10^{6}$ & $3.10\times10^{5}$ &  1,570 &   260 \\
Calorimetry    & $1.87\times10^{7}$ & $6.87\times10^{6}$ & 39,900 & 2,040 \\
Muon System    & $1.90\times10^{6}$ & 1,030              &  7,820 &   332 \\
\noalign{\smallskip}\hline\noalign{\smallskip}
Total ATLAS    & $2.24\times10^{7}$ & $7.18\times10^{6}$ & 49,300 & 2,630 \\
\noalign{\smallskip}\hline\noalign{\smallskip}
\noalign{\smallskip}\hline\noalign{\smallskip}
{\bf QGSP\-\_BERT} & {\bf Steps} & {\bf Hits in SD} & {\bf Sec. above 50~\MeV{}} & {\bf Sec. above 1~\GeV{}}\\
Inner Detector & $2.13\times10^{6}$ & $1.98\times10^{5}$ &  1,450 &   269 \\
Calorimetry    & $3.93\times10^{7}$ & $1.36\times10^{7}$ & 40,100 & 2,170 \\
Muon System    & $2.69\times10^{6}$ & 1,285              &  8,210 &   385 \\
\noalign{\smallskip}\hline\noalign{\smallskip}
Total ATLAS    & $4.41\times10^{7}$ & $1.38\times10^{7}$ & 49,700 & 2,820 \\
\noalign{\smallskip}\hline
\end{tabular}
\end{center}
\end{table*}

\begin{table*}[htb]
\caption{Number of steps for various processes and detector regions during simulation with the QGSP\-\_EMV physics list.  The average was taken of 50~$\ttbar$\ events, where the average number of primary tracks per event was 482.  The ``other processes'' in the inner detector are primarily step limitation processes.\label{table:physstepsEMV}}
\begin{center}
\begin{tabular}{llllll}
\hline\noalign{\smallskip}
{\bf Process}    		& {\bf Inner Detector} & {\bf ca\-lo\-ri\-me\-ter} 	& {\bf Muon System} \\
\noalign{\smallskip}\hline\noalign{\smallskip}
{Transportation} 		& $1.50\times10^6$ 	& $9.33\times10^6$ 	& $2.15\times10^5$ \\
{MSC}            		& 4,910 		& $1.09\times10^5$ 	& 5,200 \\
{Photoelectric Effect}      	& 6,060            	& $1.32\times10^6$ 	& $2.03\times10^5$ \\
{Compton Scattering} 		& 12,800       		& $1.43\times10^6$ 	& $4.26\times10^5$ \\
{Ionization}     		& $1.08\times10^5$ 	& $4.97\times10^6$ 	& $8.10\times10^5$ \\
{brems\-strah\-lung} 		& 6,310 		& $1.28\times10^6$ 	& $1.87\times10^5$ \\
{Conversion}     		& 434 			& 82,400 		& 17,300 \\
{Annihilation}   		& 291              	& 82,800           	& 17,800 \\
{Decay}          		& 254 			& 2,320 		& 538 \\
{Other Hadronic Interaction}	& 1,710 		& $1.23\times10^5$ 	& 21,600 \\
{Other Process}  		& $1.56\times10^5$ 	& 4,800            	& 831 \\
\noalign{\smallskip}\hline\noalign{\smallskip}
{\bf Total}          		& $1.80\times10^6$ 	& $1.87\times10^7$ 	& $1.90\times10^6$ \\
\noalign{\smallskip}\hline
\end{tabular}
\end{center}
\end{table*}

\begin{table*}[htb]
\caption{Number of steps for various processes and detector regions during simulation with the QGSP\-\_BERT physics list.  The average was taken of 50~$\ttbar$\ events, where the average number of primary tracks per event was 482.  The ``other processes'' in the ca\-lo\-ri\-me\-ter and muon system are primarily neutron killer processes.\label{table:physstepsBERT}}
\begin{center}
\begin{tabular}{llllll}
\hline\noalign{\smallskip}
{\bf Process}    		& {\bf Inner Detector} & {\bf ca\-lo\-ri\-me\-ter} 	& {\bf Muon System} \\
\noalign{\smallskip}\hline\noalign{\smallskip}
{Transportation} 		& $1.76\times10^6$ 	& $1.46\times10^7$ 	& $2.31\times10^5$ \\
{MSC}            		& $2.31\times10^5$ 	& $1.48\times10^7$ 	& 5,200 \\
{Photoelectric Effect}      	& 6,760            	& $1.37\times10^6$ 	& $2.32\times10^5$ \\
{Compton Scattering} 		& 14,800       		& $1.66\times10^6$ 	& $5.03\times10^5$ \\
{Ionization}     		& $1.03\times10^5$ 	& $4.81\times10^6$ 	& $9.71\times10^5$ \\
{brems\-strah\-lung} 		& 6,060 		& $1.22\times10^6$ 	& $1.92\times10^5$ \\
{Conversion}     		& 416 			& 86,800 		& 18,100 \\
{Annihilation}   		& 271              	& 87,000           	& 18,500 \\
{Decay}          		& 212 			& 1,670 		& 402 \\
{Other Hadronic Interaction}	& 2,190 		& $6.66\times10^5$ 	& $1.23\times10^5$ \\
{Other Process}  		& 426 			& 25,400            	& 5,720 \\
\noalign{\smallskip}\hline\noalign{\smallskip}
{\bf Total}          		& $2.13\times10^6$ 	& $3.93\times10^7$ 	& $2.69\times10^6$ \\
\noalign{\smallskip}\hline
\end{tabular}
\end{center}
\end{table*}

\subsection{Simulation Optimizations}
\label{SimOpt}

In order to optimize use of both disk space and CPU time, several other modifications are made to the standard \Geant\ simulation~\cite{SimOptGroup,SimOptGroup2}.

Comparing the QGSP\-\_BERT physics list to the QGSP\-\_EMV physics list, approximately three times as many neutrons are generated in typical hard scattering events, and they travel approximately three times further.  These neutrons cause an increase in the output hit file size of approximately 75\% as well as an increase in CPU time per event for hard scattering events.  A \Geant\ neutron time cut is, therefore, applied which removes all neutrons 150~ns after the primary interaction.  This was found to be sufficient time for the hadronic shower development and did not degrade the energy scale or energy resolution of the ca\-lo\-ri\-me\-ters.  Output files are the same size when using the QGSP\-\_BERT physics list with this cut enabled as they are when using the QGSP\-\_EMV physics list without a neutron time cut.  The simulation time required for QGSP\-\_BERT is reduced by 10-15\% when the neutron cut is enabled.

Neutrinos are also removed as soon as they are created in the simulation.  No particle is allowed by \Geant\ to step through more than one volume at a time.  Therefore, neutrinos may require several thousand steps to exit the entire ATLAS detector.  They may therefore consume a noticeable fraction of simulation time, even though their interaction probability is practically null.  The removal is done when the particles are stacked.

Range cuts are \Geant\ parameters that control the creation of secondary electrons or photons during brems\-strah\-lung and ionization processes.  If the expected range of the secondary is less than some minimum value, the energy of that secondary particle is deposited at the end of the primary particle's step and no separate secondary is produced. Effectively, this parameter defines an energy scale at which particle propagation may be ignored.  By increasing the range cuts throughout the detector one can decrease the CPU time required per event.  Particularly near boundaries and thin materials, the detector's sampling fraction may be affected if the range cuts are too large.  Range cuts can be specified separately for electrons, pos\-i\-trons, and photons, but in ATLAS the same distance is used for all three.  Range cuts are specified as a distance, and for each material the distance is translated into an energy based on the average energy loss of a particle in that material. For the majority of the ATLAS detector, range cuts take a default value of 1~mm.  Exceptions are listed in Table~\ref{table:rc}.  Deviations usually occur in sensitive volumes that are very thin, where it is important to correctly calculate the sampling fraction of the detector or model the energy deposition.  Reduced range cuts are also applied to very thin volumes that are adjacent to sensitive volumes for the same reason.  In the monitored drift tube muon chambers, for example, range cuts are only reduced in the thin aluminum tubes surrounding the sensitive detector (gas) - the gas itself takes the standard 1~mm cuts.  In some shielding volumes it may be possible to relax range cuts considerably without degrading physics performance.

\begin{table*}[htb]
\caption{Range cuts for detectors that do not take the ATLAS default of 1~mm.\label{table:rc}}
\begin{center}
\begin{tabular}{lr}
\hline\noalign{\smallskip}
{\bf Subdetector} & {Range Cut Value} \\
\noalign{\smallskip}\hline\noalign{\smallskip}
Silicon pixels and strips in the inner detector & 0.05~mm \\
Gas in the transition radiation tracker & 0.05~mm \\
Electromagnetic Barrel and Endcap ca\-lo\-ri\-me\-ters & 0.1~mm \\
Forward ca\-lo\-ri\-me\-ter (all compartments) & 0.03~mm \\
Aluminum tubing of monitored drift tube muon chambers & 0.05~mm \\
\noalign{\smallskip}\hline
\end{tabular}
\end{center}
\end{table*}

\Geant\ uses a set of parameters to control errors and accumulated biases on charged particles transportation through a magnetic field.  Because the equation of motion is solved numerically, the user must select the numerical integration method to be used, including the order of integration, and the tolerances on the errors of the step.  ATLAS has chosen to use the \Geant\ standard fourth-order Runge-Kutta method with the default error parameters for the majority of the detector.  These parameters are generally satisfactory and result in errors and biases that are less than the position resolution of the detector.  In the inner detector, however, tracks were found to be shifted sufficiently that detector residuals were affected.  Here, stepping parameters were tightened by an order of magnitude. Further optimization of the stepping algorithms of \Geant\ has been undertaken, including the configuration of the choice of stepper and stepping parameters as a function of the initial particle type, energy, and position within the detector.  Such a configuration can allow more careful stepping of muons in the calorimetry without degrading the total performance of the simulation significantly.  Muons in particular can accumulate a significant bias after passing through all the sampling layers of the ca\-lo\-ri\-me\-ter, making more accurate tracking necessary.  As a fourth-order stepper requires four values of the magnetic field to be calculated, optimization of magnetic field map access will also be key to improving the performance of the simulation's tracking.

\subsection{Hit Storage Format}
\label{HitFiles}

The output from the simulation is a hit file, containing some metadata describing the configuration of the simulation during the run, all requested truth information, and a collection of hits for each subdetector.  The hits are records of energy deposition, with position and time, during the simulation.  Each subdetector is responsible for implementing their own sensitive detector for the selection, processing, and recording of these hits.  In most subsystems, including the inner detector and muon system, this consists simply of recording all hits that occur in sensitive regions of the detector for subsequent storage.  Some ad\-di\-tion\-al manipulation is done at the end of each event to compress the output as much as possible; still, the files are typically 2 MB per event for hard scattering events (\eg\ $\ttbar$\ production).

The file size is large, mostly due to the inner detector, for which the majority of hits are independently stored.  Merging hits there is difficult, since they tend to be isolated and cannot normally be merged across readout channels.  These consume typically 60\% of the disk space in a hit file (\eg\ 65\% of the hit file for \ttbar\ events).  In the calorimetry, there are far too many hits created by electromagnetic and hadronic showers for the individual storage of a four-vector for each (see Section~\ref{PhysicsList}).  Instead, hit merging occurs at the end of each event.  By optimizing time binning, hits can be compressed to a large extent.  About 10\% of the hit file is consumed by optional ``calibration hits'' for the ca\-lo\-ri\-me\-ters, hits in dead material, stored to improve the detector calibration and missing energy calculation and to study simulation-based ca\-lo\-ri\-me\-ter calibration schemes.  Under normal circumstances, the muon systems contribute a negligible portion of the hit file.  The contributions by subdetector can be found in Table~\ref{table:hitsize} for the average of 50 simulated $\ttbar$\ events.  Here and elsewhere in this paper, file sizes are without compression and are taken from ROOT~\cite{ROOT}. In practice, compression reduces the actual disk space required for the files, but file-level metadata adds several hundred kilobytes.

\begin{table*}[htb]
\caption{Hit collection size, in kB per event, by subdetector.  The average was taken of 50 simulated $\ttbar$\ events. ca\-lo\-ri\-me\-ter calibration hits are hits in the dead material of the ca\-lo\-ri\-me\-ters stored for studying simulation-based ca\-lo\-ri\-me\-ter calibration schemes.\label{table:hitsize}}
\begin{center}
\begin{tabular}{lrr}
\hline\noalign{\smallskip}
{\bf Collection Name}                & {\bf Size [kB/event]} & {\bf Percentage of File} \\
\noalign{\smallskip}\hline\noalign{\smallskip}
Silicon pixel tracker                & 82                 & 4\% \\
Silicon strip tracker                & 356                & 16\% \\
Transition radiation tracker         & 921                & 46\% \\
Electromagnetic Barrel ca\-lo\-ri\-me\-ter   & 89                 & 4\% \\
Electromagnetic Endcap ca\-lo\-ri\-me\-ter   & 104                & 5\% \\
Hadronic Barrel ca\-lo\-ri\-me\-ter          & 29                 & 1\% \\
Hadronic Endcap ca\-lo\-ri\-me\-ter          & 22                 & 1\% \\
Forward ca\-lo\-ri\-me\-ter                  & 42                 & 2\% \\
ca\-lo\-ri\-me\-ter calibration hits         & 243                & 12\% \\
Muon system (all collections)        & 3                  & <1\% \\
Truth (all collections)              & 134                & 7\% \\
\noalign{\smallskip}\hline\noalign{\smallskip}
{\bf Total}                          & {\bf 1987}         & \bf{100\%} \\
\noalign{\smallskip}\hline
\end{tabular}
\end{center}
\end{table*}

By comparing Tables~\ref{table:physhits} and~\ref{table:hitsize}, one can understand these numbers in terms of hits in the sensitive detector region.  Although the muon system is large, the majority of it is shielding.  Therefore, it collects far fewer hits than the other subsystems and requires less disk space for the hit records.  The calorimetry produces 95\% of the hits in sensitive regions during simulation.  Because of the compression applied prior to storage, the calorimetry comprises only 25\% of the hit file.

\subsection{Visualization}
\label{Vis}

\begin{figure*}
\begin{center}
\includegraphics[height=3.6in]{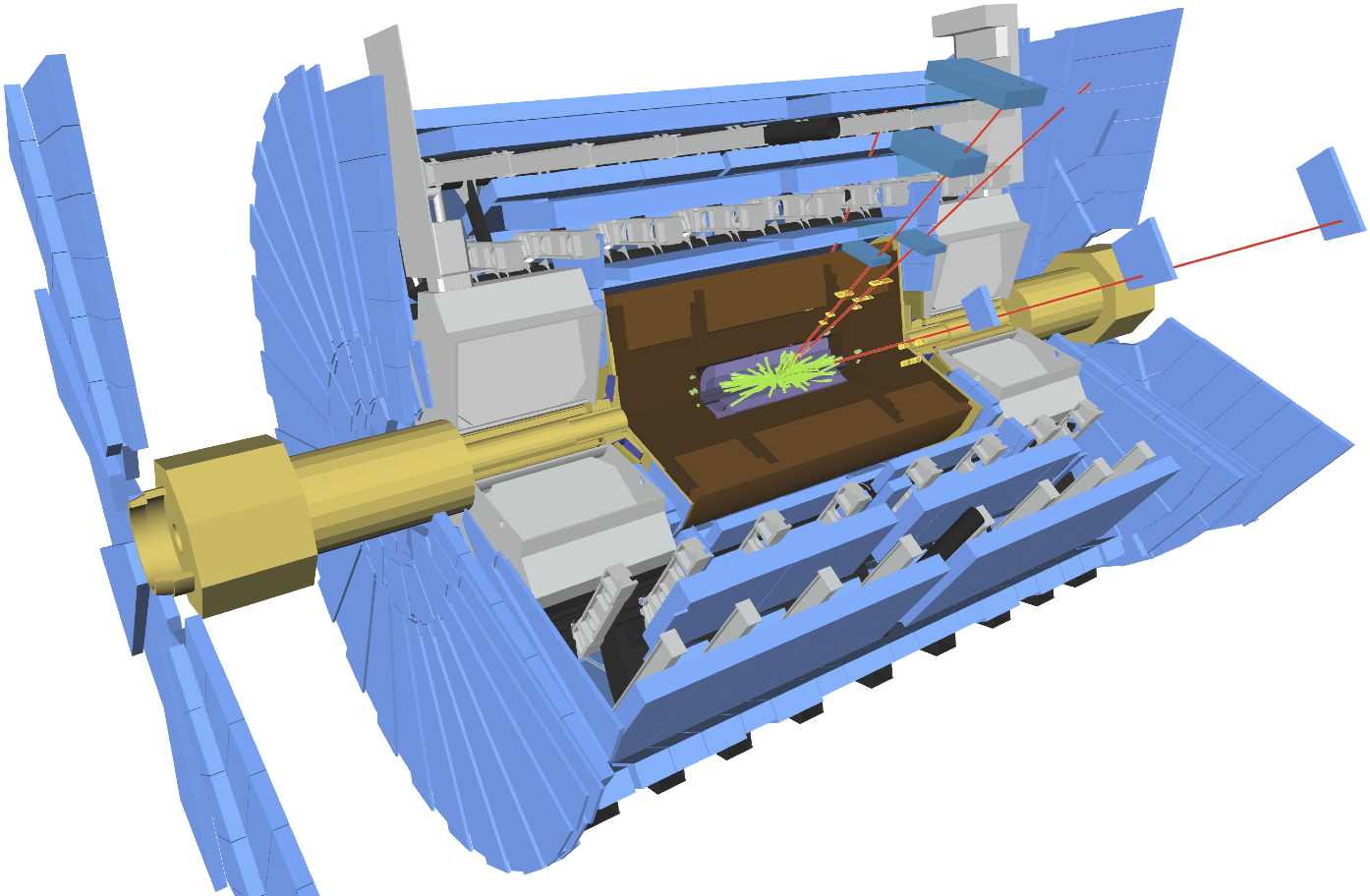}
\end{center}
\caption{An event display made with VP1.  A Higgs boson decays into four muons (shown in red).  Inner detector tracks are in green, and energy deposited in the ca\-lo\-ri\-me\-ter by the muons is shown in yellow.\label{fig:vis1}}
\end{figure*}

\begin{figure*}
\begin{center}
\includegraphics[height=3.3in]{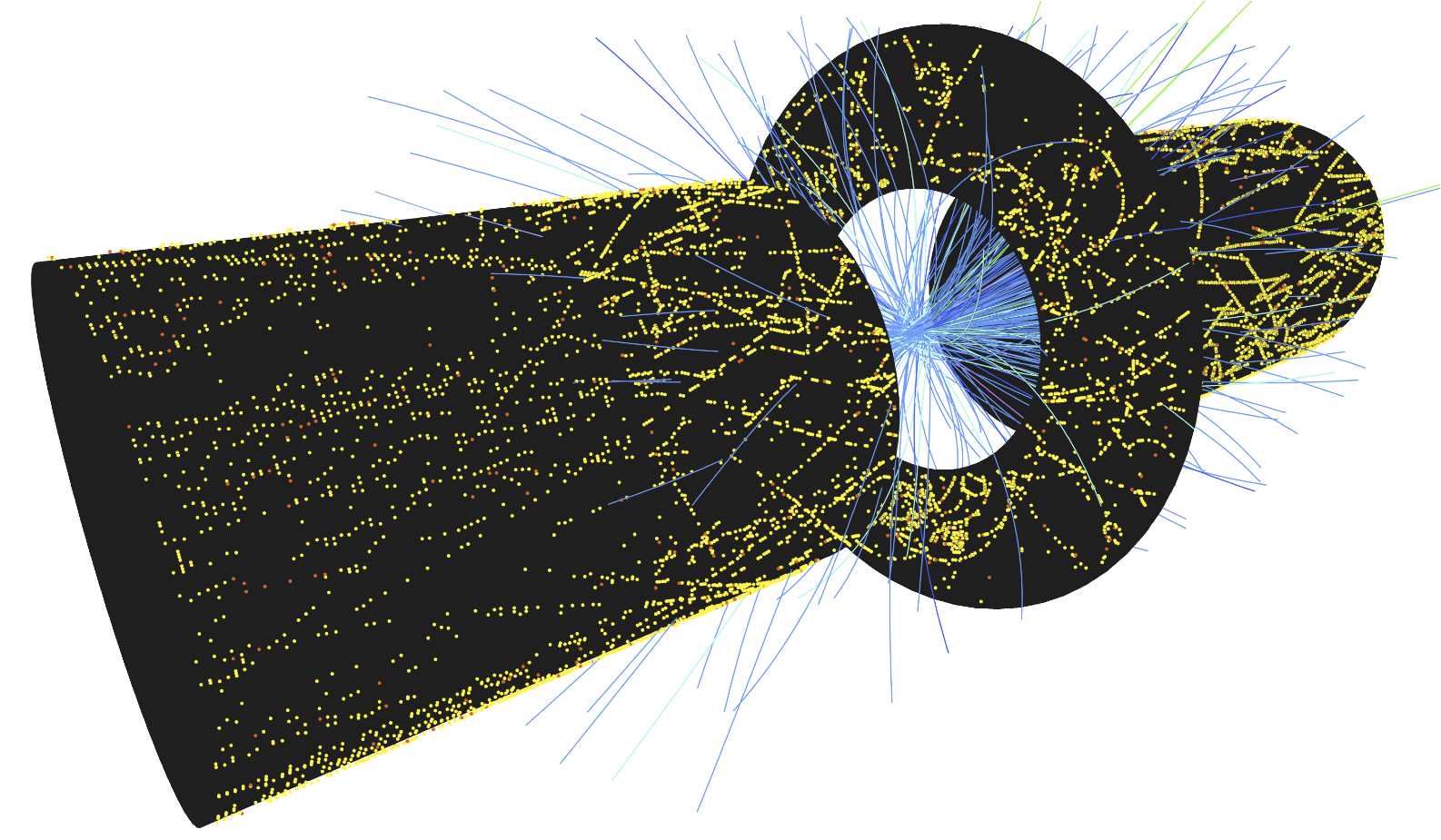}
\end{center}
\caption{A Higgs boson decaying into four muons, with only the inner detector tracks and hits in the TRT being displayed by VP1.\label{fig:vis2}}
\end{figure*}

Visualization is used to understand anomalies or features in odd events, occasionally to debug errors due to geometry, and to check for overlaps and touching volumes in the geometry that can be spotted by eye. Although \Geant\ contains viewing software of its own, because the geometry of ATLAS must be translated from GeoModel into \Geant\ format it is useful to use a viewer that can construct geometry directly from its GeoModel description. A general purpose three-dimensional event display program, VP1\footnote{ATLAS is at Point 1 of the LHC ring.  The name VP1 is short for Virtual Point 1.}~\cite{VP1}, has been developed specifically for ATLAS.  It is optimized specifically for the visualization of the ATLAS geometry and is arguably the most useful tool for understanding and debugging of detector description across all ATLAS subsystems. Two examples of VP1 event displays are in Figures~\ref{fig:vis1} and~\ref{fig:vis2}. Ray Tracer, the \Geant\ visualization utility, has also been used to visualize portions of the detector containing some exotic shapes.

VP1, as well as other event display programs used in ATLAS (\eg\ Atlantis~\cite{Atlantis} and Persint~\cite{Persint}), are mostly used for visualizing real and simulated events after they have run through the common reconstruction software. The VP1 viewer can be injected directly into the simulation job in order to visualize events immediately after the simulation step.

%%\clearpage

\section{Digitization}
\label{Digi}
\label{OverDigi}

The ATLAS digitization software converts the hits produced by the core simulation into detector responses: ``digits.'' Typically, a digit is produced when the voltage or current on a particular readout channel rises above a pre-configured threshold within a particular time-win\-dow. Some subdetector's digit formats include the signal shape in detail over this time, while others simply record that the threshold has been exceeded within the relevant time window.

The peculiarities of each subdetector's charge collection, including cross-talk, electronic noise and channel-dependent variations in detector response are modelled in subdetector-specific digitization software~\cite{PixelTest,SCTDigi,TRTSim,LArDigi,MuonDigi}. The various subdetector digitization packages are steered by a top-level \Python\ digitization package which ensures uniform and consistent configuration across the subdetectors.  The properties of the digitization algorithms were tuned to reproduce the detector response seen in lab tests, test beam data, and cosmic ray running. Dead channels and noise rates are read from database tables to reproduce conditions seen in a particular run. In some cases, dead channels are removed during the reconstruction step. 

The digits of each subdetector are written out as Raw Data Objects (RDOs). For some subdetectors this requires the digits produced to be converted to RDOs by a second algorithm during the digitization process. For others there is no intermediate digit object and RDOs are produced directly from the hits. In addition to RDOs, the digitization algorithms can also produce Simulated Data Objects (SDOs). These SDOs contain information about all the particles and noise that contributed to the signal produced in the given sensor and the amount of energy contributed to the signal by each. The relationship between RDOs and SDOs depends on the particular subdetector. For example, in the SCT each RDO represents a group of consecutive strips which recorded a hit, whereas one SDO is produced for each strip where energy was deposited by a particle in the Monte Carlo truth tree. No SDOs are created in the ca\-lo\-ri\-me\-ter. SDOs are mainly used for determining tracking efficiency and fake track rates.

Simulating the detector readout in response to a single interesting hard scattering interaction is unrealistic. In reality, for any given bunch crossing there may be multiple proton-proton interactions.  In addition to the hard scattering which triggers the detector readout, many inelastic, non-diffractive proton-proton interactions may appear. These interactions must be included in a realistic model of detector response. The effects of beam gas and beam halo interactions, as well as detector response to long-lived particles, must be incorporated. These interactions are treated separately at the event generation and simulation stages.  Within a digitization job, hits from the hard scattering are overlaid with those from the requested number of these ad\-di\-tion\-al interactions before the detector response is calculated. Because of long signal integration times, most subdetector responses are affected by interactions from neighboring bunch crossings as well. Therefore, ad\-di\-tion\-al interactions offset in time are overlaid as necessary. The overlaying off these various types of events, known collectively as ``pile-up,'' is described in Section~\ref{PileUpDigi}.

Before reconstruction can be run, byte\-stream data from the real detector must be converted into RDO format. As mentioned above, the digitization usually avoids this step by writing out RDOs directly. However, in order to do simulation studies with the High Level Trigger it is necessary to translate the RDO files into byte\-stream format. There is some loss due to truncation in the first conversion from RDO to byte\-stream, but the inverse operation is basically lossless. Having the ability to convert output in both directions also allows evaluation of the conversions themselves.

\subsection{Digitization Configuration}
\label{ConfigDigi}

The ATLAS digitization takes as input hit files produced by the ATLAS simulation. For pile-up simulation, there are also input hit files for each type of background interaction to be overlaid. In such cases it is the main hard scattering event which sets the run number and event number.  Run and event numbers from overlaid events are ignored.

The digitization steering package exists entirely in the \Python\ layer and configures how the digitization will be performed before the event loop starts. This configuration is highly flexible, but also ensures that sensible default values are given for each configurable property of the job.  In the configuration of digitization jobs, the user may specify the number of events to digitize, the number of leading events to skip in the input file, the input hit file(s), and the output file. Digitization and writing out of RDOs may be enabled or disabled by subdetector. In order to ensure consistency, the detector layout version is, by default, read from the hard scattering events' hit file metadata. 

Digitization options also include the following: 
\begin{description}
\item[Detector Noise Simulation:] Detector noise simulation can be turned off in the inner detector, ca\-lo\-ri\-me\-ter or muon spectrometer or any combination thereof. This is useful for data overlay jobs where noise is taken from real data events and for studies using a noise-free detector.
\item[Random Number Services:] The type of random number engine to be used in all digitization algorithms can be specified (Ranlux64, the default, or Ranecu~\cite{raneng}). Each algorithm has one or more random number streams. Random number seeds can be initialized from a text file or set in job options.  The user may alternately specify an offset from the default values of the seeds, to be used in all streams.
\item[Metadata:] In the default configuration, metadata from the simulation stage are used to configure the physics list (for setting the sampling fraction of the ca\-lo\-ri\-me\-ters) and the detector layout.  The metadata can be overridden.
\item[Pile-up Background Events:] The overlay of minimum bias, cavern background, beam gas and beam halo events can all be configured separately. In each case the mean number of events (if any) per bunch crossing to be overlaid and a collection of files containing the events to be overlaid onto the signal events can be specified.
\item[Beam Properties:] The LHC beam bunch spacing can be configured, as can the number of bunch crossings to overlay before and after the hard scattering event. 
\item[Detector Conditions:] Default detector conditions (including, \eg, dead electronics and noisy channels) are associated with each detector layout.  Non-default conditions may be specified globally or by subdetector for use in digitization.
\end{description}

After a check to make sure that at least one subdetector has been left switched on, the input and output streams are initialized. GeoModel is initialized using the detector layout and conditions versions read from the hits file metadata or specified by the user. Setting the detector layout version to be different from that used in the simulation is possible, but considered to be an expert action. The magnetic field service is initialized at this point.  It is necessary because the magnetic field affects charge propagation from the active regions of the detector to the readout surfaces. 

At this point, caches for pile-up events are created and configured with the appropriate collection of hits files as well as the number of events to be overlaid per bunch crossing. These caches are controlled by an overall pile-up manager service. A second pile-up service is created to hold information about the time window within which interactions can affect the response recorded by each subdetector. During the initialization stage, this information can be combined with the bunch spacing to calculate the number of bunch crossings which should be simulated for each subdetector for each event.

Subsequently, the subdetector digitization algorithms are configured and added to the sequence of algorithms to be run in the job. The collections of RDOs, hits, and truth information which are to be recorded are added to the output stream. Digitization algorithms exist for the following subdetectors: 
\begin{description}
\item[Inner Detector:] BCM, silicon pixel tracker, SCT, and TRT.
\item[ca\-lo\-ri\-me\-ter:] LAr and tile ca\-lo\-ri\-me\-ters. Separate algorithms also exist to simulate the formation of trigger towers in the ca\-lo\-ri\-me\-ters, which serve as inputs to the level one trigger.
\item[Muon Spectrometer:] Cathode Strip Chambers, Monitored Drift Tubes, Resistive Plate Chambers and Thin Gap Chambers.
\end{description} 
If requested, the level one trigger simulators are added to the algorithm sequence, provided that the digitization of the relevant parts of ATLAS have been turned on. The default mode of simulation production is to run the level one trigger simulation during the reconstruction step rather than as part of the digitization step.

As the digitization algorithm for each subdetector is configured, the names and seeds for the random number streams it requires are added to a list. In the case where seeds are to be read in from a file, the default list of stream names and their seeds are replaced by the file contents. Once all algorithms have been configured, the list is used to configure the random number service. Separate random number streams are used for each subdetector digitization algorithm and give the same result independent of what is used for the other subdetectors\footnote{Here ``digitization algorithm'' does not include the ca\-lo\-ri\-me\-ter trigger tower simulation algorithms, which require the corresponding ca\-lo\-ri\-me\-ter digitization to be performed.  Similarly, the level one trigger simulation requires the simulation and digitization of the expected trigger inputs to give meaningful results.}.

Much of the job configuration information, along with the detector layout version, is written to the output file as digitization metadata. The run number provided in the simulation metadata is used to establish a validity range for the digitization metadata corresponding to the current run only. At this point the digitization job is fully configured and the event loop begins.

\subsection{Pile-up}
\label{PileUpDigi}

To simulate pile-up, various types of events are read in, and hits from each are overlaid. The different types considered can be configured at run time, and normally comprise signal, minimum bias, cavern background, beam gas, and beam halo events. The number of events to overlay of each type per bunch crossing may also be set at run time and is a function of the luminosity to be simulated. The mean number of interactions per bunch crossing (BC), for example 23 at the design luminosity of $10^{34}$~cm$^{-2}$s$^{-1}$ with 25~ns bunch spacing, depends linearly on luminosity and bunch spacing. However, this number is Poisson-distributed, with a long tail beyond the most probable value. Thus, a substantial fraction of the bunch crossings will have more than the average number of interactions. In addition, the ATLAS subdetectors are sensitive to hits several bunch crossings before and after the BC that contains the hard scattering event (which triggers the readout). Table~\ref{tab:pileuprange} shows the simulation window for each detector along with the corresponding number of bunch crossings for 25~ns and 75~ns bunch spacing. All of these detector and electronic effects are taken into account during the pile-up event merging.

\begin{table*}
\caption{The time window (relative to the current bunch crossing) during which interactions in each subdetector are simulated during pile-up jobs, along with the corresponding numbers of bunch crossing simulated in the case of 25~ns bunch spacing and 75~ns bunch spacing.\label{tab:pileuprange}}
\begin{center}
 \begin{tabular}{lrrr}
\hline\noalign{\smallskip}
  {\bf Subdetector}   &{\bf Simulation Window [ns]}&{\bf No. Bunch Crossings}  &{\bf No. Bunch Crossings}\\
                      &                          &{\bf (25 ns bunch spacing)}&{\bf (75 ns bunch spacing)}\\
\noalign{\smallskip}\hline\noalign{\smallskip}
   BCM                &  -50, +25                &  4   	          &  1   		        \\
   Pixel trackers     &  -50, +25                &  4   		      &  1   		        \\
   SCT                &  -50, +25                &  4   		      &  1   		        \\
   TRT                &  -50, +50                &  5   		      &  1   		        \\
   LAr ca\-lo\-ri\-me\-ter    & -801,+126                &  38   		      &  12   		        \\
   Tile ca\-lo\-ri\-me\-ter   & -200,+200                &  17   		      &  5   		        \\
   Muon chambers      & -1000,+700               &  69   		      &  23   		        \\
 \noalign{\smallskip}\hline
\end{tabular}
\end{center}
\end{table*}

\subsubsection{Cavern Background}
\label{CavernBackground}

Neutrons may propagate through the ATLAS cavern for a few seconds before they are thermalized, thus producing a neutron-photon gas.  This gas produces a constant background, called ``cavern background,'' of low-energy electrons and protons from spallation. The cavern background consists mainly of thermalized slow neutrons, long-lived neutral kaons and low-energy photons escaping the ca\-lo\-ri\-me\-ter and the forward shielding elements. Muon detectors are most affected by high cavern-background rates. The radiation levels to be expected in the ATLAS cavern scale with luminosity, and they have been simulated as a function of $r$ and $z$~\cite{Shupe} for the design luminosity of $10^{34}$~cm$^{-2}$s$^{-1}$. Depending on the type of radiation, exact composition of the equipment, and sensitivity of the study, the rates sometimes have to be increased by a safety factor. Cavern background is produced in the following way:

\begin{itemize}
  \item A standalone dedicated {\sc Ge\-ant3/\-GCALOR}-\-based~\cite{GCALOR} detector simulation program with improved neutron propagation and a simplified ATLAS detector geometry is run on proton-proton collisions. The cavern walls are not included in the detector description.  The output of this program includes particle fluxes in the envelopes surrounding muon spectrometer chambers. The fluxes are provided as list of particles with all related parameters per proton-proton interaction at the entrance of each envelope.
  \item The kinematic information of all particles generated by {\sc Geant3/GCALOR} is converted to HepMC format, and the flux is modified to be uniform in the time interval of the required bunch spacing (typically $[0,25\,$ns$]$).
  \item The simulation is then carried out using the full detector geometry and \Geant, and hits are stored.
  \item The cavern events are mixed, with a safety factor of up to 10, at the digitization level with the minimum bias and signal events.
\end{itemize}

There are a number of issues with the current simulation of the cavern background. The original primary cavern events were generated in an older version of \Pythia\ where the generated particle density is a factor of two lower than in the newer versions of \Pythia~\cite{Moraes:2007rq,BadPythia2}. The statistics for the available cavern events are limited: 40,000 events are available with a safety factor of 1; 10,000 events are available with a safety factor of 2 or 5; and 5,000 events are available with a safety factor of 10. Because of the limited statistics, a number of monitored drift tubes in the muon detector fire more often than expected (\ie\ there are spikes in the hit response of the detector). Ad\-di\-tion\-ally, neutral particles are tracked through the entire detector during simulation, thus producing ad\-di\-tion\-al hits from particles that should have been removed at the edges of the muon chamber envelopes (multiple counting). 

In the short term, the problem of limited statistics of the cavern events has been alleviated by taking advantage of the $\phi$-symmetry of the muon spectrometer: the cavern events are rotated and re-simulated eight times or more (in multiples of eight). Further improvement in the available cavern statistics can be achieved by repeating the simulation of the cavern events many times with different random number seeds, since the probability of a neutral interaction is very low, of the order 1\%. 

\subsubsection{Beam Halo and Beam Gas}
\label{BeamHalo}

Beam halo is the background resulting from interactions between the beam and upstream accelerator elements. The flux from upstream (in the tunnel and collimators) is provided by the LHC Machine Division~\cite{BeamHalo1,BeamHalo2}.  Beam halo events are generated as discrete particle losses against the upstream collimators. The LHC machine division has estimated the proton loss rate in design conditions as being on the order of 1~MHz. \Fluka\ simulation of the last 150~m of the beamline indicates that daughter particles from these proton losses will reach the cavern wall (23~m from the interaction point) at a rate of $\sim 400$~kHz. This flux is input to the normal \Geant\ simulation to produce hit files. 

Beam gas includes the residual hydrogen, oxygen, and carbon gasses in the ATLAS beam pipe. Beam gas interaction events are generated with Hijing (see Section~\ref{Hijing}) with appropriate time offsets. The interactions are allowed to take place anywhere in the beam pipe of ATLAS, 23~m in either direction from the interaction point.

\subsubsection{Pile-up with Real Data}

The pile-up mechanism described above will not work with real data, because it begins at the hit level. One must instead overlay events beginning from detector electronics output (RDOs). One may collect minimum bias, cavern background, beam halo, and beam gas backgrounds from the same ``zero bias'' trigger used to understand detector electronic noise. Then, one would overlay hits from simulated hard scattering events onto the zero bias trigger data to simulate the pile-up. The zero bias trigger data needed for this type of event overlay can be selected at random from the filled-bunch crossings\footnote{The zero bias trigger is not a minimum bias trigger}. The subdetectors should be read out with as little zero-suppression as is possible and with the HLT in Òpass-throughÓ mode (\ie\ without further filtering).  One can use bunch-by-bunch luminosity information to correctly weight the event sample for pile-up studies.

In principle one needs as many zero bias events as generated events, but in practice zero bias events can be reused with independent simulated data sets without introducing any bias. During data taking, zero bias events are sampled at all times, because detector and cavern conditions are likely to vary with time. The zero-bias events could, for example, be collected exactly one orbit after a high-$p_T$ trigger has fired. Appropriately pre-scaled to the output rate needed for simulations (on the order of 1-2 Hz, or about 1\% of the recorded events), this means that the rate follows the luminosity and that the bunch structure is guaranteed to be right.

\subsection{RDO Storage Format}

The ATLAS detector electronics produce data in byte\-stream format. The RDO format can be thought of as a POOL-compatible version of the byte\-stream\footnote{POOL compatibility requires separate transient and persistent object representation.}. The file size on disk is typically around 2.5 MB per event for hard scattering events (\eg\ $t\overline{t}$ production) and increases in the presence of pile-up. Table~\ref{tab:rdofile} shows an example of the disk consumption by container for 50~$\ttbar$\ events without pile-up and with pile-up at $10^{33}$~cm$^{-2}$s$^{-1}$. In the absence of pile-up, one of the main consumers of disk space are the calibration hit collections, as described in Section~\ref{HitFiles}, which are copied directly from the hit file to the RDO. As pile-up luminosity increases however the inner detector containers become increasingly significant.

\begin{table*}
\caption{Container size on disk in RDO files. Columns two and three show the average of 50 $t\overline{t}$ events digitized in the absence of pile-up.
  These events were chosen because they produce large energy deposits throughout the detector. Columns four and five, show the average of 50
  $t\overline{t}$ events digitized in the presence of pile-up with $10^{33}\,cm^{-2}$ luminosity and 25\,ns bunch spacing. }
\label{tab:rdofile}
\begin{center}
\begin{tabular}{lrrrrr}
\hline\noalign{\smallskip}
 {\bf Category}                &{\bf No Pile-up}&{\bf No Pile-up}&&{\bf Pile-up}&{\bf Pile-up}\\
                               &{\bf Space on Disk}&{\bf Percentage }&&{\bf Space on Disk}&{\bf Percentage}\\
                               &{\bf [kB/event]}   &{\bf of File} &&{\bf [kB/event]}               &{\bf of File}\\
\noalign{\smallskip}\hline\noalign{\smallskip}
Inner Detector RDOs	           &  187              & 7.6             && 322                 	   &	11.5       \\
Inner Detector SDOs	           &  247              &10.0             && 333           		       & 	11.9       \\
ca\-lo\-ri\-me\-ter Raw Channels       &  995              &40.3             &&1006               		   &	35.9       \\
ca\-lo\-ri\-me\-ter Calibration Hits   &  601              &24.3             && 601               		   &	21.4       \\
Muon Spectrometer RDOs         &    1              & 0.04            &&  27               		   &	 1.0       \\
Muon Spectrometer SDOs         &    1              & 0.05            &&  59               		   &	 2.1       \\
Level One Trigger	           &  289              &11.7             && 300             		   &	10.7      \\
Truth			               &  147              & 6.0             && 151            			   &	 5.4       \\
Headers		                   &   $>$1              & 0.01            &&   2             		   &	 0.1      \\
\noalign{\smallskip}\hline\noalign{\smallskip}
  {\bf Total}                  &    {\bf 2469}     &{\bf 100.00}     &&{\bf 2801}              	   &{\bf 100.00}\\
\noalign{\smallskip}\hline
\end{tabular}
\end{center}
\end{table*}

%%\clearpage

\section{Fast Simulations}
\label{FastSim}

Because of the complicated detector geometry and detailed physics description used by the ATLAS \Geant\ simulation, it is impossible to achieve the required simulated statistics for many physics studies without faster simulation.  To that end, several varieties of fast simulation programs have been developed to complement the \Geant\ simulation.  In this section, the standard \Geant\ simulation will be referred to as ``full simulation.''

Almost 80\% of the full simulation time is spent simulating particles traversing the calorimetry, and about 75\% of the full simulation time is spent simulating electromagnetic particles.  The Fast G4 Simulation aims to speed up this slowest part of the full simulation~\cite{FastG4Sim,FastG4Sim2}.  The approach taken, therefore, is to remove low energy electromagnetic particles from the ca\-lo\-ri\-me\-ter and replace them with pre-simulated showers stored in memory.  Using this approach, CPU time is reduced by a factor of three in hard scattering events (\eg\ $\ttbar$\ production) with little physics penalty.  This simulation may eventually become the default simulation for all processes that do not require extremely accurate modeling of ca\-lo\-ri\-me\-ter response or electromagnetic physics.

\Atlfast-I has been developed for physics parameter space scans and studies that require very large statistics but do not require the level of detail contained in the full simulation~\cite{AtlfastI1,AtlfastI3}.  Truth objects are smeared by detector resolutions to provide physics objects similar to those of the reconstruction.  Object four-vectors are output, without any detailed simulation of efficiencies and fakes.  A factor of 1000 speed increase over full simulation is achieved with sufficient detail for many general studies.

\Atlfast-II is a fast simulation meant to provide large statistics to supplement full simulation studies. The aim is to try to simulate events as fast as possible while still being able to run the standard ATLAS reconstruction. \Atlfast-II is made up from two components: the Fast ATLAS Tracking Simulation (\Fatras) for the inner detector and muon system simulation~\cite{bib:fatras} and the Fast ca\-lo\-ri\-me\-ter Simulation (Fast\-Calo\-Sim) for the ca\-lo\-ri\-me\-ter simulation. Optionally, any subdetector can be simulated with \Geant\ to provide the higher level of accuracy without the same CPU time consumption as full simulation of the entire detector. An improvement over full simulation time of a factor of 10 is achieved with full \Geant\ inner detector and muon simulation and Fast\-Calo\-Sim, and a factor of 100 is achieved with \Fatras\ and Fast\-Calo\-Sim.  

\subsection{Fast G4 Simulation}
\label{FastG4Sim}

The Fast G4 Simulation reduces CPU time consumption without sacrificing accuracy by speeding up the slowest parts of the full simulation.  By treating (as described below) electromagnetic showers in the sampling portions of the ca\-lo\-ri\-me\-ters, a reduction in CPU time of a factor of three can be achieved even in hadronic events.  Although the calorimetry dominates simulation time for the full simulation, after treatment of the electromagnetic showers the simulation time is evenly distributed throughout all subdetectors.  One particular advantage of this fast simulation over the other varieties is that its output file matches identically the format of the output of the full \Geant\ simulation.  The data can therefore be run through the identical tests and digitization software following simulation, and the standard ATLAS trigger and reconstruction can be run.

There are three treatments applied to electromagnetic showers.  For very high energy ($>$10~\GeV{}) electrons and pos\-i\-trons, a tuned shower parameterization is available.  For medium energy (10~\MeV{} to 1~\GeV{}) electrons, pos\-i\-trons, and photons, libraries of pre-simulated showers can be applied during the event.  For very low energy ($<$10 \MeV{}) electrons and pos\-i\-trons, a single hit can be deposited to recreate detector response.  Each one of these treatments can be turned on by the user in each compartment of the electromagnetic ca\-lo\-ri\-me\-ter and the forward hadronic ca\-lo\-ri\-me\-ter.  The energy ranges can be set for each method, compartment, and particle.

For electrons and pos\-i\-trons above a sufficiently high energy, around 10~\GeV{} in the central ca\-lo\-ri\-me\-ters, the sampling ca\-lo\-ri\-me\-ter is sufficiently homogeneous to apply a shower parameterization.  Small steps are taken in the direction of the original particle, depositing energy according to several tuned functions as it traverses the detector.  The longitudinal profiles of showers are parameterized and normalized with an energy scale to approximate the sampling fraction in each subdetector.  The radial profile changes as a function of depth in the shower and is normalized by the longitudinal profile.  Energy is deposited in hits in order to mimic the full simulation.  Fluctuations are introduced in three separate places, representing the random characteristics of shower length and shape, the sampling resolution of the ca\-lo\-ri\-me\-ter, and the geometric fluctuations in the energy collected.

Particles captured by the fast simulation in the appropriate energy range (typically $<$1~\GeV{}) are replaced by a shower from a pre-simulated library, rotated and scaled to match the primary particle.  Shower libraries are generated in bins of pseu\-do\-rapidity and energy for electrons and photons. Only hits in the sensitive detectors are stored in order to save space on disk and in memory.  The binning reproduces the fine structure in the ca\-lo\-ri\-me\-ters.  The libraries are read into memory the first time they are requested by the simulation, ensuring minimal memory overhead.  They consume about 200~MB when all are in use.  Showers are randomly selected with linear weighting from the energy bin above or below the primary particle and from the pseu\-do\-rapidity bin above or below the particle.  The shower is then rotated to match the primary particle's original direction, and the energy is scaled to match the primary particle's energy.  For example, an electron at pseu\-do\-rapidity of 2.37 and with an energy of 12~\MeV{} might use a shower from the electron and positron library's 2.4 bin in pseu\-do\-rapidity and 10~\MeV{} bin in energy. The shower would then be moved and rotated to match the position and direction of the original particle, and its energy would be scaled up by 20\%.

Electrons and pos\-i\-trons with energies below about 10~\MeV{} typically deposit only one hit in the sensitive region of the ca\-lo\-ri\-me\-ter.  These particles are removed when inside the regular sampling region of any of the ca\-lo\-ri\-me\-ters (``killed''), and a single hit is placed in the ca\-lo\-ri\-me\-ter.  The position of the hit is determined by a random exponential number times the radiation length in the detector in order to approximate the particle's range.  The energy of the hit is scaled to the response of the detector and smeared by its resolution.

\begin{table*}
 \caption{The default combination of strategies used in the Fast G4 Simulation for each ca\-lo\-ri\-me\-ter compartment.\label{table:strat}}
\begin{center}
 \begin{tabular}{llll}
\hline\noalign{\smallskip}
  {\bf ca\-lo\-ri\-me\-ter}                & {\bf Parameterization} & {\bf Shower Libraries} & {\bf Killing} \\
\noalign{\smallskip}\hline\noalign{\smallskip}
   Electromagnetic Barrel  		& Not used 		& $10-1000$~MeV $\epem$ & $<10$~MeV $\epem$ \\
                                       	&          		& $<10$~MeV photons 	& \\
   Electromagnetic Endcap  		& Not used 		& $10-1000$~MeV $\epem$ & $<10$~MeV $\epem$ \\
                                       	&          		& $<10$~MeV photons 	& \\
   Electromagnetic Forward 		& $>9$~GeV $\epem$ 	& $10-1000$~MeV $\epem$ & $<10$~MeV $\epem$ \\
                                       	&          		& $<10$~MeV photons 	& \\
   Hadronic Forward        		& $>1.5$~GeV $\epem$ 	& Not used 		& $<10$~MeV $\epem$ \\
 \noalign{\smallskip}\hline
\end{tabular}
\end{center}
\end{table*}

The standard combination of strategies is shown in Table~\ref{table:strat}.  The strategies used in a particular subdetector is optimized for maximum CPU time improvement with minimal complexity.  The upper energy bound for shower libraries, 1~GeV, balances memory use with speed.  Libraries at higher energies also may not correctly reproduce the tails of electromagnetic shower shape distributions as well as low-energy libraries do.  The minimum energy for application of the parameterization model is based purely on CPU time.  In most subdetectors, it is faster to allow \Geant\ to produce 1~GeV secondaries and apply shower libraries to those secondaries than it is to apply the shower parameterization.  The same argument applies to high energy photons.  They pair produce sufficiently quickly that treating them separately only adds complexity to the models.  The speed of the parameterization is limited by random number generation and locating hits within the detector geometry.

\subsection{\Atlfast-I}
\label{ATLFAST1}

\Atlfast-I performs a fast simulation of the ATLAS detector, including object reconstruction, in order to produce high statistics samples of signal and background ev\-nets. The lowest possible CPU time per event is achieved by replacing detailed detector simulation with parameterizations of the desired detector and reconstruction effects. The high speed of simulation in \Atlfast-I makes it possible to study channels where the statistics involved would otherwise be prohibitive.  For example, the background to a \Ztautau study from fake taus in di-jet events is expected to require $O(10^9)$ events in $100$~fb$^{-1}$ of data. Some searches also require many datasets to be simulated in order to scan across parameter space for the model being tested, such as SUSY. 

\Atlfast-I is the least detailed simulation method. There is no realistic detector description, so studies of detector-based quantities, such as ca\-lo\-ri\-me\-ter sampling energies and track hit positions, are not possible. There is no simulation of reconstruction efficiency or misidentification rates, discussed later on, which means the presence of genuine physics objects are overestimated while fake objects are not modeled, with two exceptions. Because jet-flavor tagging efficiencies are applied, fake b-jets and taus are simulated. However, \Atlfast-I provides a useful method of making quick estimates of systematic uncertainties in early data analyses due to the simple process of re-parameterizing the detector and modeling reconstruction effects. The speed of operation enables datasets to be reproduced with different generator configurations, allowing quick estimates of systematic uncertainties arising from generators.

Common to the reconstruction of all objects in \Atlfast-I is that by default no reconstruction efficiencies are applied. These efficiencies can be taken from full simulation and accounted for by the user in the analysis. This applies to electrons, photons and jets as well as to \Atlfast-I tracks. It should be noted that tagging efficiency factors are implicitly taken into account in the tau- and b-tagging procedures. A system to apply a common set of efficiencies and misidentification rates at the analysis stage is in development. The misidentification rates will allow the modeling of fake objects as well.

\Atlfast-I takes input in HepMC format, enabling it to read the output of all ATLAS generators. Generator input is filtered to choose only particles that are useful in the current step.  For example, only charged particles are considered in the tracking stage, and all particles are required to be a part of the final state.

The following sections describe steps taken in \Atlfast-I.

\subsubsection{Tracks}

Charged particle tracks from the generator with $p_T > 500$~\MeV{} and  
with $|\eta| < 2.5$ are considered as reconstruct\-ed \Atlfast-I tracks,  
and five track parameters\footnote{The five parameters are: the azimuthal 
angle $\phi$; longitudinal impact parameter, $z_0$, transverse impact 
parameter, $d_0\equiv\sqrt{x^2+y^2}$; polar angle in $\theta$; and charge 
divided by momentum amplitude.} are associated to them. These  
parameters are calculated from the true particle properties by applying  
parametrized resolution functions which account for the measurement  
precision, energy loss, and multiple scattering as well as for  
hadronic interactions in the inner detector material. The resolution 
functions are taken from fully simulated events.  The non-Gauss\-i\-an 
tails resulting from hadronic interactions are taken into account by  
applying a double-Gauss\-i\-an correlated smearing to the track parameters  
of had\-rons~\cite{AtlfastI1,AtlfastI3}. No vertex smearing is applied. 
In \Atlfast-I, three types of charged particles are  
distinguished: had\-rons, electrons and muons. Due to the relatively  
large energy loss from brems\-strah\-lung, high-$p_T$ electrons are  
treated separately, and an ad\-di\-tion\-al energy loss correction is  
applied. It should be noted that while these tracks are used for  
specific studies in B physics, they are not used for lepton  
identification or b-tagging. 

\subsubsection{Track-Based Tau Identification}

Track-based tau identification is split into two distinct parts, namely reconstruction and identification of tau candidates. The reconstruction part applies a parameterized efficiency to the tracks to calculate the charged component in a tau candidate, while the neutral component is calculated directly from neutral particles in the generated event. 

Once a sample of tau candidates has been reconstruct\-ed, the identification part is carried out by separating the sample into true and fake taus. True taus are defined as those matched to a hadronic decay in the truth record with $\Delta R \equiv \sqrt{\Delta \eta^2 + \Delta \phi^2} \leq 0.2$, whereas the remainder are considered fakes. Subsequently, a parametrization of the identification efficiency is applied based on the number of tracks. 

\subsubsection{Calorimetry}

Stable charged particles from the event generators are propagated through the  
magnetic field along a simple helix. The primary vertex is assumed to  
be at the geometric origin. Using a helix model and assuming a  
perfectly homogeneous magnetic field inside the central tracking  
volume, the impact point on the ca\-lo\-ri\-me\-ter surface is calculated. To  
calculate this point, no interactions of the particle with the  
detector material (\ie\ no multiple scattering, energy loss, or  
nuclear interactions) are taken into account. In particular, this  
implies that no energy loss due to brems\-strah\-lung for electrons and no  
pair production from photons in the inner detector media are simulated  
for the energy depositions in the ca\-lo\-ri\-me\-ters. The effects of these  
interactions are, however, implicitly taken into account by the  
application of appropriate resolution functions. For the calculation  
of track parameters, the four-momenta and the starting  
point of the particles (\eg\ for stable decay products of long-lived  
particles) are taken from the generator information.

The energies of the electrons, photons, and had\-rons are deposited in a  
ca\-lo\-ri\-me\-ter cell map. The response of the ca\-lo\-ri\-me\-ter is assumed to be  
unity and uniform over the full detector. No smearing (\ie\ no  
resolution function) is applied. The energy of the particle is  
entirely deposited in the hit ca\-lo\-ri\-me\-ter cell, assuming a granularity  
of the ca\-lo\-ri\-me\-ter cell map of $\eta\times\phi = 0.1\times0.1$ up to $| 
\eta| < 3.2$ and $\eta\times\phi = 0.2\times0.2$ for $3.2 <|\eta|  <  
5.0$. Neither lateral nor longitudinal shower development is simulated.  
Therefore, the longitudinal fine structure of the  
ca\-lo\-ri\-me\-ters is not taken into account. There is  
also no separation between the electromagnetic and the hadronic  
ca\-lo\-ri\-me\-ter compartments.

Based on the map of deposited cell energies, cluster reconstruction is  
carried out using either SISCone~\cite{SISCone} or FastKt algorithms via the FastJet  
libraries~\cite{FastJet}. The default clustering routine is SISCone with  
a cone size of 0.4. The cluster transverse energy must pass a  
threshold, typically 5~\GeV{}. The clusters may get  
re-classified as electrons, photons, taus or jets in one of the  
following steps. If they are associated to one of these objects, they  
are removed from the list of clusters.

\subsubsection{Electrons and Photons}

For each true electron or photon, the reconstruct\-ed 
energy is obtained by smearing the true energy according to a resolution 
calculated by interpolating between resolutions measured in fully simulated events at precise 
values of $\eta$ and energy. If, after smearing, the candidate 
electron or photon has transverse energy exceeding a threshold value, 
typically 5~\GeV{}, and has $|\eta| < 2.5$, then it is recorded with 
the $\eta$ and $\phi$ directions of the true particle.

Electrons and photons are matched to ca\-lo\-ri\-me\-ter clusters in $(\eta,\phi)$ 
space, with a maximum allowed separation of $\Delta R = 0.15$. If there
is a matching cluster then it is removed from the list of clusters to be 
considered as jet candidates later on.

\subsubsection{Muons}
For each true muon with $p_T > 0.5$~\GeV{}, the reconstruct\-ed momentum  
is calculated from the true muon momentum. A Gauss\-i\-an resolution  
function which depends on $p_T$, $\eta$, and $\phi$ is applied. After  
smearing, muons with $p_T > 5$~\GeV{} and with $|\eta| < 2.5$ are kept.

\subsubsection{Isolation}
In order to define isolated electrons, photons, and muons, the following criteria are  
applied: the difference in the un-smeared energy in a cone of $\Delta R = 0.2$  
around the object direction and its smeared energy needs to be below 
10~\GeV{}. In addition, there should be no further clusters reconstruct\-ed 
with $\Delta R < 0.4$ around the object direction. 

\subsubsection{Jets}
All clusters that have not been assigned to a true electron or 
photon are considered jets if their transverse energy exceeds 10~\GeV{}. 
The jet energy is taken to be the cluster energy, after adding 
non-isolated muons within $\Delta R = 0.4$, 
and is smeared according to the jet energy resolution.

These functions do not account for pile-up, although there is a ``high luminosity'' mode available which adds a pile-up term to the resolution. The pile-up correction is constant with respect to jet transverse energy and is dependent on the size of the jet.

The jet direction is taken to be the cluster direction. Since the  
response function of the ca\-lo\-ri\-me\-ter is set to one, no jet calibration  
is needed to correct for the non-compensation of the ca\-lo\-ri\-me\-ter.  
However, an out-of-cone energy correction is needed. This correction  
is applied in a separate jet calibration step~\cite{AtlfastI1,AtlfastI3}.

\subsubsection{Tagging}

For each jet found, a label is attached to indicate whether the true jet
originated from a light quark, b-quark, c-quark, or tau. This label is 
based on matching b or c partons or the visible 
decay products of hadronically-decaying taus at truth-level with $\Delta R < 0.3$ 
to a reconstruct\-ed jet. In the case of hadronically-decaying taus, the 
ratio between the true visible energy and the jet energy is also
required to be larger than unity minus $2 \sigma$, where $\sigma$ is the jet
energy resolution as above.

The results of b- and tau-tagging are then simulated by applying 
identification efficiencies and fake rates to the labels. These 
efficiencies are determined from full simulation studies and 
are parameterized as a function of $p_T$ and $\eta$.

\subsubsection{Missing $E_T$}
The missing transverse energy is calculated from all re\-con\-struct\-ed  
objects: isolated electrons, photons, muons, taus, jets and non-isolated 
muons, and remaining ca\-lo\-ri\-me\-ter clusters not associated to jets. In  
addition, cells not associated to clusters are included in the missing  
$E_T$ calculation. The cell energies are smeared by applying the jet  
resolution functions.

\subsection{\Atlfast-II}
\label{ATLFAST2}

\Atlfast-II directly simulates the input to the standard Athe\-na reconstruction algorithms to mimic the full simulation. Unlike \Atlfast-I, which provides only momenta for the reconstruct\-ed objects, reconstruct\-ed \Atlfast-II output includes all the properties associated with a reconstruct\-ed object.  In the case of \Fatras\ these include the hits in the inner detector and muon system, and for Fast\-Calo\-Sim these include the energies in the ca\-lo\-ri\-me\-ter cells.  Because the standard reconstruction is run, it is possible to work with a combination of full and \Atlfast-II simulated events without modifying any analysis code.  Both \Fatras\ and Fast\-Calo\-Sim run together with the event reconstruction.  The simulation time is reduced by making use of the simplified detector description used for reconstruction~\cite{bib:trackinggeometry}.  By default, \Atlfast-II uses full simulation for the inner detector and muon system and Fast\-Calo\-Sim in the calorimetry.  \Atlfast-IIF uses Fast\-Calo\-Sim in the calorimetry and \Fatras\ in the inner detector and muon system.

As input, Fatras uses input in HepMC format, performs a smearing of the primary vertex position to represent the luminous region within ATLAS, and records truth information in a way similar to the full simulation. Fast\-Calo\-Sim uses the truth information of all interacting particles at the end of the inner detector volume as input to the ca\-lo\-ri\-me\-ter simulation.  In order to simulate pile-up, generated events must be overlaid prior to detector simulation.

\subsubsection{\Fatras}
\label{Fatras}

\Atlfast-II with the fast track simulation engine \Fatras (\Atlfast-IIF) reduces simulation time in the inner detector and muon system. \Fatras\ is an ATLAS specific development and establishes a complete simulation within the track reconstruction framework. The reconstruction geometry is a simplified description of the full detector geometry, which keeps the same descriptive accuracy for sensitive detector parts, while approximating all other detector components as simplified layers that carry a high-granularity density map. This detector material description can be sufficient.  A factor of 100 reduction in CPU time is obtained with only small physics performance degradation. The propagation of the particles through the tracking detectors is carried out by the extrapolation engine~\cite{bib:extrapolation} used in the offline track reconstruction applications. 

The interactions of the particles with the simplified detector layers are simulated using several methods.  Multiple Coulomb scattering is implemented as a Gauss\-i\-an mixture model to account for tail effects from single large-angle scattering processes; ionization and radiative energy loss are simulated according to the Bethe-Bloch and Bethe-Heitler models; conversion of a photon into an electron and positron is carried out depending on the thickness of the traversed material; hadronic interactions of particles with the detector layers are simulated from a parametric model that has been obtained from \Geant simulation results. The decay of unstable particles is enhanced by a dedicated wrapper of the associated \Geant modules~\cite{bib:fatras}. The ca\-lo\-ri\-me\-ter simulation of \Atlfast-IIF is typically Fast\-Calo\-Sim, and \Fatras\ provides the input particle collection. Energy deposition for muons in the ca\-lo\-ri\-me\-ter layers is also recorded according to the material description of the reconstruction geometry and is further used for cluster simulation in the Fast\-Calo\-Sim application.

\Fatras\ was first established as a validation tool for the newly deployed inner detector reconstruction sequence. It has already been used for noise studies in the Transition Radiation Tracker and first simulations for a potential future upgrade of the ATLAS inner detector. The validation of \Fatras\ against the full simulation results to be used for first collisions data from LHC is ongoing.  An extension of the fast track simulation within the reconstruction geometry has taken place that also allows the use of \Fatras\ in the muon spectrometer. The particles being simulated at the end of the inner detector volume are filtered.  Muons are transported through the ca\-lo\-ri\-me\-ter, and their deposited energy is stored as an input to the Fast\-Calo\-Sim module. The trajectories of the muons are then simulated in the muon spectrometer, and the hits within sensitive detector elements are recorded. Standard digitization is applied on top of the simulated hits to account for the detailed calibration that must be included for a comparison to data.

\subsubsection{FastCaloSim}
\label{FastCaloSim}

Instead of simulating the particle interactions with the detector material, the energy of single particle showers is deposited by Fast\-Calo\-Sim directly using parametrizations of the longitudinal and lateral energy profile. The distribution of active and inactive material in the ca\-lo\-ri\-me\-ter needs to be respected by the parametrization, so a fine binning of the parametrization in the particle energy and pseu\-do\-rapidity is needed. Furthermore, the energy deposition depends strongly on the origin of the shower in the ca\-lo\-ri\-me\-ter, so all parametrizations are also binned versus the longitudinal depth of the shower center. 

The parametrizations are based on a 30 million event sample of fully-simulated (\ie\ simulated with \Geant) single photons and charged pions in an energy range between 200~\MeV{} and 500~\GeV{}, evenly distributed in $|\eta|<5.0$ and $-\pi<\phi<\pi$. All electron and photon showers are approximated by the photon parametrization and all hadronic showers are approximated by the charged pion parametrization. The simplified reconstruction geometry of the ca\-lo\-ri\-me\-ter is used with details at the level of the readout cells.

The parameterization of the longitudinal energy distribution is constructed from histograms of the total energy in all ca\-lo\-ri\-me\-ter layers, the longitudinal depth of the shower center, and the energy fraction in each layer for the fully-simulated single-particle events. The dominant correlations between fractional energy deposits in each ca\-lo\-ri\-me\-ter layer (\ie\ those related to the longitudinal depth of the shower's origin) are accounted for in the parameterization binning.  Gauss\-i\-an correlations between fractional energy deposits in each ca\-lo\-ri\-me\-ter layer (\ie\ those describing shower development) are stored in a correlation matrix and are applied to improve the parameterized energy distribution. During fast simulation, the parametrization closest in energy and pseu\-do\-rapidity to the particle is taken, and then the total shower energy and the shower depth are chosen randomly from the stored histograms and rescaled to match the true particle energy. It was found that after rescaling no interpolation between parametrizations is necessary.  Afterwards, the energy fractions in all ca\-lo\-ri\-me\-ter layers are generated randomly, taking into account the correlation matrix. The lateral energy distribution inside each ca\-lo\-ri\-me\-ter layer is simulated using a symmetric average radial shape function. The shape functions are extracted from fits to fully-simulated single-particle events and are constructed for bins of particle type, primary particle energy, position in $\eta$, and shower depth in the ca\-lo\-ri\-me\-ter. The asymmetry of shower shapes for particles entering the ca\-lo\-ri\-me\-ter at large incident angles is absorbed in a shape function describing a pseu\-do\-rapidity-dependent asymmetry term. During simulation, the energy of a ca\-lo\-ri\-me\-ter cell is determined by the integral of the shape function over the cell surface area. Fluctuations derived from the intrinsic resolutions of each ca\-lo\-ri\-me\-ter are applied to the cell energy.  The total energy of all cells in one ca\-lo\-ri\-me\-ter layer is normalized to the total energy in the layer making use of the longitudinal shower shape.

The histograms and shape functions needed as input for the parametrizations use about 200MB of memory. Since no simulation of particle interactions is done, the dominant part of the simulation time is spent on the numerical integration of the lateral shape functions. Overall, the ca\-lo\-ri\-me\-ter simulation time for a single particle is a few microseconds, and a typical (\eg\ $\ttbar$) event needs a few seconds.

The parameterization of Fast\-Calo\-Sim differs in several important ways from that of the Fast G4 Simulation.  Fast\-Calo\-Sim fills the readout geometry of ATLAS and applies a parameterization from the edge of the inner detector, whereas the Fast G4 Simulation places hits like those of \Geant\ into the full ATLAS detector geometry and is only applied in the sampling portion of the ca\-lo\-ri\-me\-ter (\eg\ excluding the cryostats surrounding the calorimetry).  As a result, the Fast G4 Simulation output can be run through the standard digitization software, whereas the Fast\-Calo\-Sim output is fed directly into the reconstruction.

\subsection{Computing Performance}

\begin{figure*}
\begin{center}
\includegraphics[width=6.75in]{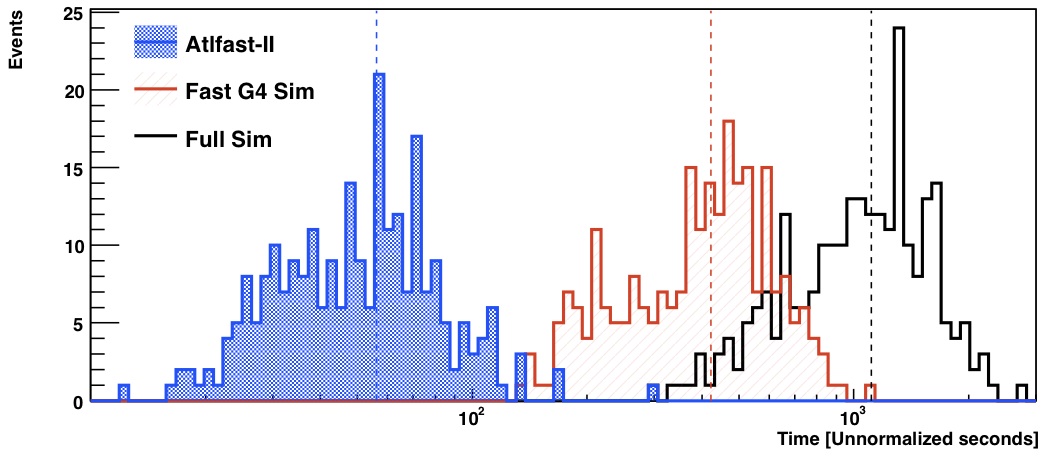}
\caption{Distributions of CPU time for 250~$\ttbar$\ events in full, Fast G4, and \Atlfast-II simulations.  Vertical dotted lines denote the averages of the distributions.\label{fig:eventTimes}}
\end{center}
\end{figure*}

Examples of simulation times in kSI2K seconds~\cite{kSI2K} for various types of events in the full and fast simulations are provided in Table~\ref{tab:fastSimTimes}\footnote{Measurements were performed on Sun Fire X2200 M2 units with dual dual-core 2.6 GHz AMD Opteron 2218 processors.  Normalization was done using the peak specmark int 2000 rating 1794.  For the same system, the peak specmark floating point 2000 rating was 3338. The normalization follows the published results, rather than the WLCG formula in~\cite{WLCG-norm}.  Details of cross-platform benchmarking can be found in~\cite{KV}.}. In single central ($|\eta|<3$) electron events the simulation time is decreased by a factor of ten or more by the fast G4 simulation, and in hard scattering events the simulation time is decreased by a factor of 2-5. \Atlfast-II without \Fatras\ decreases simulation time by a factor of 20-40, and \Atlfast-IIF decreases simulation time by a factor of 100. Fast\-Calo\-Sim accounts for about 10\% of the total simulation time in \Atlfast-II and 60-70\% of the total simulation time in \Atlfast-IIF. \Atlfast-I requires a relatively negligible amount of CPU time even for hard scattering events.  \Atlfast-I, Fast\-Calo\-Sim, and \Fatras\ run during the reconstruction step, but for these purposes the time consumed by their methods is included in ``simulation time.''  Figure~\ref{fig:eventTimes} shows the distribution of simulation times per event for full, Fast G4, and \Atlfast-II simulation of 250~$\ttbar$\ events.  Here, the average time required to run Fast\-Calo\-Sim on these events has been added to the full inner detector and muon simulation time of each event.  The distributions are similar in shape.

\begin{table*}[htb]
\caption{Simulation times per event, in kSI2K seconds, for single particles generated with $|\eta|<3.0$ and with the same transverse momentum.  All times are averaged over 500 events.  \Atlfast-II uses full simulation for the inner detector and muon system and Fast\-Calo\-Sim in the calorimetry.  \Atlfast-IIF uses Fast\-Calo\-Sim in the calorimetry and \Fatras\ in the inner detector and muon system.\label{table:fastSimSingleTimes} }
\begin{center}
\begin{tabular}{lllllll}
\hline\noalign{\smallskip}
{\bf Sample} & {\bf Full Sim} & {\bf Fast G4 Sim} & {\bf \Atlfast-II} & {\bf \Atlfast-IIF} & {\bf \Atlfast-I} \\
\noalign{\smallskip}\hline\noalign{\smallskip}
5~GeV $\mu^{\pm}$ 	& 0.879	& 0.899	& 1.28	& 0.633	& 0.011	\\
50~GeV $\mu^{\pm}$	& 1.63	& 1.15	& 2.71	& 0.606	& 0.011	\\
500~GeV $\mu^{\pm}$ 	& 12.0	& 10.4	& 11.8	& 0.615	& 0.011	\\
1~GeV $e^{\pm}$		& 3.62	& 0.734	& 0.825	& 0.513	& 0.011	\\
5~GeV $e^{\pm}$		& 17.8	& 1.64	& 1.00	& 0.542	& 0.011	\\
50~GeV $e^{\pm}$	& 179.	& 4.86	& 1.25	& 0.588	& 0.013	\\
1~GeV $\pi^{\pm}$ 	& 2.40	& 1.48	& 0.701	& 0.515	& 0.011	\\
5~GeV $\pi^{\pm}$ 	& 10.4	& 4.27	& 0.811	& 0.540	& 0.011	\\
50~GeV $\pi^{\pm}$	& 94.7	& 30.3	& 1.04	& 0.569	& 0.011	\\
\noalign{\smallskip}\hline
\end{tabular}
\end{center}
\end{table*}

%5~GeV $\mu^{\pm}$ 	& 0.490	& 0.501	& 0.714	& 0.353	& 0.006	\\
%50~GeV $\mu^{\pm}$	& 0.906	& 0.641	& 1.51	& 0.338	& 0.006	\\
%500~GeV $\mu^{\pm}$ 	& 6.71	& 5.77	& 6.56	& 0.343	& 0.006	\\
%1~GeV $e^{\pm}$	& 2.02	& 0.409	& 0.460	& 0.286	& 0.006	\\
%5~GeV $e^{\pm}$	& 9.92	& 0.914	& 0.560	& 0.302	& 0.006	\\
%50~GeV $e^{\pm}$	& 99.9	& 2.71	& 0.699	& 0.328	& 0.007	\\
%1~GeV $\pi^{\pm}$ 	& 1.34	& 0.827	& 0.391	& 0.287	& 0.006	\\
%5~GeV $\pi^{\pm}$ 	& 5.81	& 2.38	& 0.452	& 0.301	& 0.006	\\
%50~GeV $\pi^{\pm}$	& 52.8	& 16.9	& 0.578	& 0.317	& 0.006	\\

\begin{table*}[htb]
\caption{Simulation times per event, in kSI2K seconds, for the full simulation, Fast G4 simulation, \Atlfast-II, \Atlfast-IIF, and \Atlfast-I.  \Atlfast-II uses full simulation for the inner detector and muon system and Fast\-Calo\-Sim in the calorimetry.  \Atlfast-IIF uses Fast\-Calo\-Sim in the calorimetry and \Fatras\ in the inner detector and muon system.All times are averaged over 250~events, except heavy ion times which were averaged over only 50~events.  Because the memory required to reconstruct heavy ion events exceeds 3~GB and because Fast\-Calo\-Sim runs during the reconstruction step, the amount of time taken by Fast\-Calo\-Sim could not be measured in that sample.  It was estimated as 10\% of the full inner detector simulation time, consistent with the other hard scattering events.\label{tab:fastSimTimes} }
\begin{center}
\begin{tabular}{llllll}
\hline\noalign{\smallskip}
{\bf Sample} & {\bf Full Sim} & {\bf Fast G4 Sim} & {\bf \Atlfast-II} & {\bf \Atlfast-IIF} & {\bf \Atlfast-I} \\
\noalign{\smallskip}\hline\noalign{\smallskip}
Minimum Bias 	& 551.		& 246.		& 31.2	& 2.13	& 0.029	\\
\ttbar	 	& 1990		& 757.		& 101.	& 7.41	& 0.097	\\
Jets 		& 2640		& 832.		& 93.6	& 7.68	& 0.084	\\
Photon and jets	& 2850		& 639.		& 71.4	& 5.67	& 0.063	\\
\Wenu{} 	& 1150		& 447.		& 57.0	& 4.09	& 0.050	\\
\Wmunu{} 	& 1030		& 438.		& 55.1	& 4.13	& 0.047	\\
Heavy ion 	& 56,000	& 21,700	& $\sim$3050	& 203	& 5.56	\\
\noalign{\smallskip}\hline
\end{tabular}
\end{center}
\end{table*}

%Minimum Bias 	& 307.		& 137.		& 17.4	& 1.19	& 0.016	\\
%\ttbar	 	& 1110		& 422.		& 56.2	& 4.13	& 0.054	\\
%Jets 		& 1470		& 464.		& 52.1	& 4.28	& 0.047	\\
%Photon and jets& 1590		& 356.		& 39.8	& 3.16	& 0.035	\\
%\Wenu{} 	& 642.		& 249.		& 31.8	& 2.28	& 0.028	\\
%\Wmunu{} 	& 572.		& 244.		& 30.7	& 2.30	& 0.026	\\
%Heavy ion 	& 31,200	& 12,100	& 1700	& 113	& 3.1	\\

In evaluating these CPU times, it is necessary to keep in mind the ad\-di\-tion\-al steps required before analysis of the data can be performed.  For both full and fast G4 simulation, the data must be digitized and reconstruct\-ed.  For \Atlfast-II, the inner detector and muon system must be digitized\footnote{The inner detector and muon system together require about 2/3 of the total digitization time.} and reconstruct\-ed, but the ca\-lo\-ri\-me\-ter requires only reconstruction.  For \Atlfast-IIF, only the muon system must be digitized before reconstruction is performed.  The output of \Atlfast-I is in a format similar to that of the reconstruction and needs no further processing.  The CPU time required for these ad\-di\-tion\-al steps is given in Table~\ref{table:cputime}.

\subsection{Physics Performance}

The fast simulations have been compared to full simulation in both low-level analyses with single particles entering the ca\-lo\-ri\-me\-ter and high-level analyses of detector observables with jets and active hard scattering events.  The Fast G4 Simulation agrees to about 1-2\% in jet energy scale after the standard calibration procedure and agrees to within 5\% percent in electron identification efficiencies.  Due to the simplifications in the ca\-lo\-ri\-me\-ter simulation, Fast\-Calo\-Sim differs at the 5\% level from full simulation after reconstruction, especially in properties that are sensitive to the shape of hadronic showers. The jet energy scale differs by 1-2\% after recalibration, and electron identification efficiency differs by about 5\%.  Since all particles are simulated using an average lateral shape function, visible effects like electromagnetic subshowers in charged pion showers are not described. These differences can be reduced by applying ad\-di\-tion\-al object-dependent correction functions after reconstruction.  Fakes and ca\-lo\-ri\-me\-ter punch-through are not well modeled in \Atlfast-II and \Atlfast-IIF.

\begin{figure*}
\begin{center}
\subfigure % caption for subfigure a
{
    \includegraphics[width=3.2in]{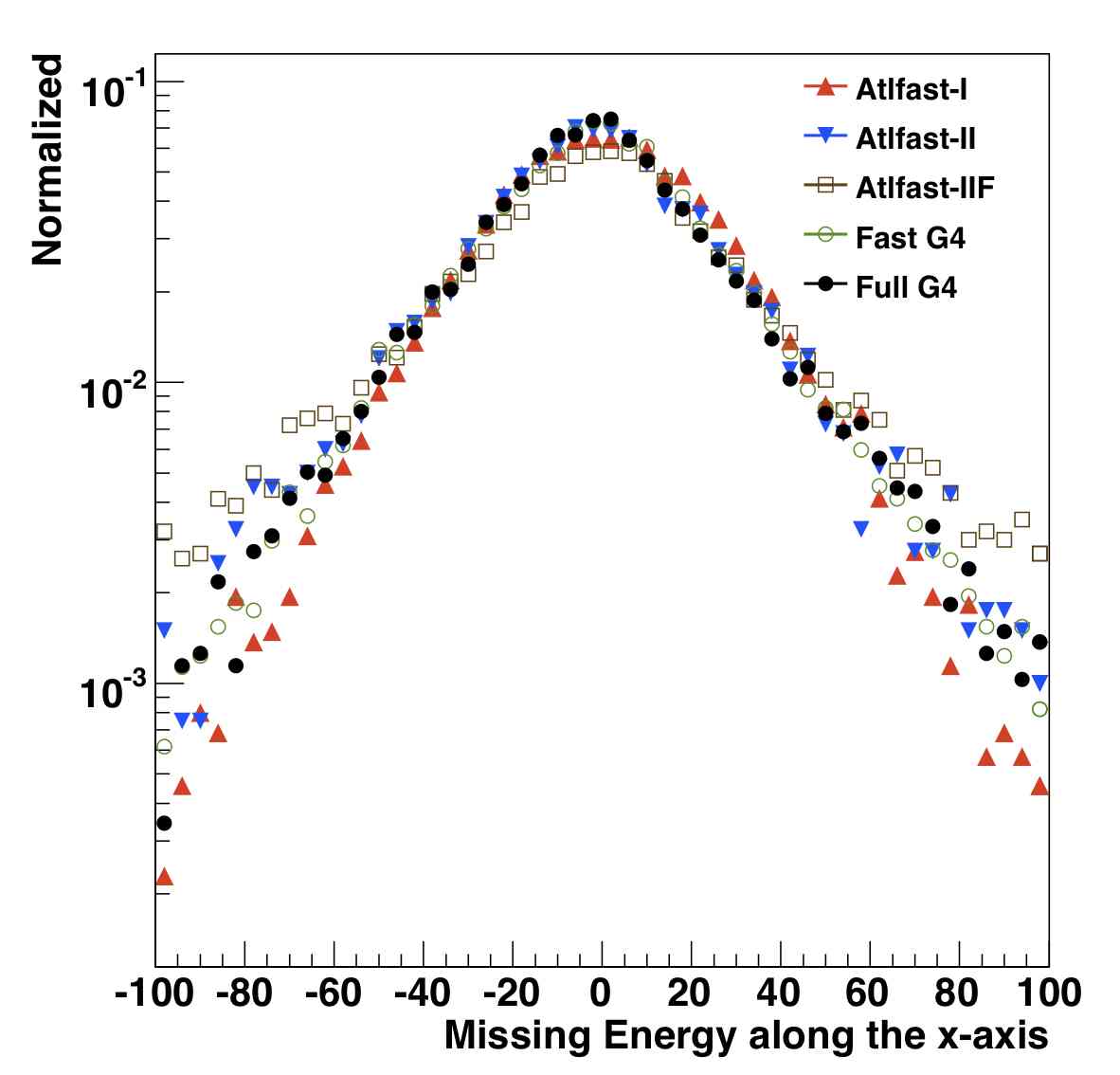}
}
\subfigure % caption for subfigure a in []
{
    \includegraphics[width=3.2in]{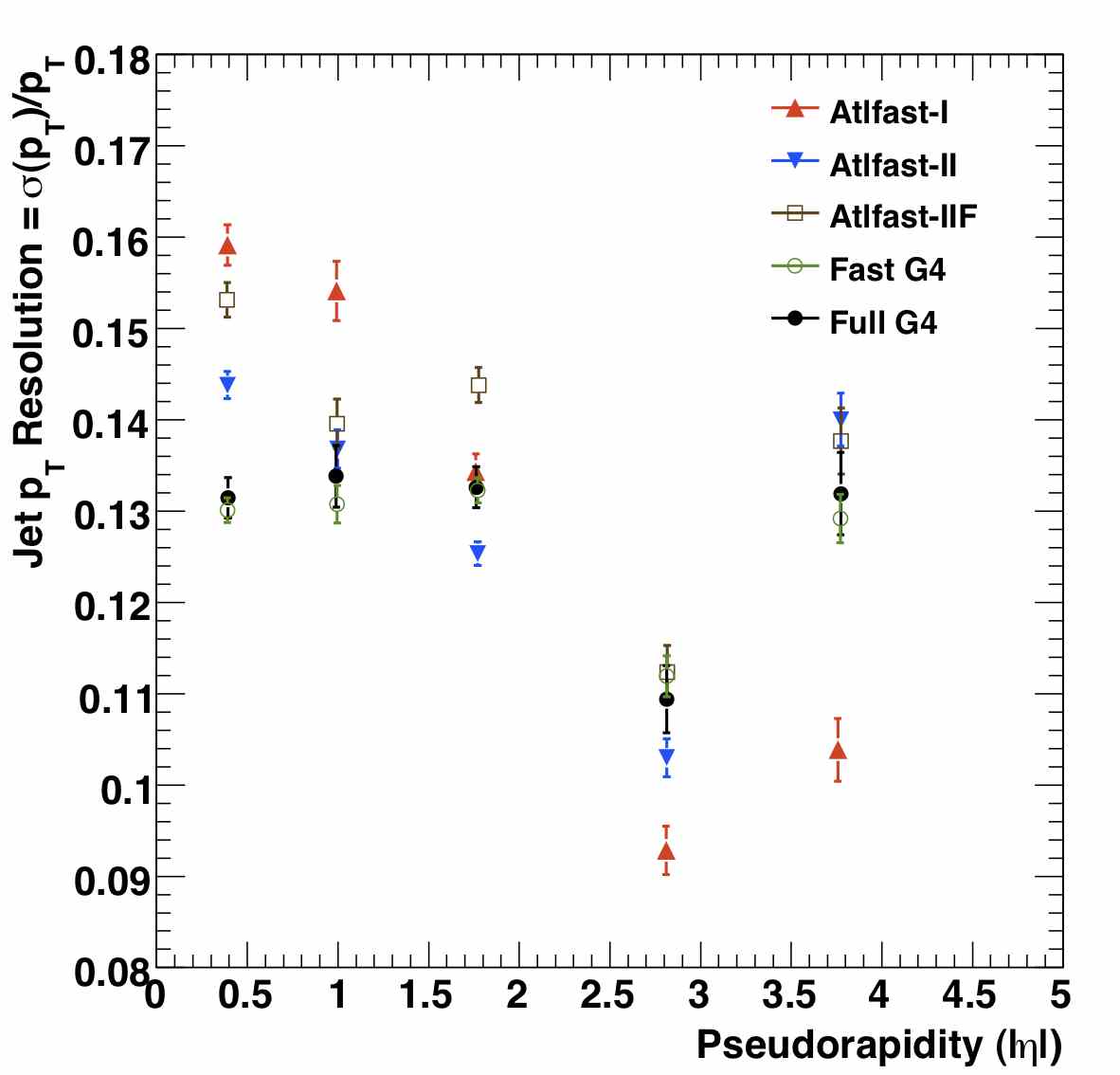}
}
\caption{Left, fast simulations (color) and full simulation (black) comparison of missing transverse energy along the $x$-axis in di-jet events with a leading parton $p_T$ between 560~and 1120~\GeV{}.  Right, a comparison of jet $p_T$ resolution as a function of pseu\-do\-rapidity in $\ttbar$\ events for jets with $20<p_T^{True}<40$~GeV.\label{fig:fastmet}}
\end{center}
\end{figure*}

\begin{figure*}
\begin{center}
\subfigure % caption for subfigure a
{
    \includegraphics[width=3.2in]{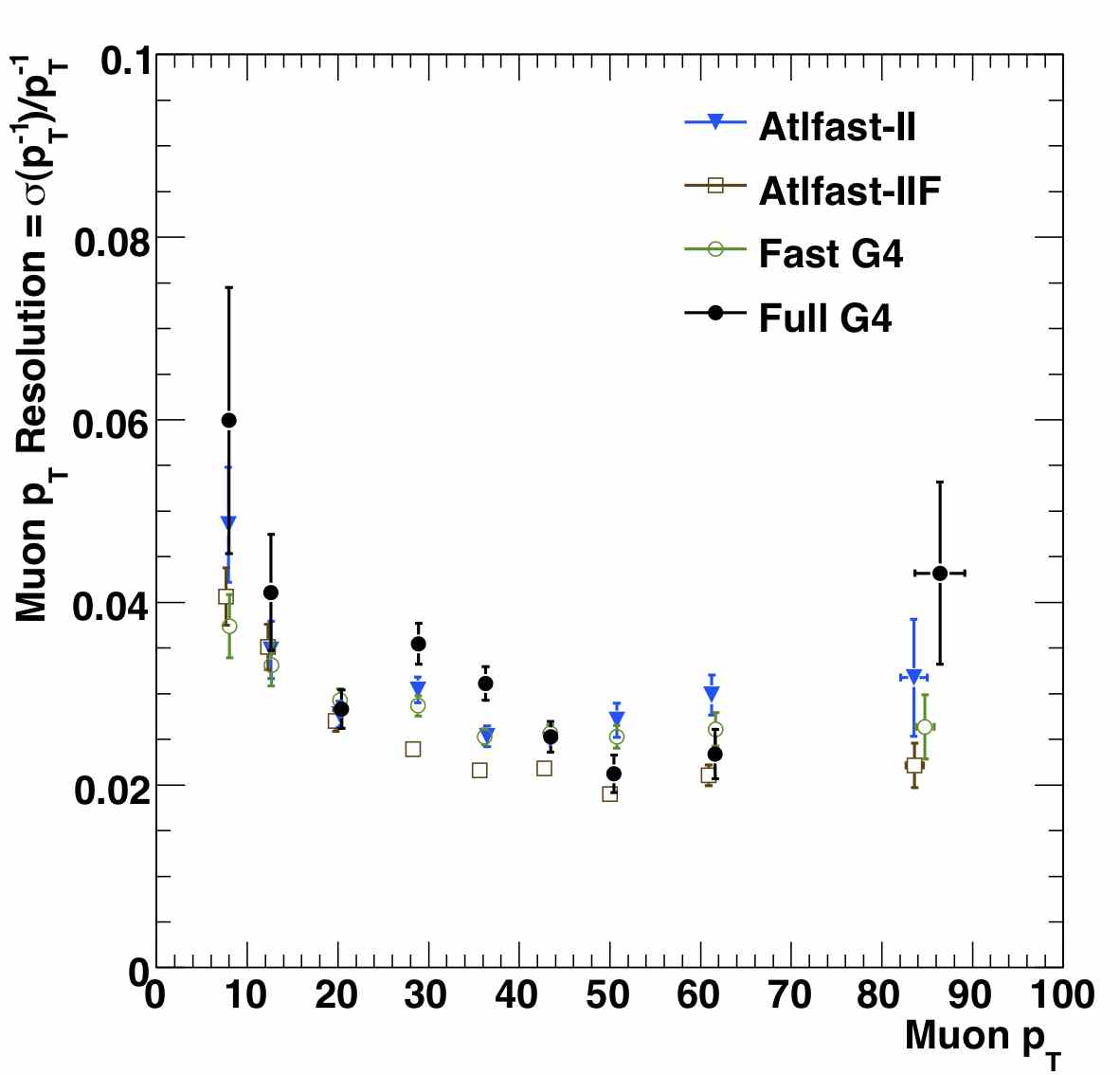}
}
\subfigure % caption for subfigure a in []
{
    \includegraphics[width=3.2in]{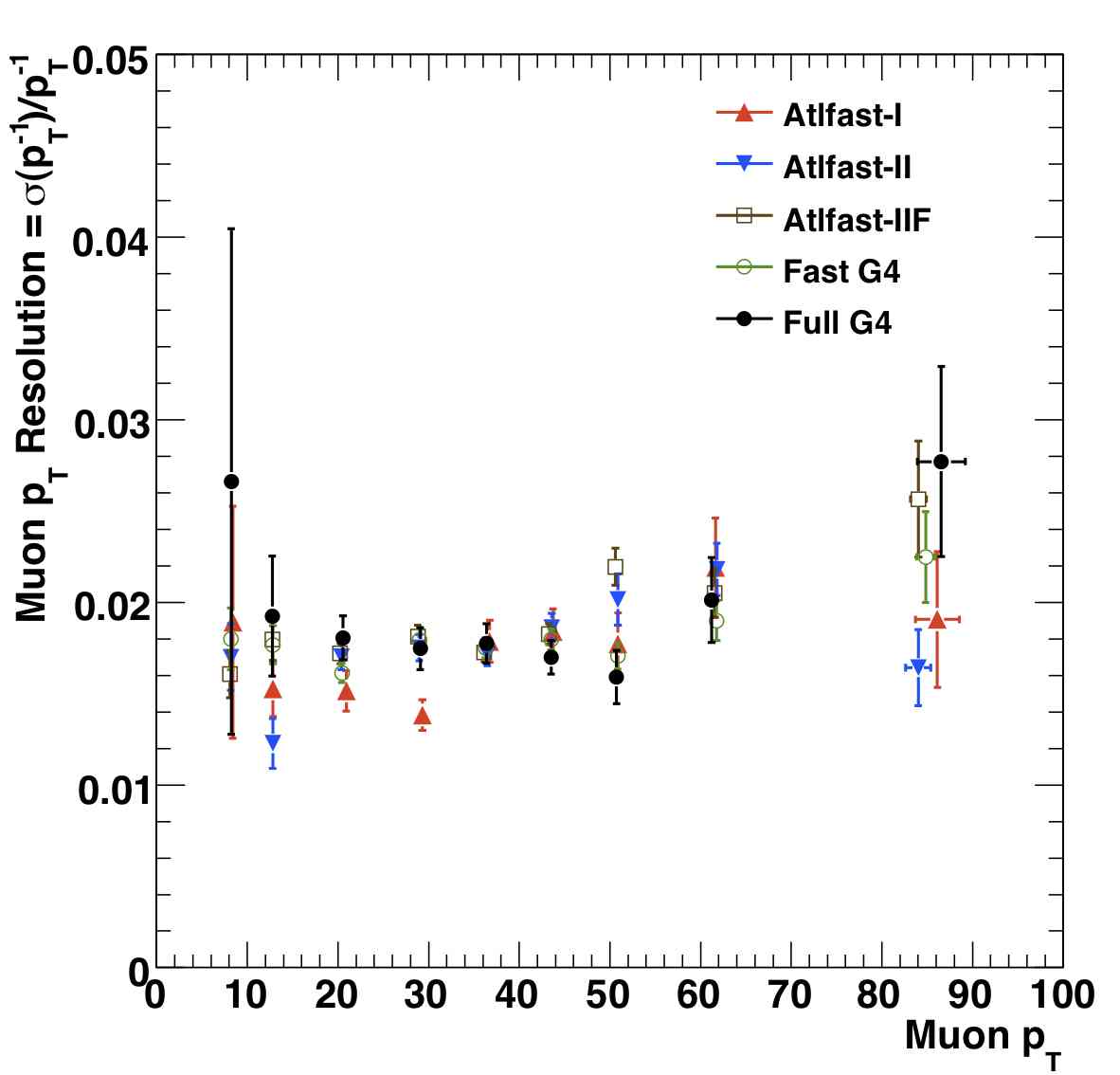}
}
\end{center}
\caption{Fast simulations (color) and full simulation (black) comparison of reconstruct\-ed muon $p_T$ resolution as a function of muon $p_T$ for central ($|\eta|<1.2$) muons in $\Zmumu$\ events for muons reconstruct\-ed using only the muon spectrometer (left) and using both the inner detector and spectrometer (right).  \Atlfast-I only provides one type of muon, which is included in the right plot.\label{fig:fastmus}}
\end{figure*}

Figure~\ref{fig:fastmet} shows missing transverse energy along the $x$-axis for the full and fast simulations in di-jet events with a leading parton $p_T$ between 560~and 1120~\GeV, as well as jet $p_T$ resolution as a function of $\eta$ in $\ttbar$\ events for jets with $20<p_T^{True}<40$~GeV.  \Atlfast-II and the Fast G4 Simulation agree well with full simulation in missing transverse energy spectrum, even in the tails of the distribution.  \Atlfast-I does not sufficiently populate the tails of the missing transverse energy distribution, and \Atlfast-IIF has too wide a distribution.  \Atlfast-I, \Atlfast-IIF, and \Atlfast-II show 10-20\% deviations from full simulation in jet transverse momentum resolution.  Fast G4 simulation is consistent with full simulation through the entire range in pseu\-do\-rapidity.  Figure~\ref{fig:fastmus} shows reconstruct\-ed muon $p_T$ resolution as a function of muon $p_T$ in $\Zmumu$\ events.  Muons reconstruct\-ed using the muon spectrometer alone and those reconstruct\-ed using both the muon spectrometer (``standalone'') and inner detector (``combined'') are shown.  Only one type of muon is provided by \Atlfast-I, so it is only included in the combined reconstruction plot.  In the cases of \Atlfast-II and the Fast G4 simulation, muon spectrometer simulation is done by \Geant and should, therefore, be identical to full simulation.  The fast simulations show generally good agreement over the entire range of $p_T$.  \Atlfast-IIF has standalone muon resolution that is 10\% better than full simulation in some bins of $p_T$, but since the muon system simulation of \Atlfast-IIF is still under development, the agreement is expected to improve.  It is generally left to the physics groups to evaluate the fast simulations with their analyses and determine which is acceptable.

%%\clearpage

\section{Validation}
\label{Validation}

Validation of the ATLAS simulation chain is done in two distinct phases.  First, the software performance must be assessed.   Then, the physics performance must be tested and compared to available data.  The first step includes testing robustness, testing software performance, and testing basic functionality.  The second step includes comparison to test beam, cosmic data, and physics results obtained from previous simulation productions.  In this section the infrastructure for each stage of validation is described.

ATLAS has a fresh software build every night for coordinating software development.  Each nightly build is run through a rigorous test cycle, and as a release deadline approaches the test results are increasingly scrutinized to evaluate stability and performance.  Thanks to the evaluation prior to release, generally only rare bugs appear in production for the first time. The automatic testing infrastructure also allows evaluation of many different versions of the Athe\-na software.  Separate bug-fixing and development branches are employed, for example, and significant interface changes or low-level code migrations take place in separate branches until they are sufficiently stable to be merged into the main branch.  Each version of the software comes in several flavors for different system architectures, operating systems, compilers, and so on.  The simulation production in 2008 used 32-bit builds with gcc 3.4.6~\cite{gcc} on CERN's Scientific Linux 4~\cite{slc}.  External dependencies include CLHEP 1.9.3.1 and WLCG 54G.

A web portal has been constructed, using the Savannah bug tracking software~\cite{Savannah}, for monitoring problems with the various facets of the simulation software.  Bugs can be reported and tracked as they are diagnosed and solutions are found, and new features can be requested.

\subsection{Automated Testing}
\label{Night}

The software performance of the simulation is monitored in three types of automated tests: ATLAS Testing Nightly (ATN) tests, Run Time Tester (RTT) tests, and Full Chain Tests (FCT).  ATN tests are run every night on every software build and are basic functionality tests.  RTT tests are run on a subset of builds and include 50 simulation tests to ensure functionality and, in some cases, consistent results.  FCT tests are run on only a few builds each night and test the entire software chain in a production-like environment. Releases are required to pass a minimal number of milestones before being declared ready for production.  For details about the ATN and RTT, refer to~\cite{SoftManagement}.

FCT tests are run daily on a small set of jobs. The aim of the FCT is to verify the readiness of a production cache release candidate for Grid production. The FCT runs jobs that test the functionalities of generation, the different flavors of fast and full simulation, digitization, byte\-stream conversion, and reconstruction of Standard Model processes, black hole production, and heavy ion collisions. In the case of Standard Model processes, a full chain\footnote{A single chain of jobs runs all steps from event generation through reconstruction, sequentially, using the output of one step as the input of the next.} of jobs are run per release, with hard scattering events that stress the software.  If successful, the output from each day's run is saved for use in the next step of the test the following day (e.g. Monday's generation provides input for Tuesday's simulation, which provides input for Wednesday's digitization). The typical number of events processed (50) is limited by the CPU requirements for the full simulation. 

As part of the FCT, 1000 events that were simulated with an old validated release are reconstruct\-ed. This long test allows better evaluation of the reconstruction's stability. Moreover, the relatively large sample is used to make a preliminary check on the quality of the reconstruction for final state objects (jets, electrons, muons, etc.). All the other tests only check for the success or failure of the job, the number of events in the output file, and unknown error messages in the log file.  If any of these checks fail, the release candidate is rejected, and an ad\-di\-tion\-al iteration of bug fixing is undertaken. Only once a release succeeds in all FCT tests is it distributed to the Grid. 

\subsection{Computing Performance Benchmarking}
\label{Perf}

Event generation jobs are typically fast enough that not a large effort is made to test their software performance.  In general, the jobs take a tens of milliseconds per event.  Generation with \Pythia\ or \Herwig\ requires about 450 MB of memory, and generation with Hijing requires about 170~MB of memory.  The files produced in generation jobs are tens of kB per event (\eg\ 40~kB/event for $\ttbar$\ events).

CPU time, memory consumption, and output file size for the simulation are tested in each stable release using a variety of physics processes.  Single muons, electrons, and charged pions are used, as well as di-jets in bins of leading parton $p_T$, a supersymmetric benchmark point\footnote{ATLAS mSUGRA benchmark point SU3: $m_0=$100~\GeV, $m_{1/2}$=300~\GeV, $A_0=-300$, $\tan{\beta}=6$, and $\mu>0$.}, minimum bias, Higgs boson decaying to four leptons, $\Zee$, $\Zmumu$, and $\Ztautau$\ events.  The same input events are used every time to ensure fair comparison of simulation results independent of changes to the event generation.

Simulation of the full detector requires typically $\sim750$~MB of memory (\ie\ VSIZE) and includes loading of almost 400 libraries into memory.  Memory consumed during simulation is also broken down into its three key components by taking snapshots of memory use during initialization: GeoModel, the ATLAS-side detector geometry, typically about $\sim100$~MB; G4Atlas, the purely \Geant\ component of the memory, typically about $\sim300$~MB; and load modules, the remaining algorithms and services loaded during the job, typically about $\sim300$~MB.  Significant changes in any of these can indicate the proper source of a change in memory.  The memory requirement is independent of the number of events in the job and varies by only a few percent for different physics processes.  Although up to 2~GB of memory may be reserved for a Grid job, by keeping the memory requirements of a typical simulation job under 1~GB, more machines can be used.  The memory required to build each piece of the detector can be found in Table~\ref{table:geomem}. Of major concern is any increase in memory (``leaks'') during the event loop once all libraries have been loaded and setup is complete.  Some increase due to caching is expected during the processing of the first few events.  However, if the memory required by the application continues to grow beyond the system limits, memory corruption and memory pressure can result in serious problems.  The memory required by the ATLAS simulation has been found to increase by less than 0.25~MB per event under normal circumstances.  The increases are not steady, but come in large ($\sim10$~MB) and sporadic jumps.  The source of these increases is not fully understood, but a 50~event simulation job still consumes well under 1~GB of memory.

The CPU time consumed by generation, full simulation, digitization, and reconstruction for various types of events is shown in Tables~\ref{table:cputimesingleevents} and~\ref{table:cputime} and is typically several minutes per hard scattering event.  All times are normalized to kSI2K seconds~\cite{kSI2K}.%\footnotemark[17].  
For the purposes of testing, logfile output was suppressed and no output files (\eg\ hit or RDO files) were created.  CPU time is also measured as a function of other simulation input parameters prior to significant changes, for example using different physics lists.  For these runs, output files were disabled; in simulating $\ttbar$\ events the time per event is increased by $\sim0.5$\% when file writing is enabled.  The hard scattering events shown in Table~\ref{table:cputime} were generated with a 14~TeV center of mass energy; for 10~TeV center of mass energy the simulation time is reduced by 17\% for $\ttbar$\ events.  The distribution of CPU time for simulation of 250~$\ttbar$\ events is shown in Figure~\ref{fig:eventTimes}.

For the samples in tables~\ref{table:cputime} and~\ref{tab:fastSimTimes}, event generation of \Wboson production, minimum bias interactions, di-jet events, and photon and jet events was done using \Pythia.  \ttbar\ production was done using MC\-@\-NLO for the hard scattering and Herwig for hadronization and showering.  Heavy ion event production was done using Hijing.

\begin{table*}[htb]
\caption{Simulation times per event, in kSI2K seconds, for single particles generated with $|\eta|<3.0$.  All times are averaged over 500 events.  For timing, logfile output and output files were suppressed.\label{table:cputimesingleevents} }
\begin{center}
\begin{tabular}{llllll}
\hline\noalign{\smallskip}
{\bf Particle} & $p_T=1$~\GeV{} & $p_T=5$~\GeV{} & $p_T=50$~\GeV{} & $p_T=500$~\GeV{} \\
\noalign{\smallskip}\hline\noalign{\smallskip}
Single electrons ($e^{\pm}$) 		& 3.62	& 17.8	& 179.	& - 	\\
Single muons ($\mu^{\pm}$)		& - 	& 0.879	& 1.63	& 12.0	\\
Single charged pions ($\pi^{\pm}$) 	& 2.40	& 10.4	& 94.7	& - 	\\
\noalign{\smallskip}\hline
\end{tabular}
\end{center}
\end{table*}

%Single electrons ($e^{\pm}$) 		& 2.02	& 9.92	& 99.9	& - 	\\
%Single muons ($\mu^{\pm}$)		& - 	& 0.490	& 0.906	& 6.71	\\
%Single charged pions ($\pi^{\pm}$) 	& 1.34	& 5.81	& 52.8	& - 	\\

\begin{table*}[htb]
\caption{Generation, simulation, digitization, and reconstruction times per event, in kSI2K seconds.  Generation times are averaged over 5000 events, except generation of heavy ion events, which were averaged over 250 events.  The generation time for $\ttbar$\ events includes only the hadronization time, not the time consumed by MC\-@\-NLO generation.  Simulation, digitization, and reconstruction times are averaged over 250 events, except simulation of heavy ion events, which were averaged over 50 events.  The heavy ion event simulation time is for events with a random impact parameter.  Central collisions require on average 3.4 times longer to simulate.  Reconstruction time can vary dramatically depending on the algorithms run and the trigger configuration.  These times should be taken as indicative of the order of magnitude, rather than as a precise measurement.  During reconstruction of heavy ion events, the testing machines ran out of memory.  Based on a previous release, heavy ion collision reconstruction is estimated to take $\sim10$ times longer than $\ttbar$\ event reconstruction.  For timing, logfile output and output files were suppressed.\label{table:cputime} }
\begin{center}
\begin{tabular}{llllll}
\hline\noalign{\smallskip}
{\bf Sample} & {\bf Generation} & {\bf Simulation} & {\bf Digitization} & {\bf Reconstruction} \\
\noalign{\smallskip}\hline\noalign{\smallskip}
Minimum Bias 		& 0.0267	& 551.		& 19.6	& 8.06	\\
\ttbar Production	& 0.226		& 1990		& 29.1	& 47.4	\\
Jets 			& 0.0457	& 2640		& 29.2	& 78.4	\\
Photon and jets		& 0.0431	& 2850		& 25.3	& 44.7	\\
\Wenu{} 		& 0.0788	& 1150		& 23.5	& 8.07	\\
\Wmunu{} 		& 0.0768	& 1030		& 23.1	& 13.6	\\
Heavy ion 		& 2.08		& 56,000	& 267	& -	\\
\noalign{\smallskip}\hline
\end{tabular}
\end{center}
\end{table*}

%trig1_misal1_mc12.005200.T1_McAtNlo_Jimmy.recon.v13003003 and ideal0_misal1_mc12_V1.008801.Hijing_PbPb_5p5TeV_MinBias.recon.v13003003 for estimate of Hijing time

%Minimum Bias 		& 0.0149	& 307.		& 10.9	& 4.49	\\
%\ttbar Production	& 0.126		& 1110		& 16.2	& 26.4	\\
%Jets 			& 0.0255	& 1470		& 16.3	& 43.7	\\
%Photon and jets	& 0.0240	& 1590		& 14.1	& 24.9	\\
%\Wenu{} 		& 0.0439	& 642.		& 13.1	& 4.50	\\
%\Wmunu{} 		& 0.0428	& 572.		& 12.9	& 7.60	\\
%Heavy ion 		& 1.16		& 31,200	& 149	& $\sim100$	\\

Digitization jobs are generally fast, but memory consumption can be a serious concern during jobs with many overlaid events.  Table~\ref{table:digiperf} shows how resource consumption during digitization of 50~\ttbar\ events scales with pile-up luminosity\footnote{The ATLAS software is under a continuous process of improvement, with improving performance in terms of calculation speed and memory profile, and problems such as memory leaks being identified and eliminated.}.  The memory required for digitizing with a luminosity of 10$^{34}$~cm$^{-2}$~s$^{-1}$ is sufficiently large that the memory limit of the testing machine was reached, and, therefore, swapping resulted in a significant increase in CPU time.  The allocated memory per event is provided for some benchmark of the change in memory over the course of a single event.

\begin{table*}[htb]
\caption{Digitization computing resources for 50~\ttbar\ events as they scale with luminosity.  CPU times are normalized to the time required by a no pile-up job.  Cavern background was overlaid during these jobs with a safety factor of one.  Beam gas and beam halo were ignored.\label{table:digiperf} }
\begin{center}
\begin{tabular}{lllll}
\hline\noalign{\smallskip}
{\bf Resource} & {\bf No Pile-Up} & {$10^{33}$~cm$^{-2}$~s$^{-1}$} & {$3.5\times10^{33}$~cm$^{-2}$~s$^{-1}$} & {10$^{34}$~cm$^{-2}$~s$^{-1}$} \\
\noalign{\smallskip}\hline\noalign{\smallskip}
CPU Time Factor			& 1.0	& 2.3	& 5.8	& 160	\\
Memory Leak [kB/event]		& 10	& 270	& 800	& 2100	\\
Virtual Memory [MB]		& 770	& 1000	& 1300	& 2000	\\
Allocated Memory [MB/event]	& 12	& 21	& 40	& 985	\\
\noalign{\smallskip}\hline
\end{tabular}
\end{center}
\end{table*}

\subsection{Physics Validation}

Once a new release is distributed to the Grid sites, a set of several physics samples is produced. Typically, a ``validation sample'' includes 10,000 events for each process, a total of 110,000 single particle events and 250,000 hard scattering events. This standard validation sample includes single muons, pions, and electrons, Standard Model processes ($\ttbar$\ production, vector boson production, B-phys\-ics), and exotic processes (\eg\ supersymmetric events and black hole production). The composition of the validation sample has been chosen to test all aspects of the event reconstruction. 

The running of the validation sample on the Grid usually exposes rare software problems in the release. It is unlikely that software bugs that appear with a frequency much lower than 1/1000 events are caught by the automatic validation procedure (1000 events is the size of the ``long'' jobs of the FCT, described in Section~\ref{Night}). This first round of production provides a feedback mechanism for the developers, who produce bug fixes before the next production cycle.

The last step before using a release for production is physics validation. A dedicated group of experts, including representatives from every detector performance (\eg\ tracking, b-tagging, and jet reconstruction groups) and physics group (\eg\ Standard Model, supersymmetry, and exotics search groups) in ATLAS, runs physics analyses on the validation samples. Their task is to verify the quality of the single object reconstruction (\eg\ jets, electrons, and muons) and the results of more complex physics analyses (\eg\ mass reconstruction in $\Zmumu$, $\Zee$, and $\ttbar$\ events). The relatively large validation samples may expose minor problems that could not be found with lower statistics, for example a shift of a few percent in the reconstruct\-ed energy. In order to properly validate each version of the software, the results from each release are typically compared to those of previous validated releases.  The software must, therefore, maintain backwards-compatibility in order to allow fair comparisons.  Shifts in file format are carefully coordinated, and maintenance of the old format is continued for as long as necessary to ensure result consistency.  The physics validation procedure is also used for checking major changes in the fast and full simulation (detector description, change in the simulation parameters, etc.).

The \Geant\ simulation has also been validated in a physics sense with all available detector data.  Combined test beam studies have proven invaluable in understanding the performance of each of the subsystems, and the standalone test beam analyses have provided crucial input towards the optimization of the simulation and choice of parameters~\cite{CTB,CTB2,MCCTB1,MCCTB2}.  In 2008, a significant sample of cosmic ray data was collected with multiple subdetectors.  The data have provided an important test of the simulation~\cite{LArCosmic}.

Although the detector simulation relies heavily on \Geant, a significant effort was put into comparing tile ca\-lo\-ri\-me\-ter test stand response with the \Fluka\ simulation tool\-kit~\cite{Flugg}.  For this comparison, the test stand geometry was translated into the \Fluka\ geometry format, and the output from the \Fluka\ simulation was translated back into a format comparable to that of \Geant.  It was eventually concluded that little would be gained by attempting a transition to \Fluka\ that could not already by gained by modifications to parameters and a different choice of physics models within \Geant.  \Fluka\ has also been used to study neutron flux and radiation levels throughout the detector~\cite{Shupe}, but many of these studies are being updated in \Geant with the high-precision neutron physics list (see Section~\ref{PhysicsList}).

Extensive efforts are underway to compare simulated data to real data and validate the output from each detector.  Thanks to the multiple detector descriptions, several analyses have already been prepared and tested to find discrepancies between the detector description of the simulation and that of the as-built detector. For example, subdetectors can be ``weighed'' in the simulation to ensure that the amount of material is within a few percent of the constructed detector.  Although the agreement with first high-energy collision data is not expected to be perfect, a great deal of experience has been gained.  The effects of modifications to \Geant\ parameters have also been studied in some detail, so that differences between data and simulation might be remedied rapidly.

Digitization algorithms have been tuned against laboratory test results, test beam data, and, where possible, cosmic ray data taken during the detector commissioning. The studies continue with the data.

%%\clearpage

\section{Summary and Conclusions}

We have presented the status of the ATLAS simulation project, including all steps from event generation to digitization.  A robust and flexible framework is required to cope with the demands of complex detector descriptions and physics models.  The software project has been prepared for data since late 2008 and is ready for data.

A variety of event generators are available to provide the user with a complete set of tools for testing new phys\-ics models.  The simulation is highly configurable to ensure maximal flexibility in the face of the uncertain challenges approaching.  The detector description itself, conditions of the detector, and many parameters used in the simulation can be modified at run time.  The digitization is also made configurable to cope with uncertainty in machine performance, detector conditions, and cavern conditions.  Three varieties of fast simulation have been made available to ease the difficulties caused by the time consumption of the full detector simulation.  They each complement the full simulation.

Generation, simulation, and digitization tasks are running continually on the Grid.  The validation program has produced a high quality simulation sample for the ATLAS experiment data.

\begin{acknowledgement}

We are greatly indebted to all CERN's departments and to the LHC project for their immense efforts not only in building the LHC, but also for their direct contributions to the construction and installation of the ATLAS detector and its infrastructure. We acknowledge equally warmly all our technical colleagues in the collaborating Institutions without whom the ATLAS detector could not have been built. Furthermore we are grateful to all the funding agencies which supported generously the construction and the commissioning of the ATLAS detector and also provided the computing infrastructure.

The ATLAS detector design and construction has taken about fifteen years, and our thoughts are with all our colleagues who sadly could not see its final realisation.

We acknowledge the support of ANPCyT, Argentina; Yerevan Physics Institute, Armenia; ARC and DEST, Australia; Bundesministerium f\"ur Wissenschaft und Forschung, Austria; National Academy of Sciences of Azerbaijan; State Committee on Science \& Technologies of the Republic of Belarus; CNPq and FINEP, Brazil; NSERC, NRC, and CFI, Canada; CERN; CONICYT, Chile; NSFC, China; COLCIENCIAS, Colombia; Ministry of Education, Youth and Sports of the Czech Republic, Ministry of Industry and Trade of the Czech Republic, and Committee for Collaboration of the Czech Republic with CERN; Danish Natural Science Research Council and the Lundbeck Foundation; European Commission, through the ARTEMIS Research Training Network; IN2P3-CNRS and Dapnia-CEA, France; Georgian Academy of Sciences; BMBF, DFG, HGF and MPG, Germany; Ministry of Education and Religion, through the EPEAEK program PYTHAGORAS II and GSRT, Greece; ISF, MINERVA, GIF, DIP, and Benoziyo Center, Israel; INFN, Italy; MEXT, Japan; CNRST, Morocco; FOM and NWO, Netherlands; The Research Council of Norway; Ministry of Science and Higher Education, Poland; GRICES and FCT, Portugal; Ministry of Education and Research, Romania; Ministry of Education and Science of the Russian Federation, Russian Federal Agency of Science and Innovations, and Russian Federal Agency of Atomic Energy; JINR; Ministry of Science, Serbia; Department of International Science and Technology Cooperation, Ministry of Education of the Slovak Republic; Slovenian Research Agency, Ministry of Higher Education, Science and Technology, Slovenia; Ministerio de Educaci\'{o}n y Ciencia, Spain; The Swedish Research Council, The Knut and Alice Wallenberg Foundation, Sweden; State Secretariat for Education and Science, Swiss National Science Foundation, and Cantons of Bern and Geneva, Switzerland; National Science Council, Taiwan; TAEK, Turkey; The Science and Technology Facilities Council and The Leverhulme Trust, United Kingdom; DOE and NSF, United States of America; Gordon and Betty Moore Foundation, United States of America.

The authors would like to thank the \Geant developers for their work and support.
\end{acknowledgement}

\bibliographystyle{atlasstylem}
\bibliography{SimEPJCPaperAtlasV6}

\end{document}